\documentclass[apj,twocolumn]{openjournal}
\usepackage{natbib} 
\usepackage[dvipsnames]{xcolor}
\usepackage{aas_macros} 
\usepackage{amssymb}
\usepackage{amsmath}
\usepackage[title]{appendix}
\usepackage{hyperref}	
\hypersetup{colorlinks=true,linkcolor=blue,citecolor=blue,filecolor=blue,urlcolor=blue}
\usepackage[caption=false]{subfig}
\usepackage{soul}                          
\usepackage{xspace}
\usepackage{booktabs}
\usepackage{savesym}
\savesymbol{tablenum}
\usepackage{siunitx}
\usepackage{pifont}
\usepackage{lipsum}
\restoresymbol{SIX}{tablenum}

\DeclareSIUnit\Msun{\ensuremath{M_\odot}}
\DeclareSIUnit\Zsun{\ensuremath{Z_\odot}}
\DeclareSIUnit\hred{\ensuremath{\textit{h}}}

\newcommand{\megatron}{{\small MEGATRON}}
\newcommand{\sphinx}{{\small SPHINX}}

\begin{document}
\title{MEGATRON: Disentangling Physical Processes and Observational Bias in the Multi-Phase ISM of High-Redshift Galaxies\vspace{-15mm}}
\author{Nicholas Choustikov$^{1*}$, Harley Katz$^{2,3}$, Alex J. Cameron$^{1}$, Aayush Saxena$^{1,4}$,\\ Julien Devriendt$^{1}$, Adrianne Slyz$^{1}$, Martin P. Rey$^{5}$, Corentin Cadiou$^{6}$, Jeremy Blaizot$^{7}$, Taysun Kimm$^{8}$,\\ Isaac Laseter$^{9}$, Kosei Matsumoto$^{10}$, and Joki Rosdahl$^{7}$}
\thanks{$^*$E-mail: \href{mailto:nicholas.choustikov@physics.ox.ac.uk}{nicholas.choustikov@physics.ox.ac.uk}}

\affiliation{$^{1}$Sub-department of Astrophysics, University of Oxford, Keble Road, Oxford OX1 3RH, United Kingdom}
\affiliation{$^{2}$Department of Astronomy \& Astrophysics, University of Chicago, 5640 S Ellis Avenue, Chicago, IL 60637, USA}
\affiliation{$^{3}$Kavli Institute for Cosmological Physics, University of Chicago, Chicago IL 60637, USA}
\affiliation{$^{4}$Department of Physics and Astronomy, University College London, Gower Street, London WC1E 6BT, United Kingdom}
\affiliation{$^{5}$University of Bath, Department of Physics, Claverton Down, Bath, BA2 7AY, UK}
\affiliation{$^{6}$Institut d'Astrophysique de Paris, Sorbonne Universites, CNRS, UMR 7095, 98 bis bd Arago, 75014 Paris, France}
\affiliation{$^{7}$Universite Claude Bernard Lyon 1, CRAL UMR5574, ENS de Lyon, CNRS, Villeurbanne, F-69622, France}
\affiliation{$^{8}$Department of Astronomy, Yonsei University, 50 Yonsei-ro, Seodaemun-gu, Seoul 03722, Republic of Korea}
\affiliation{$^{9}$Department of Astronomy, University of Wisconsin-Madison, Madison, WI 53706, USA}
\affiliation{$^{10}$Sterrenkundig Observatorium Department of Physics and Astronomy Universiteit Gent, Krijgslaan 281 S9, B-9000 Gent, Belgium}

\begin{abstract}

\noindent Now detected out to redshifts of $z\sim 14.5$, the rest-frame ultraviolet and optical spectra of galaxies encode numerous physical properties of the interstellar medium (ISM). Accurately extracting these properties from spectra remains a key challenge that numerical simulations are uniquely suited to address. We present a study of the observed ISM of galaxies in {\small MEGATRON}: a suite of cosmological radiation hydrodynamics simulations coupled to on-the-fly non-equilibrium thermochemistry, with multiple prescriptions for star formation/feedback and parsec-scale resolution; capable of directly predicting spectroscopic properties of early galaxies. We find that irrespective of feedback physics used, the ISM of high-redshift galaxies is denser, less metal enriched, and subject to higher ionization parameters and radiation fields compared to similar mass galaxies in the local Universe -- in agreement with interpretations of JWST observations. Using common observational techniques to infer bulk galaxy properties, we find that ISM gas density controls the slope of the mass-metallicity relation. Similarly, at the densities reached in some high-redshift galaxies, O32 becomes a density tracer rather than one of ionization parameter. This motivates the use of other line ratios like C43 and N43 to infer the ionization state of the gas. Finally, various feedback models populate different regions of strong-line diagnostic diagrams as the line ratios are sensitive to the feedback-modulated density-temperature structure of the ISM. Therefore, observed strong-line diagnostics can provide a strong constraint on the underlying physics of star formation and feedback in the high-redshift Universe.
\end{abstract}
\keywords{high-redshift galaxies, ISM, galaxy formation}

\section{Introduction}

The launch of JWST \citep{Gardner:2023}, has revolutionized our understanding of the interstellar medium (ISM) at high-redshift. In particular, the Near-Infrared Spectrograph (NIRSpec, \citealt{Boker:2023}) provides the capability to directly measure the spectra of galaxies in the Epoch of Reionization in the rest-frame UV-optical.

Bright emission lines in the rest-frame UV and optical can be used to understand the properties of the ISM, such as the electron density, temperature, pressure, ionization parameter, and chemical abundances \citep[see][for a review]{Kewley:2019}. Likewise, the shape of the continuum emission contains information about the age and properties of the stellar populations or accreting black holes \citep[e.g.][]{Iyer:2025}, the presence of dense neutral gas reservoirs around galaxies \citep[e.g.][]{Heintz:2025}, and electron temperatures in the ISM \citep[e.g.][]{Laseter:2024,Morishita:2024,Hayes:2025,Cataldi:2025,Pollock:2025}. The spectra can be used to infer the possible presence of exotic stellar populations \citep[e.g.][]{Cameron:2024,Cullen:2025}, reddening due to dust attenuation or nebular emission \citep[e.g.][]{Saxena:2024b,Topping:2024b,Cullen:2024}, or to discover the existence of entirely new classes of objects, such as `Little Red Dots' \citep[e.g.][]{Matthee:2024}.

Spectroscopic surveys of high-redshift galaxies  \citep[e.g.][]{Cameron:2023b,Sanders:2023,Hu:2024,Roberts-Borsani:2024,Roberts-Borsani:2025,Hayes:2025,Tang:2025,Shapley:2025,Pollock:2025} have identified four key characteristics that differentiate the ISM at high redshift from that in the local Universe:

\textbf{Harder ionizing spectra:} A large number of high ionization lines (e.g. $\rm He~{\small II}~\lambda 1640$, N~{\small IV}]~$\lambda\lambda 1483,1486$ and C~{\small IV}~$\lambda\lambda 1548,1550$) have now been observed at high-redshift \citep[e.g.][]{Bunker:2023,Castellano:2024,Topping:2024,Topping:2025b,Naidu:2025}. While these lines are often attributed to AGN \citep[e.g.][]{Feltre:2016,Laporte:2017}, this emission can also be produced by low-metallicity massive stars \citep[e.g.][]{Stark:2015}. In either case, the detection of these lines in high-redshift galaxies implies the presence of sources with hard ionizing photons \citep{Berg:2021}.

\textbf{Lower gas-phase metallicities:} Measurements of gas-phase oxygen abundance have found that galaxies of a given mass at high-redshift tend to be systematically less enriched, resulting in redshift-dependent normalizations and slopes in the mass-metallicity relation \citep[MZR, e.g.][]{Heintz:2023,Nakajima:2023,Curti:2023,Curti:2024,Morishita:2024,Chemerynska:2024,Sarkar:2025,Pollock:2025}. Furthermore, these works have found evidence for departures from the low-redshift `Fundamental Metallicity Relation', indicating that galaxies of a given stellar mass and star formation rate are also less enriched at high-redshift.

\textbf{Higher densities:} Numerous galaxies with direct determinations of electron densities have been observed \citep[e.g.][]{Isobe:2023,Reddy:2023,Abdurrouf:2024,Li:2025b,Topping:2025}. In most cases, densities are found to be higher in the early Universe. Furthermore, some objects have now been found with extreme densities $n_{\rm e}>10^5~{\rm cm}^{-3}$ \citep{Senchyna:2024,Topping:2024}, rarely seen in the local Universe except in extreme star-forming galaxies \citep{Mingozzi:2022}. Finally, recent work combining rest-frame optical and IR emission have necessitated multiple H~{\small II} regions of different density to explain the observed emission \citep{Harikane:2025,Usui:2025}. Altogether, this implies that the ISM of high-redshift galaxies is typically highly complex and inhomogeneous.

\textbf{Higher ionization parameters:} Many high-redshift galaxies are found to have line ratios consistent with large ionization parameters \citep[e.g.][]{Cameron:2023b,Mascia:2023,Nakajima:2023,Topping:2025b,Tang:2025b,Glazer:2025,Hayes:2025}, defined as the number of ionizing photons per hydrogen atom and typically traced by O32 ($\rm [O~{\small III}]\lambda5007/[O~{\small II}]\lambda\lambda3726,3728$). In particular, stacks have been used to demonstrate a correlation between redshift and O32 \citep{Sanders:2023,Roberts-Borsani:2024}, showing that these conditions become more extreme in the early Universe. Spectroscopic measurements of high-redshift galaxies reveal high emission line equivalent widths \citep{Boyett:2024,Endsley:2024} and high ionizing photon production efficiencies \citep{Atek:2024,Saxena:2024,Simmonds:2024b,Simmonds:2024,Laseter:2025,Choustikov:2025}. Together, these imply that the ISM of galaxies at high-redshift had significantly higher ionization parameters.

Finally, JWST observations reveal an over-abundance of bright galaxies in the early Universe \citep[e.g.][]{Finkelstein:2023,Donnan:2023,Leung:2023,Harikane:2024,Chemerynska:2024b} compared to what was predicted by most pre-JWST galaxy formation models \citep[see review by][]{Robertson:2022}. Possible solutions include increasing the star-formation efficiency to produce more stars (\citealt{Dekel:2023}, c.f. \citealt{Ferrara:2025}), allowing burstier star formation histories to increase scatter in the $L_{\rm UV}$-$\rm M_{\rm halo}$ relation \citep{Mason:2023,Shen:2023,Sun:2023,Kravtsov:2024}, or modifying the SFR-$L_{\rm UV}$ relation by including the nebular continuum \citep{Katz:2024b}, varying the stellar initial mass function \citep[e.g.][]{Yung:2024,Cueto:2024,Trinca:2024}, modifying the physical properties of stars \citep{Katz:2024c}, and including contributions from AGN \citep[e.g.][]{Hegde:2024}. While all of these models have been explored in terms of their effects on the UV luminosity function, their impact on the ISM remains less understood.

The most robust approach for inferring physical properties from spectra relies on using theoretical line ratios, set by quantum mechanics that are sensitive to quantities such as electron temperature or density. When this is not possible (e.g. ionization parameter), one typically relies on equilibrium photoionization models that make simplistic assumptions on star formation history and geometry \citep[see][for a review]{Kewley:2019}. One-dimensional models run with codes such as \texttt{CLOUDY} \citep{Ferland:2017,Chatzikos:2023} and \texttt{MAPPINGS} \citep{Dopita:1996,Sutherland:2018} are either used in isolation \citep[e.g.][]{Feltre:2016,Gutkin:2016}, assembled from weighted combinations of models \citep[e.g.][]{Lebouteiller:2022,Marconi:2024} or applied as a post-processing step to cosmological simulations \citep[e.g.][]{Hirschmann:2017,Vijayan:2021,Vijayan:2025,Hirschmann:2023,Katz:2019,Katz:2019b,Katz:2022c,Katz:2022d,Katz:2023,Wilkins:2023,Choustikov:2024,Nyhagen:2024,Kohandel:2024,Lovell:2024,Lovell:2025,Giovinazzo:2025}.

While this approach is very computationally efficient, it begins to break down under certain conditions. For example, the presence of turbulent gas in H~{\small II} regions can have non-negligible effects on emission line ratios, even when compared to one-dimensional models with the same average conditions \citep[see discussions in ][]{Gray:2017,Jin:2022,Xing:2026}. Secondly, non-equilibrium effects can be important in the presence of non-thermal physics such as cosmic rays \citep{Katz:2022}, or in regions where cooling timescales are sufficiently short that gas can be over-ionized \citep{Oppenheimer:2013,Katz:2022,Richings:2022,Ploeckinger:2025}. Finally, galaxy spectra integrated over multiple unresolved H~{\small II} regions can also lead to unrepresentative emission line ratios \citep{Cameron:2023}. As a result, we are motivated to explore the use of cosmological simulations with on-the-fly non-equilibrium thermochemistry that can complement photoionization models by overcoming some of these aforementioned issues.

In this work, we aim to understand the physics underpinning observed properties of the high-redshift ISM, in particular focusing on the physical and observational effects which bias our ability to extract information from spectra. To accomplish this, we use the \megatron\ simulations \citep{Katz:2025b,Rey:2025,Cadiou:2025}. These simulations employ a detailed thermochemistry network of primordial species, metals, and molecules, coupled to on-the-fly multi-frequency radiation transport and a state-of-the-art galaxy formation model, which allows us to self-consistently predict the spectra of the simulated high-redshift galaxies. In particular, we use this unique capability to conduct a direct comparison with JWST-observed high-redshift galaxies, most notably measuring all bulk galaxy properties from the mock spectra in the same way that is typically done for JWST observations. 

This paper is organized as follows. In Section~\ref{sec:full_methods}, we describe the \megatron\ simulations and outline our spectroscopic analysis approach. In Section~\ref{sec:results}, we investigate each of the four unique ISM properties of high-redshift galaxies. Finally, in Section~\ref{sec:caveats}, we discuss caveats of our approach before concluding in Section~\ref{sec:conclusions}.

\section{Methods}
\label{sec:full_methods}

\begin{figure*}
\begin{center}
    \includegraphics[width=0.85\textwidth,trim={0.0cm 0cm 0.0cm 0cm},clip]{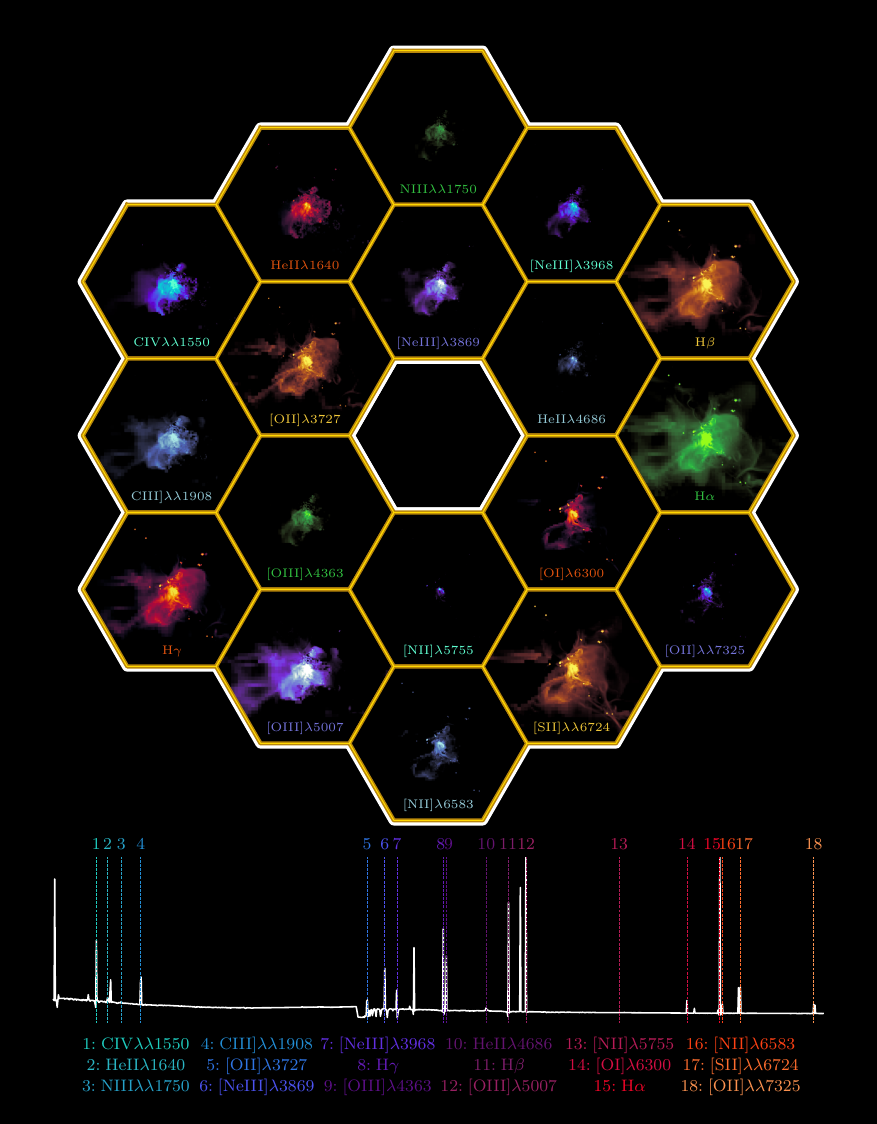}
\end{center}
    \caption{Projected UV-optical emission line maps normalized by the total $\rm H\beta$ flux for the galaxy with the largest specific star-formation rate in the bursty star formation simulation. Low surface-brightness gas has been brightened. The width of each panel is twice the Virial radius of the host halo. On the bottom, we show the total intrinsic spectrum (in arbitrary $\log\ f_{\nu}$ units), highlighting the relative strengths of each line. This galaxy has a stellar mass of $10^{7.9}M_{\odot}$ at $z=8.93$.}
    \label{fig:hero}
\end{figure*}

In this section, we provide an overview of our methodology: describing the variety of simulations and observational techniques used in the present work.

\subsection{Galaxy Evolution with Non-Equilibrium Thermochemistry}\label{sec:method}

For the purpose of this work, we use data from the high-redshift suite of the \megatron\ simulations \citep{Katz:2025b}. These are four separate high-resolution cosmological radiation hydrodynamical simulations run with {\small RAMSES-RTZ}\footnote{A fork of {\small RAMSES} \citep{Teyssier:2002} and {\small RAMSES-RT} \citep{Rosdahl:2013}.} \citep{Katz:2022b} and various models (described in Section \ref{sec:sims}) for star formation and feedback. In each simulation, we follow the evolution of an early-forming Milky Way mass halo in a relative over-density to a final redshift of $z=8.5$. The simulations use a constant-comoving resolution scheme, which refines cells based on a quasi-Lagrangian approach as well as requiring that the Jeans length is resolved by at least 4 cells. As a result, at $z\sim8.5$, the maximum physical resolution is $\sim5~{\rm pc}~h^{-1}$, compared to $\sim1~{\rm pc}~h^{-1}$ when the first stars form at $z\sim 30$. For full numerical details, we direct the interested reader to \citet{Katz:2024,Katz:2025b}. Below we summarize the numerical methods relevant to the ISM.

Radiation and hydrodynamics are coupled on-the-fly to a non-equilibrium chemistry solver \citep{Katz:2022b} for primordial species, metals, and molecules; following the evolution of H~$\rm \small I-II$, He~$\rm \small I-III$, $\rm e^-$, C~$\rm \small I-VI$, N~$\rm \small I-VII$, O~$\rm \small I-VIII$, Ne~$\rm \small I-X$, Mg~$\rm \small I-X$, Si~$\rm \small I-XI$, S~$\rm \small I-XI$, Fe~$\rm \small I-XI$, $\rm H_2$, and CO. All metal ionization states not followed are assumed to be in collisional ionization equilibrium. Gas heating and cooling processes include photo-heating, photoelectric heating, H$_2$ formation heating, H$_2$ excitation/dissociation heating, compton heating/cooling, primordial cooling, H$_2$ cooling, CO cooling, dust recombination cooling, dust-gas collisional cooling (or heating), and metal line cooling (\citealt{Katz:2022}; \citealt{Katz:2024}, \textcolor{blue}{2025}). We assume that the dust-to-gas mass ratio varies with metallicity following \cite{Remy-Ruyer:2014} using the BARE-GRS dust composition model of \cite{Zubko:2004}.

Star-formation is modeled based on the local, turbulent properties of the gas \citep{Padoan:2011,Federrath:2012,Kimm:2017,Rosdahl:2018}. Specifically, stars can form in a gas cell if the following criteria are fulfilled: it has a density $n_{\rm H}>10~{\rm cm^{-3}}$, is a local density maximum, if the fluid flow is locally convergent, and if the turbulent Jeans length is smaller than the cell width. If these criteria are fulfilled, star-formation proceeds using a Schmidt law \citep{Schmidt:1959} with an efficiency per free-fall time that depends on the turbulent properties of the gas \citep{Padoan:2011,Federrath:2012}. For Population~II star particles, we adopt a Kroupa IMF \citep{Kroupa:2001} with a maximum mass of $300~M_{\odot}$. Radiation from Population~II stars\footnote{Population~III stars are also modeled explicitly \citep{Storck:2025} However, all of the galaxies considered in this work are already completely dominated by Population~II stars.} is taken from {\small BPASS v2.2.1} \citep{Eldridge:2017,Stanway:2018}. Stellar winds, core-collapse SN, and type Ia SN are modeled following \cite{Agertz:2021} with minor modifications \citep{Katz:2025b}.

\subsubsection{Variations to Sub-Grid Physics}\label{sec:sims}

The four high-redshift \megatron\ simulations differ in their subgrid physics in ways that loosely follow proposed solutions to the over-abundance of high-redshift bright galaxies. Here we summarize the key differences between the four models.

\begin{itemize}
    \item \textbf{Efficient Star Formation:} This model represents the fiducial simulation, as described above. As this simulation has the weakest feedback compared to the others, it has the highest conversion efficiency of gas into stars.
    \item \textbf{Bursty Star Formation:} In this simulation, the energy injected by all SNe has been increased by a factor of 5 compared to the efficient star formation model. As a result, stellar masses are regulated and star-formation histories become significantly more stochastic and bursty.
    \item \textbf{Variable IMF:} In this simulation, the high-mass slope of the IMF is allowed to vary according to the local gas density and metallicity according to \cite{Marks:2012}. Furthermore, a metallicity-dependent fraction of high-mass stars \citep{Kobayashi:2006} explode as hypernovae with a mass-dependent energy \citep{Nomoto:2006}. While the IMF is allowed to vary, we cap the maximum mass at $120~{\rm M_{\odot}}$ and use SEDs from {\small STARBURST99} \citep{Leitherer:1999}. As a result, stellar populations can form with very low mass-to-light ratios.
    \item \textbf{High $\boldsymbol{\epsilon_{\rm ff}}$, HN:} In this simulation, stars are formed in lower-density gas with an efficiency per free-fall time of $100\%$. Like the variable IMF model, a fraction of high-mass stars are able to explode as hypernovae \citep{Kobayashi:2006}, with variable energy \citep{Nomoto:2006}. As a result, this model regulates stellar masses significantly better and produces a less-dense ISM with bursty star-formation histories.
\end{itemize}

\subsubsection{Self-Consistent Synthetic Observations}

\begin{figure*}
    \includegraphics[width=\textwidth]{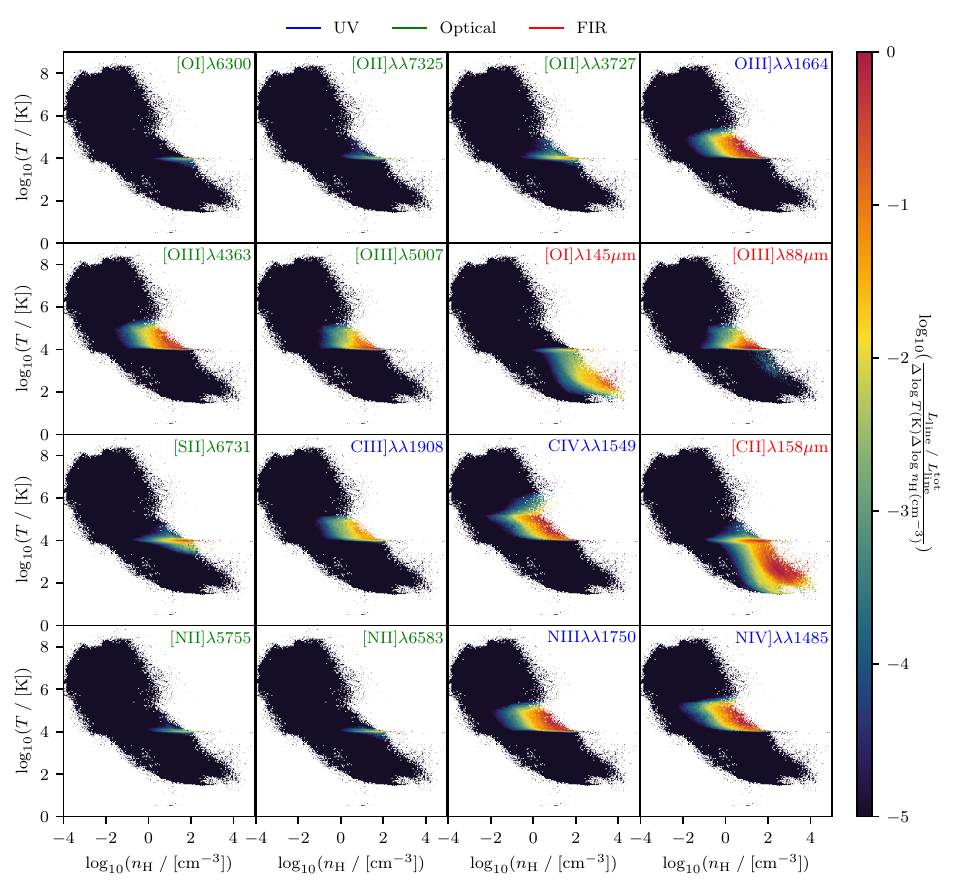}
    \caption{Temperature-density phase diagrams for an example $\sim10^{8.5}M_{\odot}$ stellar mass galaxy at $z=8.6$ in the efficient star formation run, with all cells coloured by their contribution to the total luminosity of a given emission line. We include UV, optical, and FIR lines in blue, green, and red respectively. No two emission lines are produced by the same gas, introducing potential biases when emission lines are combined to make measurements of ISM properties.}
    \label{fig:all_lines}
\end{figure*}

\begin{figure}
    \includegraphics{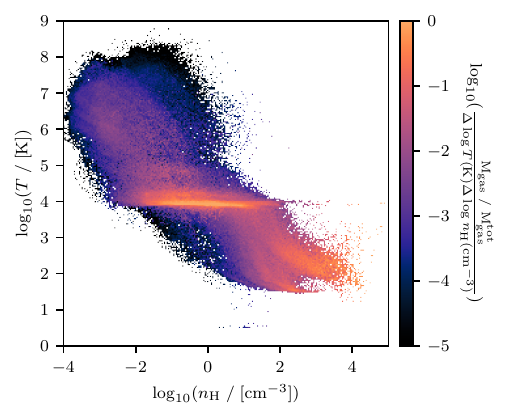}
    \caption{Temperature-density phase diagrams for an example $\sim10^{8.5}M_{\odot}$ stellar mass galaxy at $z=8.6$ in the efficient star formation run (as in Figure~\ref{fig:all_lines}), with all cells coloured by their contribution to the total gas mass.}
    \label{fig:met_pd}
\end{figure}

\begin{figure*}
\begin{center}
    \includegraphics[width=0.8\textwidth,trim={0.0cm 0cm 0.0cm 0cm},clip]{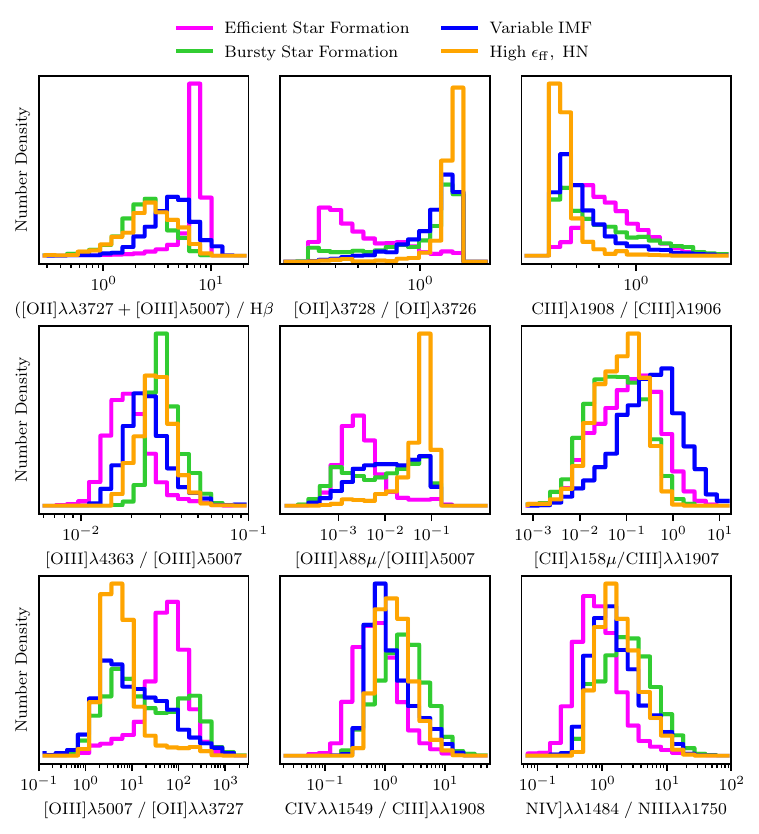}
\end{center}
    \caption{Distributions of common UV, optical, and FIR emission line ratios for each {\footnotesize MEGATRON} simulation, highlighting the impacts that different subgrid prescriptions for star-formation/feedback have on the ISM, and therefore on its spectroscopic properties.}
    \label{fig:line_ratio_hist}
\end{figure*}

The key benefit of evolving the full non-equilibrium thermochemistry of the ISM is that we are able to predict the intrinsic spectra of galaxies, accounting for the nebular continuum, nebular line emission, and the stellar continuum. Each mock spectrum considers all gas cells within $25\%$ of the virial radius of a particular halo, designed to capture the entirety of the ISM of the central galaxy (see Appendix \ref{app:rvir}). Line emissivities and the nebular continuum are computed using \texttt{PyNeb} \citep{Luridiana:2015} with atomic data for collisional emission from Version~10 of the \texttt{CHIANTI} database (\mbox{\citealt{Dere:1997};} \citealt{Dere:2019}; \citealt{DelZanna:2021}). Finally, the stellar continuum is computed for each star particle based on the SEDs used in the simulation.

Like all numerical simulations, we are limited by finite numerical resolution. At high densities, the Str{\"o}mgren spheres of star particles are sometimes unresolved \citep[see discussion in][]{Choustikov:2024}. To correct for this, we replace the emission of cells that contain unresolved Strömgren spheres with \texttt{CLOUDY} models\footnote{\texttt{RAMSES-RTZ} has been shown to agree with \texttt{CLOUDY} to within $10\%$ for idealized Str{\"o}mgren sphere models \citep{Katz:2024}.} \citep{Ferland:2017} using the local gas conditions and star particle properties \citep{Katz:2025b}. Finally, for the purpose of this work we ignore the effect of dust attenuation, scattering, and emission on the synthetic spectra of \megatron\ galaxies (see discussion in Section \ref{sec:caveats}).

Figure~\ref{fig:hero} shows example emission line maps of a highly star-forming galaxy. A 1D spectrum for the same galaxy is also shown. While each emission line primarily originates from star-forming regions, they are not entirely co-spatial. For example, the Balmer series and lines like $\rm [O~{\small III}]~\lambda 5007$ can be seen to also trace the large-scale gas distribution around the galaxy, while high-ionization lines like $\rm C~{\small IV}~\lambda\lambda1549$ and $\rm He~\small{II}~\lambda1640$ and auroral lines like $\rm [N~\small {II}]~\lambda5755$ and $\rm [O~\small {II}]~\lambda\lambda7325$ primarily trace the H~{\small II} regions.

Figure~\ref{fig:all_lines} shows temperature-density phase diagrams for a typical galaxy, weighted by emission line luminosity. UV lines are labeled in blue, optical lines are in green and FIR lines are given in red. While certain phases \citep[e.g. diffused ionised gas,][]{Mcclymont:2024} may contribute to the luminosity of each line, we find that no two emission lines are produced by gas of the same temperature/density distribution -- therefore tracing different physical conditions. For example, $\rm [O~{\small III}]~\lambda 4363$ is produced in preferentially hotter gas than $\rm [O~{\small III}]~\lambda 5007$. As a result, electron temperature measurements from these lines are typically biased high, skewing the inferred metallicity low (\citealt{Cameron:2023}, see also discussion in \citealt{Kewley:2019}). To further highlight this point, Figure~\ref{fig:met_pd} shows another temperature-density phase diagram of the same galaxy, now weighted by gas mass fraction -- displaying the multi-phase ISM structure observed in \megatron\ galaxies \citep{Katz:2024}. Here, we find that line-emitting gas often represents a small mass fraction of the ISM, and that fluctuations in physical conditions on the scale of the galaxy can be important \citep{Gray:2017,Jin:2022,Cameron:2023}.

A simple constant-density or isobaric photoionization model cannot sample the full distribution of ISM properties that we find in these simulations. To highlight differences between observed properties for each simulation, Figure \ref{fig:line_ratio_hist} presents histograms of common UV, optical, and FIR emission line ratios for each run. We find that in the case of each ratio, every simulation produces a different distribution. In particular, density- and temperature-dependent ratios show the largest variance between each run. These differences are driven by changes to sub-grid star formation/stellar feedback physics, which directly modify the density-temperature structure of the ISM \citep[see discussion in][]{Katz:2024}. While each model has its own limitations, variations to the sub-grid physics allow for an ISM parameter exploration that would be otherwise impossible with a single simulation.  These variations demonstrate that it may be possible to use an ensemble of observed emission line ratios to constrain sub-grid physics. We leave this exploration to future work. Altogether, the complex nature of galaxies in the \megatron\ simulations make them a unique tool for understanding the physics that give rise to observed galaxy spectra.

\begin{table*}
    \centering
    \caption{A list of all emission line diagnostic ratios used in this work. Throughout this work, all emission line ratios are considered in log base 10.}
    \begin{tabular}{lll}
         Diagnostic Name & Emission Line Ratio & Property Traced \\
         \hline
         R3 & $[{\rm O~{\small III}}]\lambda 5007/{\rm H\beta}$ & Metallicity \\
         R23 &  $([{\rm O~{\small III}}]\lambda\lambda 4959,5007 + [{\rm O~{\small II}}]\lambda\lambda 3726,3728)/{\rm H\beta}$ & Excitation \\
         S2 & $[{\rm S~{\small II}}]\lambda6731 / [{\rm S~{\small II}}]\lambda6719$ & Density \\
         O2 & $[{\rm O~{\small II}}]\lambda3728 / [{\rm O~{\small II}}]\lambda3726$ & Density \\
         O2O2 & $[{\rm O~{\small II}}]\lambda\lambda 7319,7320,7330,7331 / [{\rm O~{\small II}}]\lambda\lambda 3726,3728$ & Temperature \\
         O32 & $[{\rm O~{\small III}}]\lambda5007 / [{\rm O~{\small II}}]\lambda\lambda 3727$ & Ionization parameter\\
         O3O3 & $[{\rm O~{\small III}}]\lambda 4363 / [{\rm O~{\small III}}]\lambda 5007$ & Temperature \\
         O3 & $[{\rm O~{\small III}}]\lambda 88\mu / [{\rm O~{\small III}}]\lambda 52\mu$ & Density\\
         C3 & ${\rm C~{\small III}}]\lambda 1908/ {\rm C~{\small III}}]\lambda 1906$  & Density\\
         C43 & ${\rm C~{\small IV}}\lambda\lambda 1550/ {\rm C~{\small III}}]\lambda\lambda 1907$ & Ionization parameter\\
         N4 & ${\rm N~{\small IV}}]\lambda 1486/ {\rm N~{\small IV}}]\lambda 1483$  & Density\\
         N43 & ${\rm N~{\small IV}}]\lambda\lambda 1483,1486/ {\rm N~{\small III}]}\lambda 1750 $ & Ionization parameter\\
         \hline
    \end{tabular}
    \label{tab:line_ratios}
\end{table*}

\subsubsection{LyC Escape Fractions}
Lyman continuum (LyC) escape fractions ($f_{\rm esc}^{\rm LyC}$) are computed for each halo resolved by more than 1,000 dark matter particles that hosts at least one star particle. The $f_{\rm esc}^{\rm LyC}$ calculation is performed in post-processing with the Monte Carlo radiation transfer code {\small RASCAS} \citep{rascas2020}. For each star particle, we propagate monochromatic photon packets at 900~\AA\ until they reach either $0.75R_{\rm vir}$\footnote{Note that we have adopted 75\% of the virial radius as the escape radius rather than at $R_{\rm vir}$ due to a data format incompatibility between {\small RASCAS} and {\small MEGATRON} outputs. We have tested this parameter for a few haloes and found our results to be well converged as most absorption occurs in the ISM and inner CGM.}, or they are absorbed by neutral hydrogen or dust. The number of photon packets is calculated as $100\times$ the number of star particles with a minimum of $10^5$ and maximum of $10^7$ photon packets per halo. The photon packets are randomly distributed to star particles based on a multinomial distribution, weighted by the LyC luminosity of each star particle. The dust model used in post-processing follows exactly that used when running the simulation --- it is a combination of the BARE-GR-S composition of \cite{Zubko:2004} with the dust-to-gas mass ratio as a function of metallicity from \cite{Remy-Ruyer:2014}. The escape fractions quoted in this work represent angle-averaged values, i.e. the quantity needed for reionization calculations, rather than line-of-sight values, which is what an observer would see. The latter will be considered in future work.

\subsection{Observational Methods}\label{sec:obsM}

As our goal is to make the best possible comparison to high-redshift observational data, a unique aspect of our work is that we measure galaxy properties from the mock spectra and photometry, rather than relying on the true value from the simulation. Here we outline the methods that we use, inspired by typical observational approaches.

\subsubsection{ISM Properties from Emission Line Diagnostics: PyNeb}\label{sec:pyneb}

We employ a combination of strong emission line ratio diagnostics, defined in Table \ref{tab:line_ratios} \citep[see also][]{Kewley:2019}, to infer properties of the simulated galaxies and compare with observations. We focus on line ratios from the same element in order to limit uncertainties due to chemical yields that are innate to all numerical simulations \citep[e.g.][]{Buck:2021}. 

More specifically, rather than using the true electron number density ($n_{\rm e}$), electron temperature ($T_e$), gas phase metallicity, or ionization, we measure these quantities using emission line ratios. This allows us to make the most direct comparisons with observational inferences. For example $n_{\rm e}$ is measured in different ionization zones using S2, O2, O3, and N4. Likewise, $T_e$ is estimated directly with auroral lines (e.g. [O~{\small III}]~$\lambda$4363) and a corresponding strong line (e.g. [O~{\small III}]~$\lambda$5007). In practice, we use the \texttt{GetTemDen} function of \texttt{PyNeb} \citep{Luridiana:2015} with atomic data for collisional emission from Version 10 of the \texttt{CHIANTI} database \citep{Dere:1997,DelZanna:2021} to measure these quantities. As temperature and density measurements are rare for the same observed high-redshift galaxy, we follow a relatively common observational procedure of assuming a temperature when measuring a density and similarly assume a density when measuring temperature. We adopt $T_{\rm e}\sim1.5\times10^4~{\rm K}$ for density measurements and $n_{\rm e}\sim300~{\rm cm}^{-3}$ for temperature measurements.

\subsubsection{Oxygen Abundances from the Direct Method}\label{sec:DM}

To compute gas-phase oxygen abundance, we assume that the vast majority of oxygen emission is coming from the low and intermediate ionization zones, giving $\rm \frac{O}{H} = \frac{O^+}{H^+} + \frac{O^{++}}{H^+}$ as follows:
\begin{multline}
    \frac{\rm O}{\rm H} = \frac{\epsilon_{\rm H\beta}(T_{\rm [O~{\small II}]},n_{\rm e,{\rm [C~{\small III}]}})}{\epsilon_{\rm [\rm O~{\small II}]}(T_{\rm [O~{\small II}]},n_{\rm e,{\rm [C~{\small III}]}})}\cdot\frac{L([{\rm O~{\small II}}]\lambda\lambda3726,3728)}{L({\rm H\beta})} \\
    + \frac{\epsilon_{\rm H\beta}(T_{\rm [O~{\small III}]},n_{\rm e,{\rm [C~{\small III}]}})}{\epsilon_{\rm [\rm O~{\small III}]}(T_{\rm [O~{\small III}]},n_{\rm e,{\rm [C~{\small III}]}})}\cdot\frac{L([{\rm O~{\small III}}]\lambda 5007)}{L({\rm H\beta})}.
\end{multline}

Ionization corrections for metallicity are expected to be small because the neutral states of H and O are highly coupled via charge-exchange reactions and stars typically do not emit significant quantities of photons hard enough to create a substantial amount of O$^{+++}$ \citep{Berg:2021}.

Observationally, deriving $T_{\rm [O~{\small II}]}$ is difficult due to the fact that the $\rm O^+$ auroral lines at 7320,7330\AA\ drop out of NIRSpec coverage at $z\sim 6.2$ and are extremely weak for a highly ionized ISM. Therefore, while we do measure this temperature directly (see the discussion in Section \ref{sec:temps}), we choose to compute $T_{\rm [O~{\small II}]}$ from $T_{\rm [O~{\small III}]}$ based on an empirical relation. While several such relations have been presented \citep{Izotov:2006,Pilyugin:2012,Cataldi:2025}, we use Equation~3 of \cite{Cameron:2023}, which was derived from an isolated galaxy simulation also run with \texttt{RAMSES-RTZ}, and corrects for temperature inhomogeneities in the ISM. Nonetheless, this choice will not significantly impact our results because $\rm O^{++}$ typically dominates the ionization state of oxygen in galaxies with direct $\rm [O~{\small III}]~\lambda 4363$ detections \citep{Andrews:2013,Curti:2017,Curti:2020,Curti:2023,Laseter:2022}.

Finally, when inferring metallicity, we use electron number densities computed from the $\rm C^{++}$ doublet\footnote{In cases where unphysical $\rm C~III]$ ratios are measured (driven by particularly low gas densities as well as low $\rm C~III]$ and $\rm [O~III]$ equivalent widths), we assume a fiducial density of $n_{\rm e}\sim300~{\rm cm}^{-3}$ in order to be consistent with our determinations in Section \ref{sec:pyneb}.} as these can be detected at high redshift \citep{Topping:2025} and better trace the gas phases where $\rm [O~{\small III}]~\lambda5007$ originates (see Figure~\ref{fig:all_lines}). In contrast, we find that using $\rm S^{+}$ densities often give us incorrect results, due to the fact that this density measurement is tracing a different ionization zone \citep[see discussion in][]{Berg:2021}. The efficacy of gas-phase metallicity measurements is presented in Appendix~\ref{app:direct_method}.

\subsubsection{SED Fitting with BAGPIPES}\label{sec:bagpipes}

As our goal is to make a direct comparison with typical observational approaches, we derive bulk stellar population properties (i.e. total stellar mass) by performing SED fitting on the photometry of all \megatron\ galaxies with M$_{\rm UV}\leq -15$ with the public version of \texttt{BAGPIPES} \citep{bagpipes}, using an approach similar to that outlined in \cite{Choustikov:2025}. To accomplish this, we first compute photometric magnitudes in numerous standard JWST filters (F115W, F150W, F200W, F277W, F335M, F356W, F410M, and F444W) by integrating over each intrinsic SED with \texttt{SEDPY} \citep{sedpy}. Next, we follow the `best case' scenario by fixing the object redshift to the known value and ignoring the presence of dust \citep[see also][]{Narayanan:2024,Cochrane:2025}.

We use the 2016 stellar population synthesis models from the BCO3 templates\footnote{While using consistent SPS models such as \texttt{BPASS} would potentially provide more accurate results, we chose to use the default setup of \texttt{BAGPIPES} to mimic uncertainties inherent to inferring stellar masses from observed data.} \citep{Bruzual:2003,Chevallard:2016} with a \cite{Kroupa:2001} IMF. Nebular emission (both continuum and line) from these stellar populations is modeled using the 2017 version of \texttt{CLOUDY} \citep{Ferland:2017}. Next, we use a non-parametric star formation history (SFH), following the continuity prior approach presented in \cite{Leja:2019} with time bins set at $t_{\rm bins} = [0,5,10,25,50,100,{\rm min}(250,t_z)]~{\rm Myr}$ where $t_z$ is the age of the Universe at redshift $z$. Furthermore, we allow the non-parametric model to better reconstruct a more bursty SFH by using $\sigma=1\;{\rm and}\;\nu=2$ in the Student t-distribution priors \citep{Tacchella:2022}. This SFH prescription is chosen as it was recently shown to reproduce the stellar masses of {\small SPHINX$^{20}$} galaxies well \citep{Cochrane:2025}. Our remaining model priors are as follows: we use uniform priors for mass and metallicity of $\log_{10}(M_*/M_{\odot})\in [0,10]$ and $\log_{10}(Z/Z_{\odot})\in[-3,1]$ and for simplicity take a fixed ionization parameter of $\log_{10}U = -2$. This simplification will not severely impact our results, as this approach was also shown to recover the intrinsic ionizing luminosity of {\small SPHINX$^{20}$} galaxies well \citep{Choustikov:2025}.

We present the efficacy of stellar mass recovery for \megatron\ galaxies in Appendix \ref{app:bagpipes}. In general, we find that below a redshift of $z\sim10.5$, the masses of galaxies in the efficient SF, bursty SF, and high $\epsilon_{\rm ff}$ are well recovered, though with a slight tendency to over-predict the masses of the lowest mass objects \citep{Choe:2025}. Next, we find that the stellar masses of the variable IMF run are typically over-predicted by a factor of $\sim2$, in accordance with the fact that a more standard IMF was used for the SED fitting, whereas variable IMF models typically exhibit lower mass-to-light ratios for young stellar populations. However, at redshifts above $z\sim10.5$, the scatter in stellar mass increases significantly, because the Balmer break redshifts beyond the F444W filter. Here, stellar masses are typically incorrect by an order of magnitude, highlighting the need for MIRI data in studies attempting to apply SED fitting to very high-redshift galaxies \citep[e.g.][]{Leung:2024}. In all cases, we find that the residuals are correlated with the amount of burstiness in the recent SFH, with galaxies with higher ${\rm SFR}_{10}/{\rm SFR}_{10}$ having up-scattered stellar masses. While outshining can in principle be important \citep{Narayanan:2024,Harvey:2025}, these galaxies are not sufficiently massive for it to matter in this case.

\subsubsection{Ionizing Photon Production Efficiency}

To compute ionizing photon production efficiencies ($\xi_{\rm ion}$), we follow a common approach and assume Case~B recombination with $T_{\rm e}=10^4~{\rm K}$, $n_{\rm e}=100~{\rm cm}^{-3}$, $f_{\rm esc} = 0$ so that:
\begin{equation}
    \xi_{\rm ion}\;/\;[{\rm Hz}/{\rm erg}] = \frac{7.3\times 10^{11} L({\rm H\alpha})}{L_{1500,\rm int}},
\end{equation}
where $L(\rm H\alpha)$ is the flux of the $\rm H\alpha$ line and $L_{1500,\rm int}$ is the intrinsic UV luminosity density \citep[e.g.][]{Maseda:2020}. When comparing to observations at $z\gtrsim 7$ where $\rm H\alpha$ drops out, we instead use $\rm H\beta$ with the case B result that $L({\rm H\alpha}) = 2.863L({\rm H\beta})$. This efficiency should be corrected for the escape fraction of LyC radiation. However, to best compare with observational inferences, we ignore this effect (see discussion in \citealt{Saxena:2024}).

\section{Results}\label{sec:results}

In this section, we explore the physics that governs the ISM conditions in the \megatron\ simulations, exploring each of the key characteristics of the high-redshift ISM. To best compare with observations, we consider only galaxies that have M$_{\rm UV}<-15$.

\begin{figure*}
\includegraphics[width=\textwidth,trim={0.0cm 0cm 0.0cm 0cm},clip]{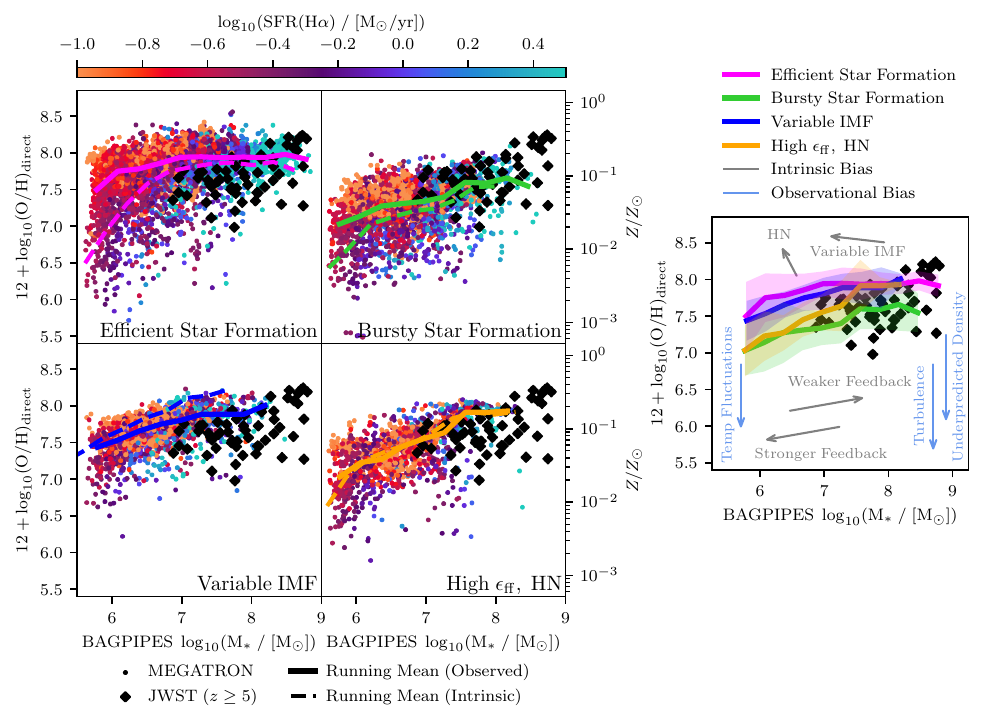}
    \caption{Mass-metallicity relations for each of the high-redshift {\footnotesize MEGATRON} simulations. 
    Left: Observed gas-phase metallicity as a function of observed stellar mass, coloured by the $\rm H\alpha$-derived star formation rate for all galaxies in {\footnotesize MEGATRON} with M$_{\rm UV}\leq-15$. We show comparisons with observational data of galaxies with direct $\rm [O~{\small III}]~\lambda 4363$ measurements above redshift $z = 5$ \protect\citep{Curti:2020,Curti:2023,Arellano-Cordova:2022,Arellano-Cordova:2025,Cameron:2023c,Morishita:2024,Pollock:2025}. Running means for the observed (solid) and intrinsic (dashed) {\footnotesize MEGATRON} data are shown as coloured lines. Observational measurements of these quantities tend to imprint a slight bias on the MZR, by flattening the slope of the relation. Right: We show comparisons between the running mean of each simulation with the same observational data. We also indicate the impact of key physical processes, highlighting intrinsic biases (e.g. different feedback models) in gray and observational biases (e.g. temperature/density fluctuations) in blue.}
    \label{fig:mzr}
\end{figure*}

\subsection{Low Metallicities}\label{sec:mets}

We begin by exploring the first key physical property of the high-redshift ISM: investigating the metallicity properties of \megatron\ galaxies.

\subsubsection{The Mass-Metallicity Relation}

The mass-metallicity relation (MZR: \citealt{Lequeux:1979,Tremonti:2004,Lee:2006}) and fundamental metallicity relation (FMR: \citealt{Ellison:2008,Mannucci:2010,Yates:2012}) have strong diagnostic power over key galaxy formation physics such as inflows, outflows, and enrichment \citep[see review by ][]{Maiolino:2019}. With JWST, we can now infer the MZR and FMR at high redshifts directly from observations.

Initially, \cite{Heintz:2023} and \cite{Nakajima:2023} both found evidence for evolution in the normalization and slope of the low-mass MZR towards higher redshifts, producing an MZR that was offset from low-redshift galaxies \citep{Curti:2020}. Since then, other studies have confirmed the apparent offset in the MZR normalization, though much debate still exists about the slope of the MZR \citep{Heintz:2023,Nakajima:2023,Curti:2024,Morishita:2024,Chemerynska:2024,Sarkar:2025}. This is of particular interest, as simulations \citep[e.g.][]{Ma:2016,Torrey:2019,Langan:2020,Dubois:2021,Pallottini:2022,Ucci:2023,D'Silva:2023,Marszewski:2024} and chemical evolution models \citep[e.g.][]{Kobayashi:2023} often struggle with the normalization and slope of the low-mass end of the high-redshift MZR \citep[e.g.][]{Curti:2024}.

We show the observed mass-metallicity relation for each of the \megatron\ simulations in Figure~\ref{fig:mzr}. On the left, we present the full dataset of mock-observed \megatron\ galaxies with M$_{\rm UV}\leq~-15$, showing direct-method metallicities as a function of \texttt{BAGPIPES} stellar mass, coloured by $\rm H\alpha$-derived star formation rates \citep{Kennicutt:1998}. We also show running means for the observed and intrinsic (i.e. true stellar mass and mass-weighted metallicity) MZRs of each simulation as solid and dashed lines respectively. Observational data for galaxies at $z\geq 5$ with direct-method metallicities are given as black diamonds \citep{Curti:2020,Curti:2023,Arellano-Cordova:2022,Arellano-Cordova:2025,Cameron:2023c,Morishita:2024,Pollock:2025}. On the right, we show how the observed running means from each simulation compare to each other alongside observed data. We also provide indications as to the effect of various intrinsic (e.g. feedback approach, gray) and  observational (e.g. temperature fluctuations, blue) processes which can affect the observed MZR. We find that each of the simulations produce a different overall mass-metallicity relation, confirming that stellar/feedback physics has a strong impact on the shape of the MZR \citep[see discussion in][]{Maiolino:2019}. Nevertheless, all simulations are able to reproduce the full scatter at the low-mass end of the observed MZR.

Furthermore, we can compare the different simulations to understand the impact of each physical prescription choice on the shape of the MZR. First, we note that the efficient SF, bursty SF, and variable IMF runs all show characteristically shallower MZRs than the high $\epsilon_{\rm ff}$ simulation. This is likely due to the fact that compared to the other three models, the high $\epsilon_{\rm ff}$ run preferentially forms stars in lower-density gas. This means that stellar feedback in the high $\epsilon_{\rm ff}$ is better coupled to the ISM, driving more powerful outflows that are capable of ejecting enriched gas further out of low-mass galaxies (and limiting subsequent star formation), thus producing a steeper MZR. Next, considering the bursty SF run, we find that a stronger feedback prescription for the same stellar populations tends to reduce the overall normalization of the MZR, by strengthening outflows and removing enriched gas -- impacting both the mass and abundance of a galaxy at later times. In contrast, the variable IMF simulation produces more massive stars for a given stellar population. Such massive stars will drive stronger feedback and reduce the final mass of the galaxy (measured after mass-loss due to winds/SNe) while also increasing the gas-phase metallicity of each galaxy since massive stars yield more metals. The presence of hypernovae can further exacerbate this process, by injecting comparatively more oxygen than supernovae \citep[e.g.][]{Nomoto:2006}, while also increasing the strength of feedback and thus reducing the final stellar masses of galaxies.

Finally, the turbulent structure of the ISM will make H~{\small II} region properties vary across a galaxy, modifying the measured oxygen abundance \citep[e.g.][]{Cameron:2023,Harikane:2025,Usui:2025}. Specifically, temperature fluctuations tend to bias the $\rm [O~{\small III}]~\lambda4363$-derived electron temperature high (see Figure~\ref{fig:all_lines}), thus driving the measured metallicity low \citep{Kewley:2019}. Similarly, density fluctuations tend to bias density measurements towards the lower-density gas\footnote{As gas in very dense regions begins to be collisionally de-excited, emission from lower-density regions begins to dominate.}, therefore skewing metallicity measurements low.

Comparing the observed running means for each \megatron\ simulation, we find that for simulations with a particularly dense ISM, the observed MZR tends to be marginally shallower than the intrinsic relation. Tests of metallicity and stellar mass measurements (see Appendices \ref{app:direct_method} and \ref{app:bagpipes} respectively) show that galaxy stellar masses are typically over-predicted at the low mass end (albeit less so for the high $\epsilon_{\rm ff}$ simulation). Furthermore, while the direct method recovers the line-weighted gas-phase metallicity well, it struggles in simulations with denser gas to accurately infer the mass-weighted metallicity, preferentially over-predicting the abundance of each galaxy. Overall, both of these processes will tend to produce a shallower MZR slope in galaxies with a dense ISM. {This may help to explain the apparent discrepancy in MZR slopes between simulations and observational results.}

\subsubsection{Observed Temperature Structure of the ISM}\label{sec:temps}
\begin{figure*}
    \includegraphics[width=\textwidth,trim={0.0cm 0cm 0.0cm 0cm},clip]{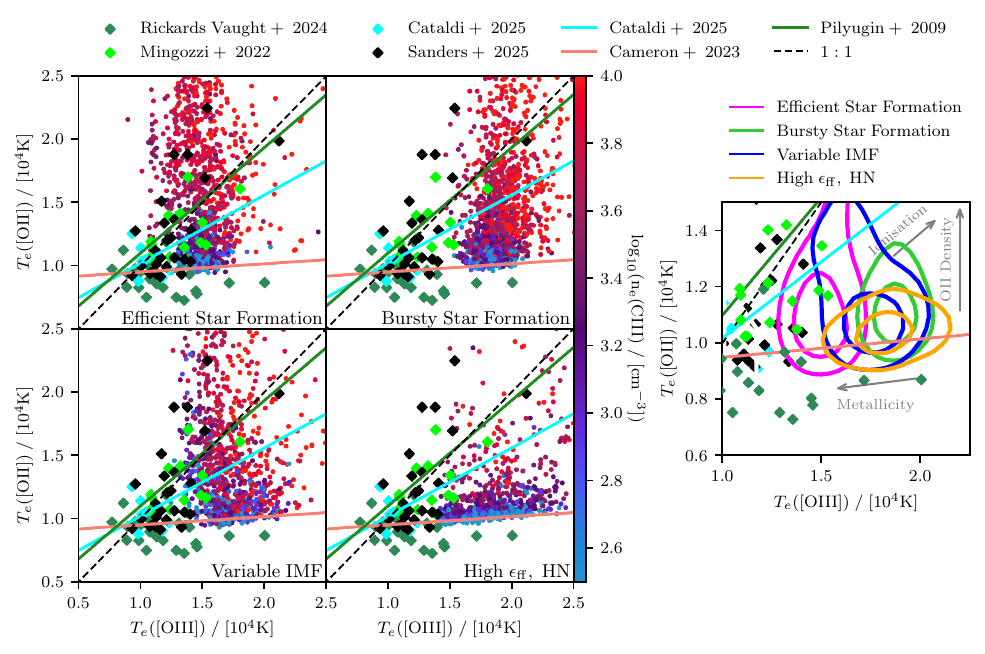}
    \caption{Observed O$^+$ and O$^{++}$ electron temperatures as measured using auroral line ratios ($\rm [O~{\small II}]~\lambda\lambda7325/[O~{\small II}]~\lambda\lambda3727$ and $\rm [O~{\small III}]~\lambda4363/[O~{\small III}]~\lambda5007$ respectively) for all galaxies in {\footnotesize MEGATRON} with M$_{\rm UV}\leq-15$. Left: $T_{\rm [O~{\small II}]}$ measurements as a function of $T_{\rm [O~{\small III}]}$ for all galaxies in each {\footnotesize MEGATRON} simulation, coloured by the electron density measured from the C3 ratio. Broadly, we find that for low-density, lower-ionization gas, $T_{\rm [O~{\small II}]}$ is fairly constant, growing rapidly with increased densities. We provide comparisons with similarly derived observational data of low-redshift analogue galaxies (lime: \protect\citealt{Mingozzi:2022}), local $\rm H~\small II$ regions (green: \protect\citealt{Rickards-Vaught:2024}), and intermediate redshift (cyan: \protect\citealt{Cataldi:2025}) and high redshift (black: \protect\citealt{Sanders:2025}) JWST samples. We also provide modeled relations (green: \protect\citealt{Pilyugin:2009}, blue: \protect\citealt{Cataldi:2025}, salmon: \protect\citealt{Cameron:2023}) and the one-to-one relation (black dashed). Right: We show comparisons between contours for each simulation, highlighting the fact that each simulation typically falls in a slightly different $T_{\rm [O~{\small III}]}-T_{\rm [O~{\small II}]}$ space.}
    \label{fig:TO2_TO3}
\end{figure*}
\begin{figure*}
\begin{center}
    \includegraphics[width=0.8\textwidth,trim={0.0cm 0cm 0.0cm 0cm},clip]{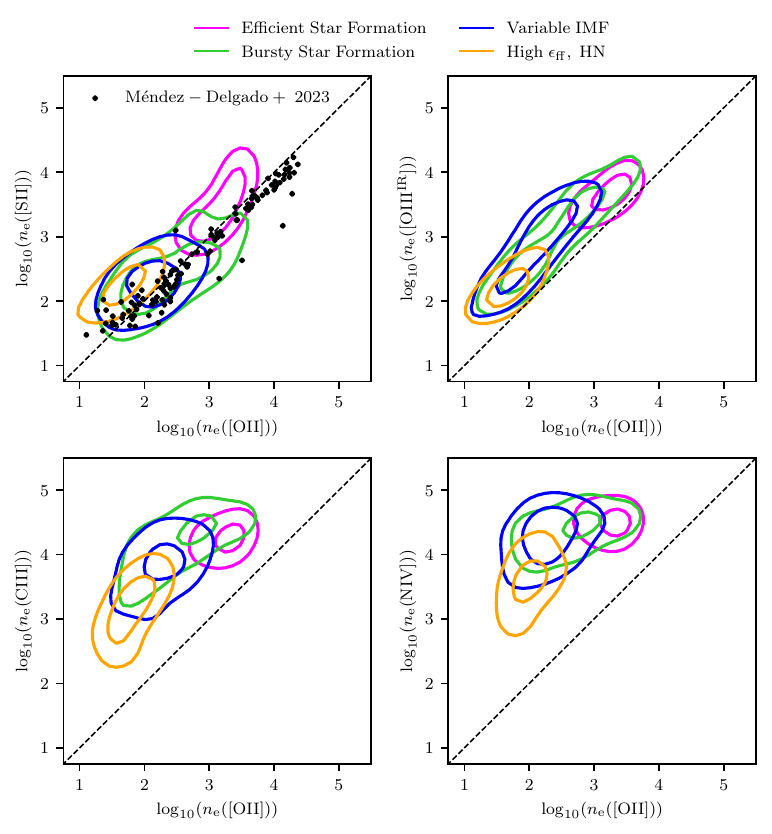}
\end{center}
    \caption{Electron number densities of galaxies with M$_{\rm UV}\leq-15$ in each {\footnotesize MEGATRON} simulation measured using a variety of emission line ratios, including those in the optical (O2 and S2), FIR (O3) and UV (C3 and N4). We find that galaxies of a given simulation produce a rich, inhomogeneous density structure, the exact properties of which are highly dependent on the star-formation/feedback prescriptions used.}
    \label{fig:ne_ne}
\end{figure*}

In an H~{\small II} region, the temperatures of different ionization zones can be inferred by various auroral line ratios. However, the relation between the temperatures in each of these zones in high-redshift galaxies is largely unknown. 

In Figure \ref{fig:TO2_TO3}, we show observed O$^+$ and O$^{++}$ electron temperatures for all galaxies in each run with M$_{\rm UV}\leq-15$. On the left, we show $T_{\rm [O~{\small II}]}$ as a function of $T_{\rm [O~{\small III}]}$, coloured by electron density as measured by the C3 ratio. We provide comparisons with observations of H~{\small II} regions in low redshift galaxies \citep{Mingozzi:2022}, local low-metallicity galaxies \citep{Rickards-Vaught:2024}, intermediate-redshift galaxies \citep{Cataldi:2025} and high-redshift galaxies \citep{Sanders:2025}. We also provide several modeled $T_{\rm [O~{\small II}]}-T_{\rm [O~{\small III}]}$ relations from observations \citep{Pilyugin:2009,Cataldi:2025} as well as from an isolated galaxy simulation also run with \texttt{RAMSES-RTZ} \citep{Cameron:2023}. On the right panel, we show contours describing the loci of \megatron\ galaxies.

We find that there is significant scatter in the $T_{\rm [O~{\small II}]}-T_{\rm [O~{\small III}]}$ relation. While low-density, low-ionization galaxies seem to follow the relation from \cite{Cameron:2023} (suggesting that this is a lower-limit irrespective of feedback prescription), galaxies with higher gas densities tend to scatter up to higher values of $T_{\rm [O~{\small II}]}$. Furthermore, galaxies with larger gas-phase metallicities tend to shift to lower values of $T_{\rm [O~{\small II}]}$ and $T_{\rm [O~{\small III}]}$, consistent with the relative normalizations of each simulation on the MZR in Figure~\ref{fig:mzr} and the expectations of stronger cooling. Galaxies with dense gas are measured to have very high values of $T_{\rm [O~{\small II}]}$, often greater than $T_{\rm [O~{\small III}]}$. This is due to the fact that the emission line ratio used to determine this temperature (O2O2) is also density dependent, due to the fact that both lines in the $\rm [O~{\small II}]\lambda\lambda 3727$ doublet have low critical densities\footnote{The same effect will impact the O3O3 ratio, though at significantly larger densities.} $\lesssim10^{3.5}~{\rm cm}^{-3}$ (see discussion in Section \ref{sec:ion}). The density range within which this doublet is a suitable density diagnostic will also produce larger values of O2O2 for a given temperature - thus over-estimating the true temperature (see Figure~2 of \citealt{Mendez-Delgado:2023}). This underscores the need to carefully fold in density diagnostics when using emission line ratios to infer properties of the ISM at high redshift.

Altogether, we find that the observed temperature structure of \megatron\ galaxies with low densities is entirely consistent with low-metallicity objects measured at low-redshift and cosmic noon \citep[see discussion in ][]{Cataldi:2025}. In contrast, the majority of UV-faint galaxies in \megatron\ have a sufficiently dense ISM that O2O2 becomes an unreliable temperature diagnostic without having an associated density indicator \citep{Mendez-Delgado:2023}. This points to the fact that low-ionization temperature diagnostics (based around lines with low critical densities) used for high-redshift galaxies should be treated with care in the cases where an associated density diagnostic is unavailable \citep[e.g.][]{Martinez+2025}. Nevertheless, the fact that oxygen abundances at high-redshift are typically dominated by $\rm O^{++}$ \citep{Laseter:2022,Curti:2023} means that any uncertainties due to density are generally subdominant compared to those due to temperature fluctuations \citep[e.g.][]{Cameron:2023}.

\subsection{High Densities}\label{sec:dens}

\begin{figure*}
\begin{center}
    \includegraphics[width=0.9\textwidth]{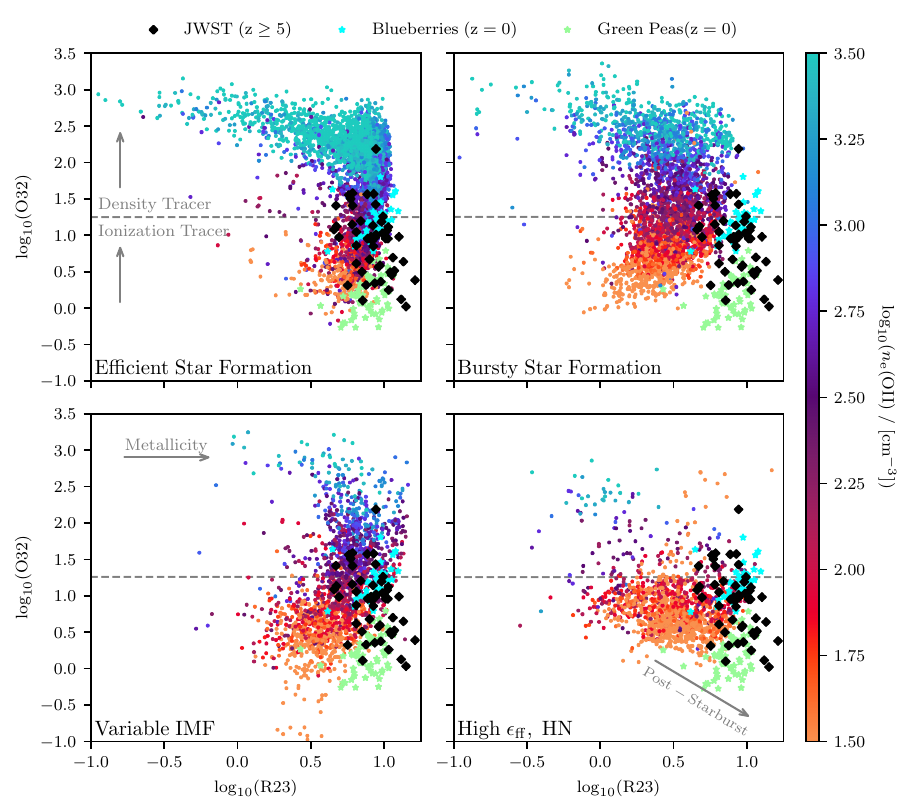}
\end{center}
    \caption{O32-R23 ionization-excitation diagnostic diagram for galaxies in each {\footnotesize MEGATRON} simulation, with each point coloured by the $\rm O~{\small II}$-derived electron density. A compilation of observational data is provided including high redshift  \protect\citep[black:][]{Cameron:2023b,Mascia:2023,Sanders:2023,Topping:2024} and low-redshift blueberries \citep[cyan:][]{Yang:2017} and green peas \citep[green:][]{Yang:2017b}. Accounting for metallicity and mass biases (see text), {\footnotesize MEGATRON} galaxies reproduce the observed scatter well, particularly for galaxies with high values of O32. Interestingly, a large number of galaxies with high O32, low R23, and a dense ISM are predicted, though these tend to be UV-faint. We have highlighted the expected region within which O32 behaves as a density tracer.}
    \label{fig:O32-R23}
\end{figure*}

\begin{figure}
    \includegraphics[width=\columnwidth,trim={0.0cm 0cm 0.0cm 0cm},clip]{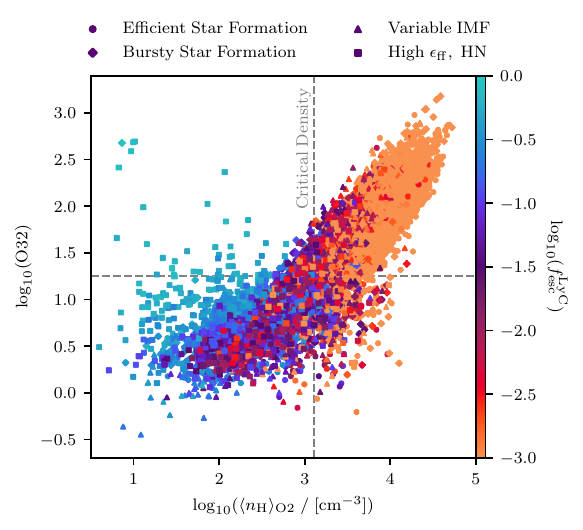}
    \caption{O32 as a function of O2-weighted gas density for all four simulations in the high-redshift {\footnotesize MEGATRON} suite and coloured by the LyC escape fraction. For values of $\rm \log_{10}(O32)\gtrsim1.25$, we find a strong correlation between the two, owing to the low critical density of the $\rm [O~{\small II}]~\lambda\lambda3727$ lines (indicated by a vertical dashed line). Systems with high values of O32 at low densities are LyC leakers.}
    \label{fig:O32_nH}
\end{figure}

\begin{figure*}
    \includegraphics[width=\textwidth,trim={0.0cm 0cm 0.0cm 0cm},clip]{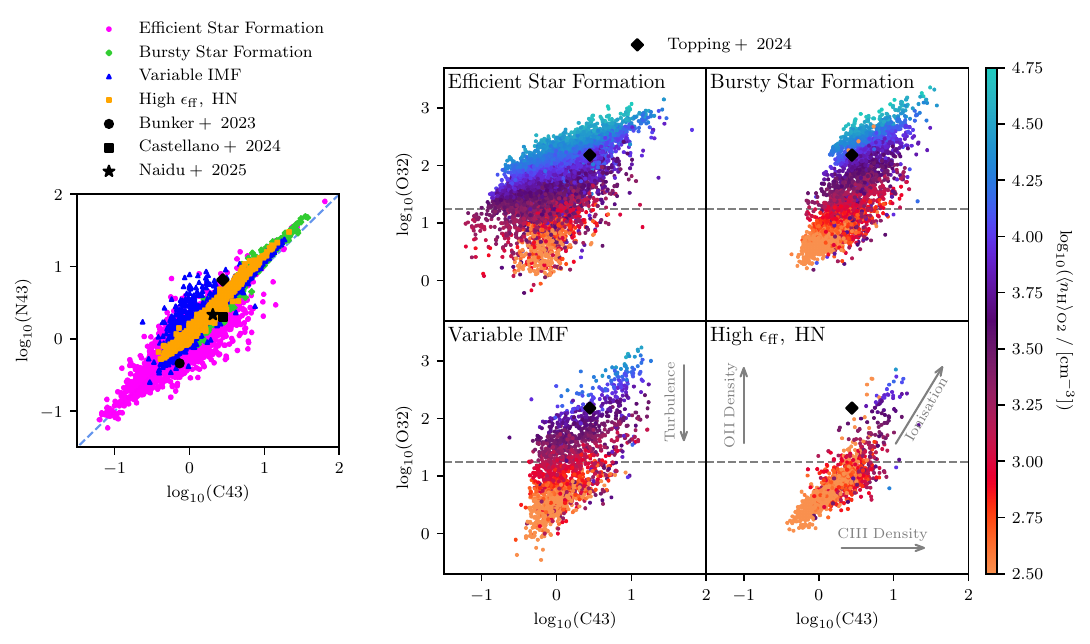}
    \caption{Left: N43 as a function of C43 for all galaxies in each {\footnotesize MEGATRON} simulation. This acts as a calibration, showing that each of these line ratios depends primarily on the ionization parameter, consistent with observational data \protect\citep{Bunker:2023,Topping:2024,Castellano:2024,Naidu:2025}. Right: O32 as a function of C43 for each galaxy in {\footnotesize MEGATRON}, coloured by the O2-weighted gas density taken from the simulation. Strong correlations between O32 and C43 only exist for a given gas density, suggesting that in the high-redshift Universe, the dependence of O32 on ionization parameter is dominated by a dependence on gas density, due to the low critical densities of the $\rm [O~{\small II}]~\lambda\lambda3727$ lines.}
    \label{fig:C43_O32}
\end{figure*}

\begin{figure*}
\begin{center}
    \includegraphics[width=0.7\textwidth,trim={0.0cm 0cm 0.0cm 0cm},clip]{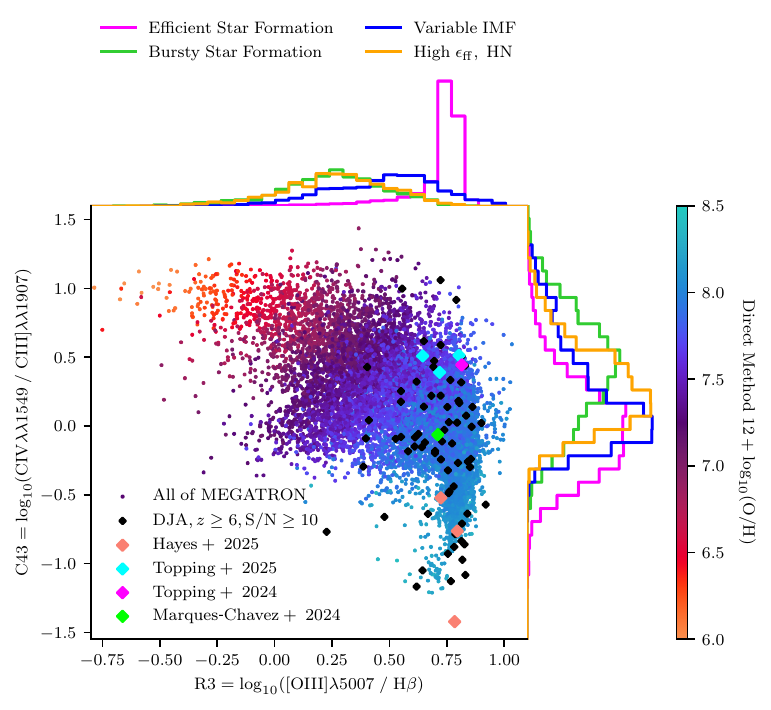}
\end{center}
    \caption{C43-R3 excitation-ionization diagnostic diagram. We show data for galaxies with M$_{\rm UV}\leq-15$ from all {\footnotesize MEGATRON} simulations coloured by the direct method oxygen abundance. On the top and right, we show histograms of the emission line ratios in question, demonstrating the impact of sub-grid star-formation/feedback prescriptions on these diagnostics. For a comparison, we include all galaxies at $z\geq6$ in the \texttt{DJA} \protect\citep{Heintz:2024} with signal-to-noise ratios greater than 10 for all lines as well as data from individual objects \protect\citep{Marques-Chaves:2024,Topping:2024,Topping:2025b} and stacks \protect\citep{Hayes:2025}.}
    \label{fig:C43_R3}
\end{figure*}

In this section, we explore the second key physical property of the high-redshift ISM: investigating the gas density properties of \megatron\ galaxies.

Traditionally, low-ionization state optical lines (e.g. $\rm [O~{\small II}]\lambda\lambda3726,3728$ and $\rm [S~{\small II}]\lambda\lambda6719,6731$) have been used to measure the densities of star-forming regions due to their strength and doublet separation. However, the information that these lines provide is inherently incomplete, as this low-ionization gas is not representative of the ISM as a whole, particularly tracing the edge of $\rm H~\small II$ regions and diffuse ionized gas (see Figure~\ref{fig:all_lines}). In contrast, the $\rm C~{\small III}]\lambda\lambda1906,1908$ and $\rm N~{\small IV}]\lambda\lambda1483,1486$ doublets are sensitive to increasingly higher energies, tracing denser gas closer to the young, massive stars \citep[e.g.][]{James:2014,Berg:2021,Mingozzi:2022}.

Observations of H~{\small II} regions at $z\gtrsim1$ with optical lines suggest that they are considerably denser than in the local Universe \citep[e.g.][]{Erb:2006,Hainline:2009,Steidel:2014,Sanders:2016,Berg:2022}. JWST now provides access to the rest UV-optical spectra of galaxies at higher redshifts, allowing multi-tracer determinations of electron densities \citep[][]{Isobe:2023,Reddy:2023,Abdurrouf:2024,Topping:2025,Li:2025}. Such observations have found elevated gas densities in the early Universe suggesting a strong evolution of gas density with redshift. Likewise a few objects with exceptionally high densities \citep[$\gtrsim10^5~{\rm cm}^{-3}$,][]{Senchyna:2024,Topping:2024}, were also discovered.

Electron densities measured from O2, S2, O3, C3, and N4 for all \megatron\ galaxies with M$_{\rm UV}<-15$ are shown in Figure~\ref{fig:ne_ne}. Each density diagnostic traces a different region of the ISM --- therefore, relations between different density measurements are not one-to-one \citep[e.g.][]{Kewley:2019,Mendez-Delgado:2023}. Focusing on optical density tracers, we find that $n_{\rm e}([{\rm O~{\small II}}]) \sim n_{\rm e}([{\rm S~{\small II}}]) $ \citep[e.g.][]{Kewley:2019,Berg:2021}. Next, we find that the O3 FIR tracer generally reports higher densities than that of O2. Given the fact that O3 is sensitive to a lower density range than O2, this suggests that $\rm O^{++}$ exists in systematically denser gas than $\rm O^+$. We note that in more massive metal-enriched galaxies than what is modeled in \megatron, observations show that the reverse may be true \cite{Harikane:2025}.

Considering UV density diagnostics, we find that both C3- and N4-based density measurements are significantly greater than those traced by O2, in line with observational results \citep{Berg:2022,Mingozzi:2022}. In particular, the $n_{\rm e}({\rm C~{\small III}}])-n_{\rm e}([{\rm O~{\small II}}])$ space yields the greatest differences between each simulation. For instance, we find that while the bursty model exhibits similar C3 and N4 densities, the variable IMF, efficient and high $\epsilon_{\rm ff}$ simulations all find that N4 traces systematically denser regions.

Overall, it is clear that the efficient star formation run produces the densest galaxies, due to the `weaker' fiducial feedback. Nevertheless, it can be instructive to highlight differences between the other models. Beginning with the high $\epsilon_{\rm ff}$ simulation, the key point is that here, stars are encouraged to form in less dense regions. As a result, these stars will illuminate lower density gas (producing emission lines that trace more diffuse gas) and produce SNe which couple more effectively to the gas\footnote{The presence of hypernovae in this run will also aid to better clear the stellar birth clouds.}, again reducing the density of the ISM. In contrast, while the bursty and variable IMF runs both feature stronger feedback, these events will occur in denser gas, making the momentum coupling to the ISM comparatively weaker \citep{Blondin:1998,Thornton:1998}.

\subsection{High Ionization Parameters}\label{sec:ion}

Here, we explore the final key physical property of the high-redshift ISM: investigating the observed ionization parameters of \megatron\ galaxies.

\subsubsection{Ionization-Excitation with the O32-R23 Diagram}

Numerous galaxies that have been spectroscopically confirmed at high redshift have been found to produce strong emission lines. Of these, the vast majority appear to have line ratios consistent with high ionization parameters \citep[e.g.][]{Cameron:2023b,Mascia:2023,Sanders:2023,Nakajima:2023,Roberts-Borsani:2024,Topping:2025b,Tang:2025,Glazer:2025,Hayes:2025}, as measured from O32 \citep{Kewley:2019}. Furthermore, various studies of stacked spectra have suggested a redshift trend, with galaxies in the earlier Universe having ever higher values of O32 \citep[e.g.][]{Sanders:2023,Roberts-Borsani:2024}. Finally, these high ionization parameters are also consistent with high EWs \citep[e.g.][]{Boyett:2024,Endsley:2024}, high ionizing photon production efficiencies \citep[e.g.][]{Atek:2024,Saxena:2024,Simmonds:2024b,Simmonds:2024,Laseter:2025,Choustikov:2025}, and high ionization-state lines \citep[e.g.][]{Bunker:2023,Topping:2024,Calabro:2024,Tang:2025b}.

The most commonly used emission line ratio diagram to diagnose ionization and excitation is the O32-R23 diagram. In Figure~\ref{fig:O32-R23} we show O32 as a function of R23 for all galaxies with M$_{\rm UV}\leq -15$ in each \megatron\ simulation, coloured by the O2-measured electron number density. We also provide a compilation of observed $z\geq 5$ galaxies \citep{Cameron:2023b,Mascia:2023,Sanders:2023,Topping:2024} as well as the population of low-redshift blueberries \citep{Yang:2017} and green peas \citep{Yang:2017b} to provide a comparison. First, visual inspection suggests that diagnostic diagrams like O32-R23 depend fundamentally on star-formation/feedback prescription, suggesting an exciting avenue to constrain stellar physics in the early Universe \citep{Katz:2024}. We find that the efficient and variable IMF runs are best able to reproduce the observed scatter. In contrast, the bursty and high $\epsilon_{\rm ff}$ simulations produce galaxies with lower apparent metallicities (see discussion in Section \ref{sec:mets}), which has the effect of shifting points to lower R23. In all cases, \megatron\ struggles to produce a large number of galaxies with high R23 and low O32. Recently, \cite{Witten:2025} used data from the \sphinx\ public data release \citep{Katz:2023} to show that these are likely to be massive post-starburst systems. While \megatron\ does contain post-starburst galaxies, these are not sufficiently massive or metal-enriched (due to the small simulation volume and high redshift) to fill this parameter space. We find that all four \megatron\ simulations are capable of producing the scatter of high-O32-high-R23 points observed both in the local universe and at high redshift, including the galaxy with the highest observed O32: RXCJ2248-ID \citep{Topping:2024}. Interestingly, all four \megatron\ simulations predict that there is a large population of low-metallicity, UV faint galaxies with low R23 values and extreme values of O32 ($\sim1000$). While these are yet to be observed in the high-redshift universe (and may be ultimately entirely obscured), their unique electron density properties ($n_{\rm e}({\rm [O~{\small II}]})\gtrsim1000$) motivate future surveys to search for them.

\subsubsection{Dependence of O32 on Density}

O32 is the most common diagnostics of ionization parameter at high-redshift due tot he strength of the lines and their rest-frame wavelengths. However, the critical density of the O2 lines is lower than often found in \megatron\ galaxies and sometimes measured at high-redshift. Figure~\ref{fig:O32_nH} shows O32 as a function of O2-weighted gas density for galaxies in each \megatron\ simulation, coloured by their angle-averaged LyC escape fraction. Broadly, we find two regimes of behavior. For values of $\log_{10}({\rm O32})\lesssim 1.25$, there is no strong relationship between O32 and gas density. In contrast, when $\log_{10}({\rm O32})\gtrsim 1.25$, we find a strong correlation between O32 and $\langle n_{\rm H} \rangle_{\rm O2}$. This is a clear manifestation of the low critical density (indicated by a vertical line) of the O2 lines biasing O32 towards higher values \citep[see also discussion in][]{Martinez+2025}. Finally, we find that galaxies with a large value of O32 at low gas densities tend to be strong LyC leakers, in contrast to those with the same values of O32 at high densities. This is a demonstration of the complex relationship between O32 and $f_{\rm esc}^{\rm LyC}$ \citep{Nakajima:2014,Katz:2020,Barrow:2020,Flury:2022,Choustikov:2024}, highlighting the dependence of $f_{\rm esc}^{\rm LyC}$ on the density of the ISM \citep{Choustikov:2024,Choustikov:2024b}.

Another approach to constrain the ionization parameter might be to use C43 and N43, since these ratios are less sensitive to density. To illustrate this, on the left of Figure~\ref{fig:C43_O32}, we show N43 as a function of C43 for galaxies in each \megatron\ simulation. We find that for the most part, these two ratios are directly proportional, suggesting that both are dependent only on the ionization parameter, to first order. Each simulation is consistent with measured line ratios from the lensed galaxy RXCJ2248-ID \citep{Topping:2024} as well as other extreme sources \citep{Bunker:2023,Castellano:2024,Naidu:2025}. In the right panel of Figure~\ref{fig:C43_O32} we show O32 as a function of C43 coloured by the O2-weighted gas density taken from each simulation, compared again to RXCJ2248-ID. We find that strong correlations between O32 and C43 exist only in a particular gas density range, again confirming that in the vast majority of \megatron\ galaxies, high values of O32 are partially driven by high gas density. 

Finally, it is important to consider the effects of turbulence in the ISM. Due to the fact that $\rm [O~{\small III}]~\lambda5007$ and $\rm [O~{\small II}]~\lambda\lambda3727$ are produced by volume and boundary species respectively, the O32 ratio can be significantly impacted by turbulence \citep[see discussions in ][]{Gray:2017,Jin:2022,Xing:2026}. For instance, introducing turbulence into an initially spherically symmetric Str{\"o}mgren sphere will increase the surface area of the ionized region for a given volume, thereby boosting $\rm [O~{\small II}]~\lambda\lambda3727$ flux for a given $\rm [O~{\small III}]~\lambda5007$ luminosity and therefore reducing the O32 ratio at fixed (mean) ionization parameter \citep{Jin:2022} and metallicity \citep{Xing:2026}.

\subsubsection{Ionization-Excitation with the C43-R3 Diagram}

Given the impact of density, temperature, and turbulence on O32, we are inspired to consider a different ionization diagnostic for the ionization-excitation diagram. Ideally, for this purpose we would use N43; due to the involved lines having a higher critical density than those used in O32 or C43; however, significantly more strong carbon emitters have been detected at high-redshift \citep[e.g.][]{Roberts-Borsani:2025}, despite the high N/O abundances that have also been measured in the early Universe \citep[e.g.][]{Cameron:2023c,Isobe:2023b,Isobe:2025,Schaerer:2024,Marques-Chaves:2024,Arellano-Cordova:2025}. Therefore, we propose the use of the C43-R3 diagram, which we show in Figure~\ref{fig:C43_R3} for all \megatron\ galaxies considered, coloured by the direct method gas-phase metallicity. Histograms are provided to show differences between each of the simulations in each diagnostic. For comparisons, we provide observational data from the \texttt{DJA} \citep[\texttt{DAWN JWST Archive\footnote{\href{https://dawn-cph.github.io/dja/}{https://dawn-cph.github.io/dja/}}:}][]{Heintz:2024} as well as for other specific objects \citep{Marques-Chaves:2024,Topping:2024,Topping:2025b} and stacks \citep{Hayes:2025}. 

We find very good agreement between observations and \megatron\ galaxies with $12+\log({\rm O}/{\rm H})\gtrsim 7.5$, though we find that the bursty and high $\epsilon_{\rm ff}$ simulations tend to produce lower values of R3. This may however be a selection effect, as galaxies to the upper-left of this diagram tend to have significantly fainter UV magnitudes. As a result, even without the density effects discussed above, high-redshift galaxies exhibit gas with high ionization parameters. This effect is further compounded with metallicity, where galaxies with low oxygen abundances tend to have higher values of C43 for a given value of R3. Moreover, we find that for the lowest values of R3, two tails appear at $\rm C43\sim 0\;and\;1$. These correspond to systems with ionizing photon production efficiencies of $\lesssim25$ and $\gtrsim26$ respectively. Finally, many galaxies with the highest ionization parameters tend to have very low values of R3, due to the fact that these systems have low metallicities and/or are sufficiently dense to begin collisionally de-exciting $\rm [O~{\small III}]~\lambda5007$ emission. This may help explain the particularly weak $\rm O~\small III$ and strong $\rm C~\small IV$ emission of \texttt{LAP1-B} \citep{Nakajima:2025}.

However, it is important to note that there remain complexities with interpreting C43. First, while the critical densities of the $\rm C~\small III$ doublet are significantly higher than those of $\rm O~\small II$, in principle at sufficiently high densities a similar effect will take place, with C43 increasing as a function of density. Furthermore, temperature can also have an impact on C43, boosting C43 by a maximum of $\sim 0.5~$dex at the most extreme H~{\small II} region temperatures. Moreover, the $\rm C~\small IV$ doublet is a resonant line, meaning that high optical depths can bias the line ratio in non-trivial ways \citep[e.g.][]{Hayes:2025} including potentially preferentially tracing outflows \citep[e.g.][]{Steidel:2010}. Finally, there is uncertainty in the shape of the attenuation law in high-redshift galaxies \citep{Reddy:2020,Reddy:2025,McKinney:2025,Sanders:2025b}, which may have an impact on C43. Nevertheless, for ISM conditions more representative of those observed in galaxies at high-redshift, this diagram should prove very helpful to discern the properties of galaxies below redshift $\sim 9$ where $\rm [O~{\small III}]~\lambda5007$ redshifts out of the NIRSpec range. In those cases, it is possible to use similar diagrams with $\rm O~{\small III}]~\lambda\lambda 1663/H\gamma$, or $\rm [O~{\small III}]~\lambda 4363]/H\gamma$, though the first introduces systematics due to dust attenuation and the latter can be rare to detect.

\section{Caveats}\label{sec:caveats}

As with any simulation, there are numerous caveats that should be kept in mind when interpreting our results.

\textbf{Numerical Resolution:} \megatron\ is subject to a finite spatial and mass resolution. For the purpose of this work, this can have two main effects. The first is that we are unable to always completely resolve every single Str{\"o}mgren sphere (and importantly the detailed physics that occurs at the ionization front). These cells will have an incorrect temperature \citep[this can affect predicted emission line luminosities, see discussion in][]{Choustikov:2024} which is corrected according to the strategy described in \cite{Katz:2024}. Secondly, our resolution limits our ability to accurately follow the effects of radiation feedback \citep[see discussions in][]{Kimm:2019,Kimm:2022} which may pre-process dense regions of the ISM ahead of SNe. This will be addressed in upcoming work.

\textbf{Dust Attenuation:} In producing mock spectroscopic observations of \megatron\ galaxies, we have ignored the effects of dust -- in terms of absorption, emission, and scattering. This is likely a good assumption, as we are predominantly focusing on low-mass, low-metallicity objects which are expected to have dust-to-gas mass ratios of $\lesssim10^{-3.5}$ for metallicities of $12+\log({\rm O}/{\rm H})\sim8$ \citep[e.g.][]{Popping:2017,Garaldi:2025}. Indeed, photometric \citep{Cullen:2024} and spectroscopic \citep{Saxena:2024b} surveys suggest that observed UV slopes at redshifts $z>9$ are consistent with pure stellar and nebular continuum emission that is not absorbed by dust (though this does not preclude the possibility of a gray attenuation curve in high-redshift galaxies, e.g. \citealt{Markov:2025,Markov:2025b}). We leave such discussions to future work.

\textbf{Missing Physics:} At present, it is important to note that the \megatron\ simulations have not included certain physical processes such as magnetic fields, cosmic rays, and on-the-fly dust formation and destruction -- all of which have the potential to impact the properties of the ISM as extra heating/cooling and/or pressure terms. These will be addressed in an upcoming work. Similarly, we have not included a prescription for AGN, which are a source of radiation and feedback. While we expect these not to dominate at the mass ranges considered in \megatron, we do note that some observations have suggested an abundance of over-massive rapidly-accreting black holes in high-redshift galaxies (\citealt{Maiolino:2024}, c.f. \citealt{Li:2025b}).

\textbf{Sub-Grid Uncertainties:} As with any physical simulation which needs to account for a high dynamic range of scales, we have had to use certain sub-grid prescriptions for stellar populations, formation and feedback. While prescriptions in the efficient star formation run have been shown to match the stellar mass - halo mass relation at $z=0$ (\citealt{Agertz:2021}, see also Section 3 of \citealt{Katz:2025b}), the remaining three simulations use prescriptions that are also physically plausible. Given the dependence of ISM properties on low-metallicity sources of ionizing radiation, in future work it will be necessary to include detailed prescriptions of x-ray binaries \citep[e.g.][]{Saxena:2021}, Wolf-Rayet stars \citep[e.g.][]{Sander:2025}, and AGN \citep[e.g.][]{Kubota:2018,Trebitsch:2021}. All of these will contribute by increasing the number of hard ionizing photons, leading to an even more extreme ISM. 

A more in-depth discussion of the caveats is provided in \cite{Katz:2025b}. Similarly, an extensive discussion on the impacts of sub-grid modeling on galaxy properties in \megatron-like simulations is given in \cite{Katz:2024}. 

\section{Conclusions}\label{sec:conclusions}

We have presented early results from the \megatron\ simulations \citep{Katz:2025b,Rey:2025,Cadiou:2025}, investigating the physics behind observed properties of the ISM of galaxies in the high-redshift Universe. The \megatron\ simulations represent an attempt to employ a detailed non-equilibrium thermochemical network of primordial species, metals, and molecules, coupled to on-the-fly radiation transport and galaxy formation physics in a cosmological context. Due to this approach, we are able to predict spectra on a cell-by-cell basis, accounting for the stellar continuum, nebular continuum and nebular emission lines.

We have approached the problem as observers: beginning with the intrinsic spectrum and photometry of each galaxy and measuring galaxy properties using techniques often applied to JWST observations. For example, electron temperatures and densities are measured from emission line ratios using \texttt{PyNeb} \citep{Luridiana:2015}, gas-phase metallicities are measured using the direct (auroral line) method \citep{Cameron:2023}, and stellar masses are measured by SED-fitting with \texttt{BAGPIPES} \citep{bagpipes}. This is a unique approach to comparing simulations with observations, that exploits some of the key strengths of the \megatron\ simulations.

Our results can be summarized as follows:

\begin{enumerate}
    \item The ISM of \megatron\ galaxies is fundamentally different to that of observed low-redshift galaxies. It is denser, less metal-enriched, subject to harder radiation fields, and has a higher ionization parameter. This is in agreement with most JWST observations \citep[e.g.][]{Bunker:2023,Cameron:2023b,Sanders:2023,Mascia:2023,Curti:2024,Hayes:2025,Tang:2025,Pollock:2025}.
    \item No two emission lines are produced by gas with the same density and temperature distributions. As a direct consequence of the turbulent, multi-phase structure of high-redshift galaxies, this can significantly bias measurements of ISM properties \citep[e.g.][]{Cameron:2023,Harikane:2025,Usui:2025}.
    \item Simulations that produce a very dense ISM are observed to have a shallow MZR slope, consistent with what is observed at high-redshift \citep[e.g.][]{Heintz:2023,Nakajima:2023,Curti:2023,Curti:2024,Morishita:2024,Chemerynska:2024,Sarkar:2025}. Observational techniques designed to measure galaxy masses and metallicities systematically bias the MZR to be flatter at the low-mass end.
    \item ISM densities inferred by UV tracers are systematically larger than those measured by optical lines, in agreement with observations \citep[][]{Isobe:2023,Reddy:2023,Abdurrouf:2024,Li:2025,Topping:2025}. Interestingly, we find that O~{\small III} FIR lines also trace denser gas than O~{\small II}. The exact density structure is sensitive to the star-formation/feedback model assumed via the strength of feedback, IMF choice, and the ambient density during stellar feedback.
    \item At the densities reached in the ISM of high-redshift galaxies, $\rm [O~{\small III}]~\lambda 5007/[O~{\small II}]~\lambda\lambda3727$ becomes a density tracer as [$\rm \small O~{\small II}$] is collisionally de-excited. If not properly taken into account, measurements of ionization parameter from O32 will be artificially high. We instead propose the use of $\rm C~{\small IV}~\lambda\lambda 1549/C~{\small III}]~\lambda\lambda1907$ or $\rm N~{\small IV}]~\lambda\lambda 1484/N~{\small III}~\lambda\lambda1750$. When considering these diagnostics, \megatron\ galaxies match observations very well, suggesting that the ionization parameter of observed galaxies is indeed very high \citep[e.g.][]{Marques-Chaves:2024,Topping:2024,Topping:2025b,Hayes:2025}.
    \item Galaxies simulated with various feedback models populate different regions of strong-line diagnostic diagrams as line ratios are sensitive to the density-temperature structure of the ISM \citep{Katz:2024}. This suggests a possible avenue for constraining high-redshift feedback physics with line ratios directly.
\end{enumerate}

This work has demonstrated the ability of simulations with on-the-fly out-of-equilibrium thermochemistry like \megatron\ to help interpret the diversity of emission line properties observed by JWST. Likewise, when large samples of JWST galaxies with a well-defined selection function become public, comparisons with simulations such as \megatron\ will be ideal for to constraining the physics of star formation as well as the impact of stellar feedback on the ISM of high-redshift galaxies. 

\vspace{0.3cm}
\section*{Acknowledgments}

NC acknowledges support from the Science and Technology Facilities Council (STFC) for a PhD studentship. AJC and AS acknowledge funding from the “FirstGalaxies” Advanced Grant from the European Research Council (ERC) under the European Union’s Horizon 2020 research and innovation programme (Grant agreement No.789056). TK is supported by the National Research Foundation of Korea (RS-2022-NR070872 and RS-2025-00516961) and by the Yonsei Fellowship, funded by Lee Youn Jae. KM acknowledges the Flemish Fund for Scientific Research (FWO-Vlaanderen), Grant number 1169822N.

This work was performed using the DiRAC Data Intensive service at Leicester, operated by the University of Leicester IT Services, which forms part of the STFC DiRAC HPC Facility (www.dirac.ac.uk). The equipment was funded by BEIS capital funding via STFC capital grants ST/K000373/1 and ST/R002363/1 and STFC DiRAC Operations grant ST/R001014/1. This work used the DiRAC@Durham facility managed by the Institute for Computational Cosmology on behalf of the STFC DiRAC HPC Facility (www.dirac.ac.uk). The equipment was funded by BEIS capital funding via STFC capital grants ST/P002293/1, ST/R002371/1 and ST/S002502/1, Durham University and STFC operations grant ST/R000832/1. This work was performed using resources provided by the Cambridge Service for Data Driven Discovery (CSD3) operated by the University of Cambridge Research Computing Service (www.csd3.cam.ac.uk), provided by Dell EMC and Intel using Tier-2 funding from the Engineering and Physical Sciences Research Council (capital grant EP/T022159/1), and DiRAC funding from the Science and Technology Facilities Council (www.dirac.ac.uk). DiRAC is part of the National e-Infrastructure. The authors thank Jonathan Patterson for smoothly running the Glamdring Cluster hosted by the University of Oxford, where part of the data processing was performed. The authors also acknowledge financial support from Oriel College’s Research Fund. The material in this manuscript is based upon work supported by NASA under award No. 80NSSC25K7009.

The authors thank Romain Teyssier and Leo Michel-Dansac for both developing and open-sourcing {\small RAMSES} and {\small RASCAS}, respectively. We thank the developers and maintainers of \textsc{pynbody} (\citealt{Pontzen2013}), \textsc{yt} (\citealt{Turk2011}), \textsc{NumPy} (\citealt{vanderWalt2011, Harris2020}), \textsc{SciPy} (\citealt{Virtanen2020}), \textsc{jupyter} (\citealt{Ragan-Kelley2014}), \textsc{matplotlib} (\citealt{Hunter2007}), \textsc{cmasher} (\citealt{van-der-Velden:2020}), \textsc{pandas} (\citealt{reback2020pandas,mckinney-proc-scipy-2010}), the Astrophysics Data Service, and the arXiv pre-print repository for providing open-source software and services that were used extensively in this work.


\bibliographystyle{mn2e}
\bibliography{References}

@ARTICLE{rascas2020,
       author = {{Michel-Dansac}, L. and {Blaizot}, J. and {Garel}, T. and {Verhamme}, A. and {Kimm}, T. and {Trebitsch}, M.},
        title = "{RASCAS: RAdiation SCattering in Astrophysical Simulations}",
      journal = {\aap},
         year = 2020,
        month = mar,
       volume = {635},
          eid = {A154},
        pages = {A154},
          doi = {10.1051/0004-6361/201834961},
archivePrefix = {arXiv},
       eprint = {2001.11252},
 primaryClass = {astro-ph.GA},
       adsurl = {https://ui.adsabs.harvard.edu/abs/2020A&A...635A.154M}
}

@ARTICLE{Maiolino:2019,
       author = {{Maiolino}, R. and {Mannucci}, F.},
        title = "{De re metallica: the cosmic chemical evolution of galaxies}",
      journal = {\aapr},
         year = 2019,
        month = feb,
       volume = {27},
       number = {1},
          eid = {3},
        pages = {3},
          doi = {10.1007/s00159-018-0112-2},
archivePrefix = {arXiv},
       eprint = {1811.09642},
 primaryClass = {astro-ph.GA},
       adsurl = {https://ui.adsabs.harvard.edu/abs/2019A&ARv..27....3M}
}

@ARTICLE{Maseda:2020,
       author = {{Maseda}, Michael V. and {Bacon}, Roland and {Lam}, Daniel and {Matthee}, Jorryt and {Brinchmann}, Jarle and {Schaye}, Joop and {Labbe}, Ivo and {Schmidt}, Kasper B. and {Boogaard}, Leindert and {Bouwens}, Rychard and {Cantalupo}, Sebastiano and {Franx}, Marijn and {Hashimoto}, Takuya and {Inami}, Hanae and {Kusakabe}, Haruka and {Mahler}, Guillaume and {Nanayakkara}, Themiya and {Richard}, Johan and {Wisotzki}, Lutz},
        title = "{Elevated ionizing photon production efficiency in faint high-equivalent-width Lyman-{\ensuremath{\alpha}} emitters}",
      journal = {\mnras},
         year = 2020,
        month = apr,
       volume = {493},
       number = {4},
        pages = {5120-5130},
          doi = {10.1093/mnras/staa622},
archivePrefix = {arXiv},
       eprint = {2002.11117},
 primaryClass = {astro-ph.GA},
       adsurl = {https://ui.adsabs.harvard.edu/abs/2020MNRAS.493.5120M}
}

@ARTICLE{Saxena:2024,
       author = {{Saxena}, Aayush and {Bunker}, Andrew J. and {Jones}, Gareth C. and {Stark}, Daniel P. and {Cameron}, Alex J. and {Witstok}, Joris and {Arribas}, Santiago and {Baker}, William M. and {Baum}, Stefi and {Bhatawdekar}, Rachana and {Bowler}, Rebecca and {Boyett}, Kristan and {Carniani}, Stefano and {Charlot}, Stephane and {Chevallard}, Jacopo and {Curti}, Mirko and {Curtis-Lake}, Emma and {Eisenstein}, Daniel J. and {Endsley}, Ryan and {Hainline}, Kevin and {Helton}, Jakob M. and {Johnson}, Benjamin D. and {Kumari}, Nimisha and {Looser}, Tobias J. and {Maiolino}, Roberto and {Rieke}, Marcia and {Rix}, Hans-Walter and {Robertson}, Brant E. and {Sandles}, Lester and {Simmonds}, Charlotte and {Smit}, Renske and {Tacchella}, Sandro and {Williams}, Christina C. and {Willmer}, Christopher N.~A. and {Willott}, Chris},
        title = "{JADES: The production and escape of ionizing photons from faint Lyman-alpha emitters in the epoch of reionization}",
      journal = {\aap},
         year = 2024,
        month = apr,
       volume = {684},
          eid = {A84},
        pages = {A84},
          doi = {10.1051/0004-6361/202347132},
archivePrefix = {arXiv},
       eprint = {2306.04536},
 primaryClass = {astro-ph.GA},
       adsurl = {https://ui.adsabs.harvard.edu/abs/2024A&A...684A..84S}
}

@ARTICLE{bagpipes,
       author = {{Carnall}, A.~C. and {McLure}, R.~J. and {Dunlop}, J.~S. and {Dav{\'e}}, R.},
        title = "{Inferring the star formation histories of massive quiescent galaxies with BAGPIPES: evidence for multiple quenching mechanisms}",
      journal = {\mnras},
         year = 2018,
        month = nov,
       volume = {480},
       number = {4},
        pages = {4379-4401},
          doi = {10.1093/mnras/sty2169},
archivePrefix = {arXiv},
       eprint = {1712.04452},
 primaryClass = {astro-ph.GA},
       adsurl = {https://ui.adsabs.harvard.edu/abs/2018MNRAS.480.4379C}
}

@ARTICLE{Cochrane:2025,
       author = {{Cochrane}, R.~K. and {Katz}, H. and {Begley}, R. and {Hayward}, C.~C. and {Best}, P.~N.},
        title = "{High-z Stellar Masses Can Be Recovered Robustly with JWST Photometry}",
      journal = {\apjl},
         year = 2025,
        month = jan,
       volume = {978},
       number = {2},
          eid = {L42},
        pages = {L42},
          doi = {10.3847/2041-8213/ad9a4d},
archivePrefix = {arXiv},
       eprint = {2412.02622},
 primaryClass = {astro-ph.GA},
       adsurl = {https://ui.adsabs.harvard.edu/abs/2025ApJ...978L..42C}
}

@ARTICLE{Narayanan:2024,
       author = {{Narayanan}, Desika and {Lower}, Sidney and {Torrey}, Paul and {Brammer}, Gabriel and {Cui}, Weiguang and {Dav{\'e}}, Romeel and {Iyer}, Kartheik G. and {Li}, Qi and {Lovell}, Christopher C. and {Sales}, Laura V. and {Stark}, Daniel P. and {Marinacci}, Federico and {Vogelsberger}, Mark},
        title = "{Outshining by Recent Star Formation Prevents the Accurate Measurement of High-z Galaxy Stellar Masses}",
      journal = {\apj},
         year = 2024,
        month = jan,
       volume = {961},
       number = {1},
          eid = {73},
        pages = {73},
          doi = {10.3847/1538-4357/ad0966},
archivePrefix = {arXiv},
       eprint = {2306.10118},
 primaryClass = {astro-ph.GA},
       adsurl = {https://ui.adsabs.harvard.edu/abs/2024ApJ...961...73N}
}

@software{sedpy,
       author = {{Johnson}, Benjamin D.},
        title = "{SEDPY: Modules for storing and operating on astronomical source spectral energy distribution}",
 howpublished = {Astrophysics Source Code Library, record ascl:1905.026},
         year = 2019,
        month = may,
          eid = {ascl:1905.026},
       adsurl = {https://ui.adsabs.harvard.edu/abs/2019ascl.soft05026J}
}

@ARTICLE{Choustikov:2025,
       author = {{Choustikov}, Nicholas and {Stiskalek}, Richard and {Saxena}, Aayush and {Katz}, Harley and {Devriendt}, Julien and {Slyz}, Adrianne},
        title = "{Inferring the ionizing photon contributions of high-redshift galaxies to reionization with JWST NIRCam photometry}",
      journal = {\mnras},
         year = 2025,
        month = mar,
       volume = {537},
       number = {3},
        pages = {2273-2290},
          doi = {10.1093/mnras/staf126},
archivePrefix = {arXiv},
       eprint = {2405.09720},
 primaryClass = {astro-ph.GA},
       adsurl = {https://ui.adsabs.harvard.edu/abs/2025MNRAS.537.2273C}
}

@ARTICLE{Bruzual:2003,
       author = {{Bruzual}, G. and {Charlot}, S.},
        title = "{Stellar population synthesis at the resolution of 2003}",
      journal = {\mnras},
         year = 2003,
        month = oct,
       volume = {344},
       number = {4},
        pages = {1000-1028},
          doi = {10.1046/j.1365-8711.2003.06897.x},
archivePrefix = {arXiv},
       eprint = {astro-ph/0309134},
 primaryClass = {astro-ph},
       adsurl = {https://ui.adsabs.harvard.edu/abs/2003MNRAS.344.1000B}
}

@ARTICLE{Chevallard:2016,
       author = {{Chevallard}, Jacopo and {Charlot}, St{\'e}phane},
        title = "{Modelling and interpreting spectral energy distributions of galaxies with BEAGLE}",
      journal = {\mnras},
         year = 2016,
        month = oct,
       volume = {462},
       number = {2},
        pages = {1415-1443},
          doi = {10.1093/mnras/stw1756},
archivePrefix = {arXiv},
       eprint = {1603.03037},
 primaryClass = {astro-ph.GA},
       adsurl = {https://ui.adsabs.harvard.edu/abs/2016MNRAS.462.1415C}
}

@ARTICLE{Kroupa:2001,
       author = {{Kroupa}, Pavel},
        title = "{On the variation of the initial mass function}",
      journal = {\mnras},
         year = 2001,
        month = apr,
       volume = {322},
       number = {2},
        pages = {231-246},
          doi = {10.1046/j.1365-8711.2001.04022.x},
archivePrefix = {arXiv},
       eprint = {astro-ph/0009005},
 primaryClass = {astro-ph},
       adsurl = {https://ui.adsabs.harvard.edu/abs/2001MNRAS.322..231K}
}

@ARTICLE{Ferland:2017,
       author = {{Ferland}, G.~J. and {Chatzikos}, M. and {Guzm{\'a}n}, F. and {Lykins}, M.~L. and {van Hoof}, P.~A.~M. and {Williams}, R.~J.~R. and {Abel}, N.~P. and {Badnell}, N.~R. and {Keenan}, F.~P. and {Porter}, R.~L. and {Stancil}, P.~C.},
        title = "{The 2017 Release Cloudy}",
      journal = {\rmxaa},
         year = 2017,
        month = oct,
       volume = {53},
        pages = {385-438},
          doi = {10.48550/arXiv.1705.10877},
archivePrefix = {arXiv},
       eprint = {1705.10877},
 primaryClass = {astro-ph.GA},
       adsurl = {https://ui.adsabs.harvard.edu/abs/2017RMxAA..53..385F}
}

@ARTICLE{Leja:2019,
       author = {{Leja}, Joel and {Carnall}, Adam C. and {Johnson}, Benjamin D. and {Conroy}, Charlie and {Speagle}, Joshua S.},
        title = "{How to Measure Galaxy Star Formation Histories. II. Nonparametric Models}",
      journal = {\apj},
         year = 2019,
        month = may,
       volume = {876},
       number = {1},
          eid = {3},
        pages = {3},
          doi = {10.3847/1538-4357/ab133c},
archivePrefix = {arXiv},
       eprint = {1811.03637},
 primaryClass = {astro-ph.GA},
       adsurl = {https://ui.adsabs.harvard.edu/abs/2019ApJ...876....3L}
}

@ARTICLE{Tacchella:2022,
       author = {{Tacchella}, Sandro and {Finkelstein}, Steven L. and {Bagley}, Micaela and {Dickinson}, Mark and {Ferguson}, Henry C. and {Giavalisco}, Mauro and {Graziani}, Luca and {Grogin}, Norman A. and {Hathi}, Nimish and {Hutchison}, Taylor A. and {Jung}, Intae and {Koekemoer}, Anton M. and {Larson}, Rebecca L. and {Papovich}, Casey and {Pirzkal}, Norbert and {Rojas-Ruiz}, Sof{\'\i}a and {Song}, Mimi and {Schneider}, Raffaella and {Somerville}, Rachel S. and {Wilkins}, Stephen M. and {Yung}, L.~Y. Aaron},
        title = "{On the Stellar Populations of Galaxies at z = 9-11: The Growth of Metals and Stellar Mass at Early Times}",
      journal = {\apj},
         year = 2022,
        month = mar,
       volume = {927},
       number = {2},
          eid = {170},
        pages = {170},
          doi = {10.3847/1538-4357/ac4cad},
archivePrefix = {arXiv},
       eprint = {2111.05351},
 primaryClass = {astro-ph.GA},
       adsurl = {https://ui.adsabs.harvard.edu/abs/2022ApJ...927..170T}
}

@ARTICLE{Harvey:2025,
       author = {{Harvey}, Thomas and {Conselice}, Christopher J. and {Adams}, Nathan J. and {Austin}, Duncan and {Li}, Qiong and {Rusakov}, Vadim and {Westcott}, Lewi and {Goolsby}, Caio M. and {Lovell}, Christopher C. and {Cochrane}, Rachel K. and {Vijayan}, Aswin P. and {Trussler}, James},
        title = "{Behind the Spotlight: A systematic assessment of outshining using NIRCam medium-bands in the JADES Origins Field}",
      journal = {arXiv e-prints},
         year = 2025,
        month = apr,
          eid = {arXiv:2504.05244},
        pages = {arXiv:2504.05244},
          doi = {10.48550/arXiv.2504.05244},
archivePrefix = {arXiv},
       eprint = {2504.05244},
 primaryClass = {astro-ph.GA},
       adsurl = {https://ui.adsabs.harvard.edu/abs/2025arXiv250405244H}
}

@ARTICLE{Luridiana:2015,
       author = {{Luridiana}, V. and {Morisset}, C. and {Shaw}, R.~A.},
        title = "{PyNeb: a new tool for analyzing emission lines. I. Code description and validation of results}",
      journal = {\aap},
         year = 2015,
        month = jan,
       volume = {573},
          eid = {A42},
        pages = {A42},
          doi = {10.1051/0004-6361/201323152},
archivePrefix = {arXiv},
       eprint = {1410.6662},
 primaryClass = {astro-ph.IM},
       adsurl = {https://ui.adsabs.harvard.edu/abs/2015A&A...573A..42L}
}

@ARTICLE{Kewley:2019,
       author = {{Kewley}, Lisa J. and {Nicholls}, David C. and {Sutherland}, Ralph S.},
        title = "{Understanding Galaxy Evolution Through Emission Lines}",
      journal = {\araa},
         year = 2019,
        month = aug,
       volume = {57},
        pages = {511-570},
          doi = {10.1146/annurev-astro-081817-051832},
archivePrefix = {arXiv},
       eprint = {1910.09730},
 primaryClass = {astro-ph.GA},
       adsurl = {https://ui.adsabs.harvard.edu/abs/2019ARA&A..57..511K}
}

@ARTICLE{Berg:2021,
       author = {{Berg}, Danielle A. and {Chisholm}, John and {Erb}, Dawn K. and {Skillman}, Evan D. and {Pogge}, Richard W. and {Olivier}, Grace M.},
        title = "{Characterizing Extreme Emission-line Galaxies. I. A Four-zone Ionization Model for Very High-ionization Emission}",
      journal = {\apj},
         year = 2021,
        month = dec,
       volume = {922},
       number = {2},
          eid = {170},
        pages = {170},
          doi = {10.3847/1538-4357/ac141b},
archivePrefix = {arXiv},
       eprint = {2105.12765},
 primaryClass = {astro-ph.GA},
       adsurl = {https://ui.adsabs.harvard.edu/abs/2021ApJ...922..170B}
}

@ARTICLE{Topping:2025,
       author = {{Topping}, Michael W. and {Sanders}, Ryan L. and {Shapley}, Alice E. and {Pahl}, Anthony J. and {Reddy}, Naveen A. and {Stark}, Daniel P. and {Berg}, Danielle A. and {Clarke}, Leonardo and {Cullen}, Fergus and {Dunlop}, James S. and {Ellis}, Richard S. and {F{\"o}rster Schreiber}, N.~M. and {Illingworth}, Garth D. and {Jones}, Tucker and {Narayanan}, Desika and {Pettini}, Max and {Schaerer}, Daniel},
        title = "{The AURORA Survey: The Evolution of Multi-phase Electron Densities at High Redshift}",
      journal = {arXiv e-prints},
         year = 2025,
        month = feb,
          eid = {arXiv:2502.08712},
        pages = {arXiv:2502.08712},
          doi = {10.48550/arXiv.2502.08712},
archivePrefix = {arXiv},
       eprint = {2502.08712},
 primaryClass = {astro-ph.GA},
       adsurl = {https://ui.adsabs.harvard.edu/abs/2025arXiv250208712T}
}

@ARTICLE{Harikane:2025,
       author = {{Harikane}, Yuichi and {Sanders}, Ryan L. and {Ellis}, Richard and {Jones}, Tucker and {Ouchi}, Masami and {Laporte}, Nicolas and {Roberts-Borsani}, Guido and {Katz}, Harley and {Nakajima}, Kimihiko and {Ono}, Yoshiaki and {Gupta}, Mansi},
        title = "{JWST \& ALMA Joint Analysis with [OII]$λλ$3726,3729, [OIII]$λ$4363, [OIII]88$μ$m, and [OIII]52$μ$m: Multi-Zone Evolution of Electron Densities at $\mathbf{z\sim0-14}$ and Its Impact on Metallicity Measurements}",
      journal = {arXiv e-prints},
         year = 2025,
        month = may,
          eid = {arXiv:2505.09186},
        pages = {arXiv:2505.09186},
          doi = {10.48550/arXiv.2505.09186},
archivePrefix = {arXiv},
       eprint = {2505.09186},
 primaryClass = {astro-ph.GA},
       adsurl = {https://ui.adsabs.harvard.edu/abs/2025arXiv250509186H}
}

@ARTICLE{Mingozzi:2022,
       author = {{Mingozzi}, Matilde and {James}, Bethan L. and {Arellano-C{\'o}rdova}, Karla Z. and {Berg}, Danielle A. and {Senchyna}, Peter and {Chisholm}, John and {Brinchmann}, Jarle and {Aloisi}, Alessandra and {Amor{\'\i}n}, Ricardo O. and {Charlot}, St{\'e}phane and {Feltre}, Anna and {Hayes}, Matthew and {Heckman}, Timothy and {Henry}, Alaina and {Hernandez}, Svea and {Kumari}, Nimisha and {Leitherer}, Claus and {Llerena}, Mario and {Martin}, Crystal L. and {Nanayakkara}, Themiya and {Ravindranath}, Swara and {Skillman}, Evan D. and {Sugahara}, Yuma and {Wofford}, Aida and {Xu}, Xinfeng},
        title = "{CLASSY IV. Exploring UV Diagnostics of the Interstellar Medium in Local High-z Analogs at the Dawn of the JWST Era}",
      journal = {\apj},
         year = 2022,
        month = nov,
       volume = {939},
       number = {2},
          eid = {110},
        pages = {110},
          doi = {10.3847/1538-4357/ac952c},
archivePrefix = {arXiv},
       eprint = {2209.09047},
 primaryClass = {astro-ph.GA},
       adsurl = {https://ui.adsabs.harvard.edu/abs/2022ApJ...939..110M}
}

@ARTICLE{Peimbert:1967,
       author = {{Peimbert}, Manuel},
        title = "{Temperature Determinations of H II Regions}",
      journal = {\apj},
         year = 1967,
        month = dec,
       volume = {150},
        pages = {825},
          doi = {10.1086/149385},
       adsurl = {https://ui.adsabs.harvard.edu/abs/1967ApJ...150..825P}
}

@ARTICLE{Cameron:2023,
       author = {{Cameron}, Alex J. and {Katz}, Harley and {Rey}, Martin P.},
        title = "{A novel approach to correcting T$_{e}$-based mass-metallicity relations}",
      journal = {\mnras},
         year = 2023,
        month = jun,
       volume = {522},
       number = {1},
        pages = {L89-L94},
          doi = {10.1093/mnrasl/slad046},
archivePrefix = {arXiv},
       eprint = {2210.14234},
 primaryClass = {astro-ph.GA},
       adsurl = {https://ui.adsabs.harvard.edu/abs/2023MNRAS.522L..89C}
}

@ARTICLE{Pilyugin:2012,
       author = {{Pilyugin}, L.~S. and {V{\'\i}lchez}, J.~M. and {Mattsson}, L. and {Thuan}, T.~X.},
        title = "{Abundance determination from global emission-line SDSS spectra: exploring objects with high N/O ratios}",
      journal = {\mnras},
         year = 2012,
        month = apr,
       volume = {421},
       number = {2},
        pages = {1624-1634},
          doi = {10.1111/j.1365-2966.2012.20420.x},
archivePrefix = {arXiv},
       eprint = {1201.1554},
 primaryClass = {astro-ph.CO},
       adsurl = {https://ui.adsabs.harvard.edu/abs/2012MNRAS.421.1624P}
}

@ARTICLE{Dere:1997,
       author = {{Dere}, K.~P. and {Landi}, E. and {Mason}, H.~E. and {Monsignori Fossi}, B.~C. and {Young}, P.~R.},
        title = "{CHIANTI - an atomic database for emission lines}",
      journal = {\aaps},
         year = 1997,
        month = oct,
       volume = {125},
        pages = {149-173},
          doi = {10.1051/aas:1997368},
       adsurl = {https://ui.adsabs.harvard.edu/abs/1997A&AS..125..149D}
}

@ARTICLE{DelZanna:2021,
       author = {{Del Zanna}, G. and {Dere}, K.~P. and {Young}, P.~R. and {Landi}, E.},
        title = "{CHIANTI{\textemdash}An Atomic Database for Emission Lines. XVI. Version 10, Further Extensions}",
      journal = {\apj},
         year = 2021,
        month = mar,
       volume = {909},
       number = {1},
          eid = {38},
        pages = {38},
          doi = {10.3847/1538-4357/abd8ce},
archivePrefix = {arXiv},
       eprint = {2011.05211},
 primaryClass = {physics.atom-ph},
       adsurl = {https://ui.adsabs.harvard.edu/abs/2021ApJ...909...38D}
}

@ARTICLE{Laseter:2024,
       author = {{Laseter}, Isaac H. and {Maseda}, Michael V. and {Curti}, Mirko and {Maiolino}, Roberto and {D'Eugenio}, Francesco and {Cameron}, Alex J. and {Looser}, Tobias J. and {Arribas}, Santiago and {Baker}, William M. and {Bhatawdekar}, Rachana and {Boyett}, Kristan and {Bunker}, Andrew J. and {Carniani}, Stefano and {Charlot}, Stephane and {Chevallard}, Jacopo and {Curtis-lake}, Emma and {Egami}, Eiichi and {Eisenstein}, Daniel J. and {Hainline}, Kevin and {Hausen}, Ryan and {Ji}, Zhiyuan and {Kumari}, Nimisha and {Perna}, Michele and {Rawle}, Tim and {Rix}, Hans-Walter and {Robertson}, Brant and {Rodr{\'\i}guez Del Pino}, Bruno and {Sandles}, Lester and {Scholtz}, Jan and {Smit}, Renske and {Tacchella}, Sandro and {{\"U}bler}, Hannah and {Williams}, Christina C. and {Willott}, Chris and {Witstok}, Joris},
        title = "{JADES: Detecting [OIII]{\ensuremath{\lambda}}4363 emitters and testing strong line calibrations in the high-z Universe with ultra-deep JWST/NIRSpec spectroscopy up to z {\ensuremath{\sim}} 9.5}",
      journal = {\aap},
         year = 2024,
        month = jan,
       volume = {681},
          eid = {A70},
        pages = {A70},
          doi = {10.1051/0004-6361/202347133},
archivePrefix = {arXiv},
       eprint = {2306.03120},
 primaryClass = {astro-ph.GA},
       adsurl = {https://ui.adsabs.harvard.edu/abs/2024A&A...681A..70L}
}

@ARTICLE{Izotov:2006,
       author = {{Izotov}, Y.~I. and {Stasi{\'n}ska}, G. and {Meynet}, G. and {Guseva}, N.~G. and {Thuan}, T.~X.},
        title = "{The chemical composition of metal-poor emission-line galaxies in the Data Release 3 of the Sloan Digital Sky Survey}",
      journal = {\aap},
         year = 2006,
        month = mar,
       volume = {448},
       number = {3},
        pages = {955-970},
          doi = {10.1051/0004-6361:20053763},
archivePrefix = {arXiv},
       eprint = {astro-ph/0511644},
 primaryClass = {astro-ph},
       adsurl = {https://ui.adsabs.harvard.edu/abs/2006A&A...448..955I}
}

@ARTICLE{Curti:2023,
       author = {{Curti}, Mirko and {D'Eugenio}, Francesco and {Carniani}, Stefano and {Maiolino}, Roberto and {Sandles}, Lester and {Witstok}, Joris and {Baker}, William M. and {Bennett}, Jake S. and {Piotrowska}, Joanna M. and {Tacchella}, Sandro and {Charlot}, Stephane and {Nakajima}, Kimihiko and {Maheson}, Gabriel and {Mannucci}, Filippo and {Amiri}, Amirnezam and {Arribas}, Santiago and {Belfiore}, Francesco and {Bonaventura}, Nina R. and {Bunker}, Andrew J. and {Chevallard}, Jacopo and {Cresci}, Giovanni and {Curtis-Lake}, Emma and {Hayden-Pawson}, Connor and {Jones}, Gareth C. and {Kumari}, Nimisha and {Laseter}, Isaac and {Looser}, Tobias J. and {Marconi}, Alessandro and {Maseda}, Michael V. and {Scholtz}, Jan and {Smit}, Renske and {{\"U}bler}, Hannah and {Wallace}, Imaan E.~B.},
        title = "{The chemical enrichment in the early Universe as probed by JWST via direct metallicity measurements at z {\ensuremath{\sim}} 8}",
      journal = {\mnras},
         year = 2023,
        month = jan,
       volume = {518},
       number = {1},
        pages = {425-438},
          doi = {10.1093/mnras/stac2737},
archivePrefix = {arXiv},
       eprint = {2207.12375},
 primaryClass = {astro-ph.GA},
       adsurl = {https://ui.adsabs.harvard.edu/abs/2023MNRAS.518..425C}
}

@ARTICLE{Cataldi:2025,
       author = {{Cataldi}, E. and {Belfiore}, F. and {Curti}, M. and {Moreschini}, B. and {Mannucci}, F. and {D'Amato}, Q. and {Cresci}, G. and {Feltre}, A. and {Ginolfi}, M. and {Marconi}, A. and {Amiri}, A. and {Arnaboldi}, M. and {Bertola}, E. and {Bracci}, C. and {Carniani}, S. and {Ceci}, M. and {Chakraborty}, A. and {Cirasuolo}, M. and {Cullen}, F. and {Kobayashi}, C. and {Kumari}, N. and {Maiolino}, R. and {Marconcini}, C. and {Scialpi}, M. and {Ulivi}, L.},
        title = "{MARTA: Temperature-temperature relationships and strong-line metallicity calibrations from multiple auroral lines detections at cosmic noon}",
      journal = {arXiv e-prints},
         year = 2025,
        month = apr,
          eid = {arXiv:2504.03839},
        pages = {arXiv:2504.03839},
          doi = {10.48550/arXiv.2504.03839},
archivePrefix = {arXiv},
       eprint = {2504.03839},
 primaryClass = {astro-ph.GA},
       adsurl = {https://ui.adsabs.harvard.edu/abs/2025arXiv250403839C}
}

@ARTICLE{Andrews:2013,
       author = {{Andrews}, Brett H. and {Martini}, Paul},
        title = "{The Mass-Metallicity Relation with the Direct Method on Stacked Spectra of SDSS Galaxies}",
      journal = {\apj},
         year = 2013,
        month = mar,
       volume = {765},
       number = {2},
          eid = {140},
        pages = {140},
          doi = {10.1088/0004-637X/765/2/140},
archivePrefix = {arXiv},
       eprint = {1211.3418},
 primaryClass = {astro-ph.CO},
       adsurl = {https://ui.adsabs.harvard.edu/abs/2013ApJ...765..140A}
}

@ARTICLE{Curti:2017,
       author = {{Curti}, M. and {Cresci}, G. and {Mannucci}, F. and {Marconi}, A. and {Maiolino}, R. and {Esposito}, S.},
        title = "{New fully empirical calibrations of strong-line metallicity indicators in star-forming galaxies}",
      journal = {\mnras},
         year = 2017,
        month = feb,
       volume = {465},
       number = {2},
        pages = {1384-1400},
          doi = {10.1093/mnras/stw2766},
archivePrefix = {arXiv},
       eprint = {1610.06939},
 primaryClass = {astro-ph.GA},
       adsurl = {https://ui.adsabs.harvard.edu/abs/2017MNRAS.465.1384C}
}

@ARTICLE{Curti:2020,
       author = {{Curti}, Mirko and {Mannucci}, Filippo and {Cresci}, Giovanni and {Maiolino}, Roberto},
        title = "{The mass-metallicity and the fundamental metallicity relation revisited on a fully T$_{e}$-based abundance scale for galaxies}",
      journal = {\mnras},
         year = 2020,
        month = jan,
       volume = {491},
       number = {1},
        pages = {944-964},
          doi = {10.1093/mnras/stz2910},
archivePrefix = {arXiv},
       eprint = {1910.00597},
 primaryClass = {astro-ph.GA},
       adsurl = {https://ui.adsabs.harvard.edu/abs/2020MNRAS.491..944C}
}

@ARTICLE{Laseter:2022,
       author = {{Laseter}, Isaac H. and {Barger}, Amy J. and {Cowie}, Lennox L. and {Taylor}, Anthony J.},
        title = "{Properties of the Lowest-metallicity Galaxies over the Redshift Range z = 0.2 to z = 1}",
      journal = {\apj},
         year = 2022,
        month = aug,
       volume = {935},
       number = {2},
          eid = {150},
        pages = {150},
          doi = {10.3847/1538-4357/ac81c7},
archivePrefix = {arXiv},
       eprint = {2207.09463},
 primaryClass = {astro-ph.GA},
       adsurl = {https://ui.adsabs.harvard.edu/abs/2022ApJ...935..150L}
}

@ARTICLE{Rosdahl:2018,
       author = {{Rosdahl}, Joakim and {Katz}, Harley and {Blaizot}, J{\'e}r{\'e}my and {Kimm}, Taysun and {Michel-Dansac}, L{\'e}o and {Garel}, Thibault and {Haehnelt}, Martin and {Ocvirk}, Pierre and {Teyssier}, Romain},
        title = "{The SPHINX cosmological simulations of the first billion years: the impact of binary stars on reionization}",
      journal = {\mnras},
         year = 2018,
        month = sep,
       volume = {479},
       number = {1},
        pages = {994-1016},
          doi = {10.1093/mnras/sty1655},
archivePrefix = {arXiv},
       eprint = {1801.07259},
 primaryClass = {astro-ph.GA},
       adsurl = {https://ui.adsabs.harvard.edu/abs/2018MNRAS.479..994R}
}

@ARTICLE{Katz:2023,
       author = {{Katz}, Harley and {Rosdahl}, Joki and {Kimm}, Taysun and {Blaizot}, Jeremy and {Choustikov}, Nicholas and {Farcy}, Marion and {Garel}, Thibault and {Haehnelt}, Martin G. and {Michel-Dansac}, Leo and {Ocvirk}, Pierre},
        title = "{The SPHINX Public Data Release: Forward Modelling High-Redshift JWST Observations with Cosmological Radiation Hydrodynamics Simulations}",
      journal = {The Open Journal of Astrophysics},
         year = 2023,
        month = dec,
       volume = {6},
          eid = {44},
        pages = {44},
          doi = {10.21105/astro.2309.03269},
archivePrefix = {arXiv},
       eprint = {2309.03269},
 primaryClass = {astro-ph.GA},
       adsurl = {https://ui.adsabs.harvard.edu/abs/2023OJAp....6E..44K}
}

@ARTICLE{Choustikov:2024,
       author = {{Choustikov}, Nicholas and {Katz}, Harley and {Saxena}, Aayush and {Cameron}, Alex J. and {Devriendt}, Julien and {Slyz}, Adrianne and {Rosdahl}, Joki and {Blaizot}, Jeremy and {Michel-Dansac}, Leo},
        title = "{The Physics of Indirect Estimators of Lyman Continuum Escape and their Application to High-Redshift JWST Galaxies}",
      journal = {\mnras},
         year = 2024,
        month = apr,
       volume = {529},
       number = {4},
        pages = {3751-3767},
          doi = {10.1093/mnras/stae776},
archivePrefix = {arXiv},
       eprint = {2304.08526},
 primaryClass = {astro-ph.GA},
       adsurl = {https://ui.adsabs.harvard.edu/abs/2024MNRAS.529.3751C}
}

@ARTICLE{Marks:2012,
       author = {{Marks}, Michael and {Kroupa}, Pavel and {Dabringhausen}, J{\"o}rg and {Pawlowski}, Marcel S.},
        title = "{Evidence for top-heavy stellar initial mass functions with increasing density and decreasing metallicity}",
      journal = {\mnras},
         year = 2012,
        month = may,
       volume = {422},
       number = {3},
        pages = {2246-2254},
          doi = {10.1111/j.1365-2966.2012.20767.x},
archivePrefix = {arXiv},
       eprint = {1202.4755},
 primaryClass = {astro-ph.GA},
       adsurl = {https://ui.adsabs.harvard.edu/abs/2012MNRAS.422.2246M}
}

@ARTICLE{Cameron:2024,
       author = {{Cameron}, Alex J. and {Katz}, Harley and {Witten}, Callum and {Saxena}, Aayush and {Laporte}, Nicolas and {Bunker}, Andrew J.},
        title = "{Nebular dominated galaxies: insights into the stellar initial mass function at high redshift}",
      journal = {\mnras},
         year = 2024,
        month = oct,
       volume = {534},
       number = {1},
        pages = {523-543},
          doi = {10.1093/mnras/stae1547},
archivePrefix = {arXiv},
       eprint = {2311.02051},
 primaryClass = {astro-ph.GA},
       adsurl = {https://ui.adsabs.harvard.edu/abs/2024MNRAS.534..523C}
}

@ARTICLE{Topping:2024,
       author = {{Topping}, Michael W. and {Stark}, Daniel P. and {Senchyna}, Peter and {Plat}, Adele and {Zitrin}, Adi and {Endsley}, Ryan and {Charlot}, St{\'e}phane and {Furtak}, Lukas J. and {Maseda}, Michael V. and {Smit}, Renske and {Mainali}, Ramesh and {Chevallard}, Jacopo and {Molyneux}, Stephen and {Rigby}, Jane R.},
        title = "{Metal-poor star formation at z > 6 with JWST: new insight into hard radiation fields and nitrogen enrichment on 20 pc scales}",
      journal = {\mnras},
         year = 2024,
        month = apr,
       volume = {529},
       number = {4},
        pages = {3301-3322},
          doi = {10.1093/mnras/stae682},
archivePrefix = {arXiv},
       eprint = {2401.08764},
 primaryClass = {astro-ph.GA},
       adsurl = {https://ui.adsabs.harvard.edu/abs/2024MNRAS.529.3301T}
}

@ARTICLE{Pollock:2025,
       author = {{Pollock}, Clara L. and {Gottumukkala}, Rashmi and {Heintz}, Kasper E. and {Brammer}, Gabriel B. and {Roberts-Borsani}, Guido and {Oesch}, Pascal A. and {Witstok}, Joris and {Arellano-C{\'o}rdova}, Karla Z. and {Cullen}, Fergus and {Scholte}, Dirk and {Terp}, Chamilla and {Rowland}, Lucie and {Sneppen}, Albert and {Ito}, Kei and {Valentino}, Francesco and {Matthee}, Jorryt and {Watson}, Darach and {Toft}, Sune},
        title = "{Novel $z\sim~10$ auroral line measurements extend the gradual offset of the FMR deep into the first Gyr of cosmic time}",
      journal = {arXiv e-prints},
         year = 2025,
        month = jun,
          eid = {arXiv:2506.15779},
        pages = {arXiv:2506.15779},
          doi = {10.48550/arXiv.2506.15779},
archivePrefix = {arXiv},
       eprint = {2506.15779},
 primaryClass = {astro-ph.GA},
       adsurl = {https://ui.adsabs.harvard.edu/abs/2025arXiv250615779P}
}

@ARTICLE{Hayes:2025,
       author = {{Hayes}, Matthew J. and {Saldana-Lopez}, Alberto and {Citro}, Annalisa and {James}, Bethan L. and {Mingozzi}, Matilde and {Scarlata}, Claudia and {Martinez}, Zorayda and {Berg}, Danielle A.},
        title = "{On the Average Ultraviolet Emission-line Spectra of High-redshift Galaxies: Hot and Cold, Carbon-poor, Nitrogen Modest, and Oozing Ionizing Photons}",
      journal = {\apj},
         year = 2025,
        month = mar,
       volume = {982},
       number = {1},
          eid = {14},
        pages = {14},
          doi = {10.3847/1538-4357/adaea1},
archivePrefix = {arXiv},
       eprint = {2411.09262},
 primaryClass = {astro-ph.GA},
       adsurl = {https://ui.adsabs.harvard.edu/abs/2025ApJ...982...14H}
}

@ARTICLE{Bunker:2023,
       author = {{Bunker}, Andrew J. and {Saxena}, Aayush and {Cameron}, Alex J. and {Willott}, Chris J. and {Curtis-Lake}, Emma and {Jakobsen}, Peter and {Carniani}, Stefano and {Smit}, Renske and {Maiolino}, Roberto and {Witstok}, Joris and {Curti}, Mirko and {D'Eugenio}, Francesco and {Jones}, Gareth C. and {Ferruit}, Pierre and {Arribas}, Santiago and {Charlot}, Stephane and {Chevallard}, Jacopo and {Giardino}, Giovanna and {de Graaff}, Anna and {Looser}, Tobias J. and {L{\"u}tzgendorf}, Nora and {Maseda}, Michael V. and {Rawle}, Tim and {Rix}, Hans-Walter and {Del Pino}, Bruno Rodr{\'\i}guez and {Alberts}, Stacey and {Egami}, Eiichi and {Eisenstein}, Daniel J. and {Endsley}, Ryan and {Hainline}, Kevin and {Hausen}, Ryan and {Johnson}, Benjamin D. and {Rieke}, George and {Rieke}, Marcia and {Robertson}, Brant E. and {Shivaei}, Irene and {Stark}, Daniel P. and {Sun}, Fengwu and {Tacchella}, Sandro and {Tang}, Mengtao and {Williams}, Christina C. and {Willmer}, Christopher N.~A. and {Baker}, William M. and {Baum}, Stefi and {Bhatawdekar}, Rachana and {Bowler}, Rebecca and {Boyett}, Kristan and {Chen}, Zuyi and {Circosta}, Chiara and {Helton}, Jakob M. and {Ji}, Zhiyuan and {Kumari}, Nimisha and {Lyu}, Jianwei and {Nelson}, Erica and {Parlanti}, Eleonora and {Perna}, Michele and {Sandles}, Lester and {Scholtz}, Jan and {Suess}, Katherine A. and {Topping}, Michael W. and {{\"U}bler}, Hannah and {Wallace}, Imaan E.~B. and {Whitler}, Lily},
        title = "{JADES NIRSpec Spectroscopy of GN-z11: Lyman-{\ensuremath{\alpha}} emission and possible enhanced nitrogen abundance in a z = 10.60 luminous galaxy}",
      journal = {\aap},
         year = 2023,
        month = sep,
       volume = {677},
          eid = {A88},
        pages = {A88},
          doi = {10.1051/0004-6361/202346159},
archivePrefix = {arXiv},
       eprint = {2302.07256},
 primaryClass = {astro-ph.GA},
       adsurl = {https://ui.adsabs.harvard.edu/abs/2023A&A...677A..88B}
}

@ARTICLE{Stanway:2018,
       author = {{Stanway}, E.~R. and {Eldridge}, J.~J.},
        title = "{Re-evaluating old stellar populations}",
      journal = {\mnras},
         year = 2018,
        month = sep,
       volume = {479},
       number = {1},
        pages = {75-93},
          doi = {10.1093/mnras/sty1353},
archivePrefix = {arXiv},
       eprint = {1805.08784},
 primaryClass = {astro-ph.GA},
       adsurl = {https://ui.adsabs.harvard.edu/abs/2018MNRAS.479...75S}
}

@ARTICLE{Eldridge:2017,
       author = {{Eldridge}, J.~J. and {Stanway}, E.~R. and {Xiao}, L. and {McClelland}, L.~A.~S. and {Taylor}, G. and {Ng}, M. and {Greis}, S.~M.~L. and {Bray}, J.~C.},
        title = "{Binary Population and Spectral Synthesis Version 2.1: Construction, Observational Verification, and New Results}",
      journal = {\pasa},
         year = 2017,
        month = nov,
       volume = {34},
          eid = {e058},
        pages = {e058},
          doi = {10.1017/pasa.2017.51},
archivePrefix = {arXiv},
       eprint = {1710.02154},
 primaryClass = {astro-ph.SR},
       adsurl = {https://ui.adsabs.harvard.edu/abs/2017PASA...34...58E}
}

@ARTICLE{Leitherer:1999,
       author = {{Leitherer}, Claus and {Schaerer}, Daniel and {Goldader}, Jeffrey D. and {Delgado}, Rosa M. Gonz{\'a}lez and {Robert}, Carmelle and {Kune}, Denis Foo and {de Mello}, Du{\'\i}lia F. and {Devost}, Daniel and {Heckman}, Timothy M.},
        title = "{Starburst99: Synthesis Models for Galaxies with Active Star Formation}",
      journal = {\apjs},
         year = 1999,
        month = jul,
       volume = {123},
       number = {1},
        pages = {3-40},
          doi = {10.1086/313233},
archivePrefix = {arXiv},
       eprint = {astro-ph/9902334},
 primaryClass = {astro-ph},
       adsurl = {https://ui.adsabs.harvard.edu/abs/1999ApJS..123....3L}
}

@ARTICLE{Curti:2024,
       author = {{Curti}, Mirko and {Maiolino}, Roberto and {Curtis-Lake}, Emma and {Chevallard}, Jacopo and {Carniani}, Stefano and {D'Eugenio}, Francesco and {Looser}, Tobias J. and {Scholtz}, Jan and {Charlot}, Stephane and {Cameron}, Alex and {{\"U}bler}, Hannah and {Witstok}, Joris and {Boyett}, Kristian and {Laseter}, Isaac and {Sandles}, Lester and {Arribas}, Santiago and {Bunker}, Andrew and {Giardino}, Giovanna and {Maseda}, Michael V. and {Rawle}, Tim and {Rodr{\'\i}guez Del Pino}, Bruno and {Smit}, Renske and {Willott}, Chris J. and {Eisenstein}, Daniel J. and {Hausen}, Ryan and {Johnson}, Benjamin and {Rieke}, Marcia and {Robertson}, Brant and {Tacchella}, Sandro and {Williams}, Christina C. and {Willmer}, Christopher and {Baker}, William M. and {Bhatawdekar}, Rachana and {Egami}, Eiichi and {Helton}, Jakob M. and {Ji}, Zhiyuan and {Kumari}, Nimisha and {Perna}, Michele and {Shivaei}, Irene and {Sun}, Fengwu},
        title = "{JADES: Insights into the low-mass end of the mass-metallicity-SFR relation at 3 < z < 10 from deep JWST/NIRSpec spectroscopy}",
      journal = {\aap},
         year = 2024,
        month = apr,
       volume = {684},
          eid = {A75},
        pages = {A75},
          doi = {10.1051/0004-6361/202346698},
archivePrefix = {arXiv},
       eprint = {2304.08516},
 primaryClass = {astro-ph.GA},
       adsurl = {https://ui.adsabs.harvard.edu/abs/2024A&A...684A..75C}
}

@ARTICLE{Choustikov:2024b,
       author = {{Choustikov}, Nicholas and {Katz}, Harley and {Saxena}, Aayush and {Garel}, Thibault and {Devriendt}, Julien and {Slyz}, Adrianne and {Kimm}, Taysun and {Blaizot}, Jeremy and {Rosdahl}, Joki},
        title = "{The great escape: understanding the connection between Ly {\ensuremath{\alpha}} emission and LyC escape in simulated JWST analogues}",
      journal = {\mnras},
         year = 2024,
        month = aug,
       volume = {532},
       number = {2},
        pages = {2463-2484},
          doi = {10.1093/mnras/stae1586},
archivePrefix = {arXiv},
       eprint = {2401.09557},
 primaryClass = {astro-ph.GA},
       adsurl = {https://ui.adsabs.harvard.edu/abs/2024MNRAS.532.2463C}
}

@ARTICLE{Giovinazzo:2025,
       author = {{Giovinazzo}, Emma and {Oesch}, Pascal A. and {Weibel}, Andrea and {Meyer}, Romain A. and {Witten}, Callum and {Bhagwat}, Aniket and {Brammer}, Gabriel and {Chisholm}, John and {de Graaff}, Anna and {Gottumukkala}, Rashmi and {Jecmen}, Michelle and {Katz}, Harley and {Leja}, Joel and {Marques-Chaves}, Rui and {Maseda}, Michael and {Shivaei}, Irene and {Trebitsch}, Maxime and {Verhamme}, Anne},
        title = "{Breaking Through the Cosmic Fog: JWST/NIRSpec Constraints on Ionizing Photon Escape in Reionization-Era Galaxies}",
      journal = {arXiv e-prints},
         year = 2025,
        month = jul,
          eid = {arXiv:2507.01096},
        pages = {arXiv:2507.01096},
          doi = {10.48550/arXiv.2507.01096},
archivePrefix = {arXiv},
       eprint = {2507.01096},
 primaryClass = {astro-ph.GA},
       adsurl = {https://ui.adsabs.harvard.edu/abs/2025arXiv250701096G}
}

@ARTICLE{Boyett:2024,
       author = {{Boyett}, Kit and {Bunker}, Andrew J. and {Curtis-Lake}, Emma and {Chevallard}, Jacopo and {Cameron}, Alex J. and {Jones}, Gareth C. and {Saxena}, Aayush and {Charlot}, St{\'e}phane and {Curti}, Mirko and {Wallace}, Imaan E.~B. and {Arribas}, Santiago and {Carniani}, Stefano and {Willott}, Chris and {Alberts}, Stacey and {Eisenstein}, Daniel J. and {Hainline}, Kevin and {Hausen}, Ryan and {Johnson}, Benjamin D. and {Rieke}, Marcia and {Robertson}, Brant and {Stark}, Daniel P. and {Tacchella}, Sandro and {Williams}, Christina C. and {Chen}, Zuyi and {Egami}, Eiichi and {Endsley}, Ryan and {Kumari}, Nimisha and {Laseter}, Isaac and {Looser}, Tobias J. and {Maseda}, Michael V. and {Scholtz}, Jan and {Shivaei}, Irene and {Simmonds}, Charlotte and {Smit}, Renske and {{\"U}bler}, Hannah and {Witstok}, Joris},
        title = "{Extreme emission line galaxies detected in JADES JWST/NIRSpec - I. Inferred galaxy properties}",
      journal = {\mnras},
         year = 2024,
        month = dec,
       volume = {535},
       number = {2},
        pages = {1796-1828},
          doi = {10.1093/mnras/stae2430},
archivePrefix = {arXiv},
       eprint = {2401.16934},
 primaryClass = {astro-ph.GA},
       adsurl = {https://ui.adsabs.harvard.edu/abs/2024MNRAS.535.1796B}
}

@ARTICLE{Wilkins:2023,
       author = {{Wilkins}, Stephen M. and {Lovell}, Christopher C. and {Vijayan}, Aswin P. and {Irodotou}, Dimitrios and {Adams}, Nathan J. and {Roper}, William J. and {Caruana}, Joseph and {Matthee}, Jorryt and {Seeyave}, Louise T.~C. and {Conselice}, Christopher J. and {P{\'e}rez-Gonz{\'a}lez}, Pablo G. and {Turner}, Jack C. and {Donnellan}, James M.~S. and {Verma}, Aprajita and {Trussler}, J.~A.~A.},
        title = "{First light and reionization epoch simulations (FLARES) XI: [O III] emitting galaxies at 5 < z < 10}",
      journal = {\mnras},
         year = 2023,
        month = jul,
       volume = {522},
       number = {3},
        pages = {4014-4027},
          doi = {10.1093/mnras/stad1126},
archivePrefix = {arXiv},
       eprint = {2301.13038},
 primaryClass = {astro-ph.GA},
       adsurl = {https://ui.adsabs.harvard.edu/abs/2023MNRAS.522.4014W}
}

@ARTICLE{Tang:2025,
       author = {{Tang}, Mengtao and {Stark}, Daniel P. and {Mason}, Charlotte A. and {Gelli}, Viola and {Chen}, Zuyi and {Topping}, Michael W.},
        title = "{The JWST Spectroscopic Properties of Galaxies at $z=9-14$}",
      journal = {arXiv e-prints},
         year = 2025,
        month = jul,
          eid = {arXiv:2507.08245},
        pages = {arXiv:2507.08245},
          doi = {10.48550/arXiv.2507.08245},
archivePrefix = {arXiv},
       eprint = {2507.08245},
 primaryClass = {astro-ph.GA},
       adsurl = {https://ui.adsabs.harvard.edu/abs/2025arXiv250708245T}
}

@ARTICLE{Cameron:2023b,
       author = {{Cameron}, Alex J. and {Saxena}, Aayush and {Bunker}, Andrew J. and {D'Eugenio}, Francesco and {Carniani}, Stefano and {Maiolino}, Roberto and {Curtis-Lake}, Emma and {Ferruit}, Pierre and {Jakobsen}, Peter and {Arribas}, Santiago and {Bonaventura}, Nina and {Charlot}, Stephane and {Chevallard}, Jacopo and {Curti}, Mirko and {Looser}, Tobias J. and {Maseda}, Michael V. and {Rawle}, Tim and {Rodr{\'\i}guez Del Pino}, Bruno and {Smit}, Renske and {{\"U}bler}, Hannah and {Willott}, Chris and {Witstok}, Joris and {Egami}, Eiichi and {Eisenstein}, Daniel J. and {Johnson}, Benjamin D. and {Hainline}, Kevin and {Rieke}, Marcia and {Robertson}, Brant E. and {Stark}, Daniel P. and {Tacchella}, Sandro and {Williams}, Christina C. and {Willmer}, Christopher N.~A. and {Bhatawdekar}, Rachana and {Bowler}, Rebecca and {Boyett}, Kristan and {Circosta}, Chiara and {Helton}, Jakob M. and {Jones}, Gareth C. and {Kumari}, Nimisha and {Ji}, Zhiyuan and {Nelson}, Erica and {Parlanti}, Eleonora and {Sandles}, Lester and {Scholtz}, Jan and {Sun}, Fengwu},
        title = "{JADES: Probing interstellar medium conditions at z {\ensuremath{\sim}} 5.5-9.5 with ultra-deep JWST/NIRSpec spectroscopy}",
      journal = {\aap},
         year = 2023,
        month = sep,
       volume = {677},
          eid = {A115},
        pages = {A115},
          doi = {10.1051/0004-6361/202346107},
archivePrefix = {arXiv},
       eprint = {2302.04298},
 primaryClass = {astro-ph.GA},
       adsurl = {https://ui.adsabs.harvard.edu/abs/2023A&A...677A.115C}
}

@ARTICLE{Katz:2024,
       author = {{Katz}, Harley and {Rey}, Martin P. and {Cadiou}, Corentin and {Kimm}, Taysun and {Agertz}, Oscar},
        title = "{The Impact of Star Formation and Feedback Recipes on the Stellar Mass and Interstellar Medium of High-Redshift Galaxies}",
      journal = {arXiv e-prints},
         year = 2024,
        month = nov,
          eid = {arXiv:2411.07282},
        pages = {arXiv:2411.07282},
          doi = {10.48550/arXiv.2411.07282},
archivePrefix = {arXiv},
       eprint = {2411.07282},
 primaryClass = {astro-ph.GA},
       adsurl = {https://ui.adsabs.harvard.edu/abs/2024arXiv241107282K}
}

@ARTICLE{Witten:2025,
       author = {{Witten}, Callum and {McClymont}, William and {Laporte}, Nicolas and {Roberts-Borsani}, Guido and {Sijacki}, Debora and {Tacchella}, Sandro and {Simmonds}, Charlotte and {Katz}, Harley and {Ellis}, Richard S. and {Witstok}, Joris and {Maiolino}, Roberto and {Ji}, Xihan and {Hayes}, Billy R. and {Looser}, Tobias J. and {D'Eugenio}, Francesco},
        title = "{Rising from the ashes: evidence of old stellar populations and rejuvenation events in the very early Universe}",
      journal = {\mnras},
         year = 2025,
        month = feb,
       volume = {537},
       number = {1},
        pages = {112-126},
          doi = {10.1093/mnras/staf001},
archivePrefix = {arXiv},
       eprint = {2407.07937},
 primaryClass = {astro-ph.GA},
       adsurl = {https://ui.adsabs.harvard.edu/abs/2025MNRAS.537..112W}
}

@ARTICLE{Jin:2022,
       author = {{Jin}, Yifei and {Kewley}, Lisa J. and {Sutherland}, Ralph S.},
        title = "{Theoretically Modeling Photoionized Regions with Fractal Geometry in Three Dimensions}",
      journal = {\apjl},
         year = 2022,
        month = jul,
       volume = {934},
       number = {1},
          eid = {L8},
        pages = {L8},
          doi = {10.3847/2041-8213/ac80f3},
archivePrefix = {arXiv},
       eprint = {2207.06672},
 primaryClass = {astro-ph.GA},
       adsurl = {https://ui.adsabs.harvard.edu/abs/2022ApJ...934L...8J}
}

@ARTICLE{Usui:2025,
       author = {{Usui}, Mitsutaka and {Mawatari}, Ken and {{\'A}lvarez-M{\'a}rquez}, Javier and {Hashimoto}, Takuya and {Sugahara}, Yuma and {Marques-Chaves}, Rui and {Inoue}, Akio K. and {Colina}, Luis and {Arribas}, Santiago and {Blanco-Prieto}, Carmen and {Nakazato}, Yurina and {Yoshida}, Naoki and {Bakx}, Tom J.~L.~C. and {Ceverino}, Daniel and {Costantin}, Luca and {Crespo G{\'o}mez}, Alejandro and {Hagimoto}, Masato and {Matsuo}, Hiroshi and {Osone}, Wataru and {Ren}, Yi W. and {Fudamoto}, Yoshinobu and {Hashigaya}, Takeshi and {Pereira-Santaella}, Miguel and {Tamura}, Yoichi},
        title = "{RIOJA. JWST and ALMA unveil the inhomogeneous and complex ISM structure in a star-forming galaxy at $z=6.81$}",
      journal = {arXiv e-prints},
         year = 2025,
        month = jul,
          eid = {arXiv:2507.02054},
        pages = {arXiv:2507.02054},
          doi = {10.48550/arXiv.2507.02054},
archivePrefix = {arXiv},
       eprint = {2507.02054},
 primaryClass = {astro-ph.GA},
       adsurl = {https://ui.adsabs.harvard.edu/abs/2025arXiv250702054U}
}

@ARTICLE{Saxena:2021,
       author = {{Saxena}, A. and {Ellis}, R.~S. and {F{\"o}rster}, P.~U. and {Calabr{\`o}}, A. and {Pentericci}, L. and {Carnall}, A.~C. and {Castellano}, M. and {Cullen}, F. and {Fontana}, A. and {Franco}, M. and {Fynbo}, J.~P.~U. and {Gargiulo}, A. and {Garilli}, B. and {Hathi}, N.~P. and {McLeod}, D.~J. and {Amor{\'\i}n}, R. and {Zamorani}, G.},
        title = "{The VANDELS Survey: new constraints on the high-mass X-ray binary populations in normal star-forming galaxies at 3 < z < 5.5}",
      journal = {\mnras},
         year = 2021,
        month = aug,
       volume = {505},
       number = {4},
        pages = {4798-4812},
          doi = {10.1093/mnras/stab1575},
archivePrefix = {arXiv},
       eprint = {2104.02624},
 primaryClass = {astro-ph.GA},
       adsurl = {https://ui.adsabs.harvard.edu/abs/2021MNRAS.505.4798S}
}

@ARTICLE{Kubota:2018,
       author = {{Kubota}, Aya and {Done}, Chris},
        title = "{A physical model of the broad-band continuum of AGN and its implications for the UV/X relation and optical variability}",
      journal = {\mnras},
         year = 2018,
        month = oct,
       volume = {480},
       number = {1},
        pages = {1247-1262},
          doi = {10.1093/mnras/sty1890},
archivePrefix = {arXiv},
       eprint = {1804.00171},
 primaryClass = {astro-ph.HE},
       adsurl = {https://ui.adsabs.harvard.edu/abs/2018MNRAS.480.1247K}
}

@ARTICLE{Trebitsch:2021,
       author = {{Trebitsch}, Maxime and {Dubois}, Yohan and {Volonteri}, Marta and {Pfister}, Hugo and {Cadiou}, Corentin and {Katz}, Harley and {Rosdahl}, Joakim and {Kimm}, Taysun and {Pichon}, Christophe and {Beckmann}, Ricarda S. and {Devriendt}, Julien and {Slyz}, Adrianne},
        title = "{The OBELISK simulation: Galaxies contribute more than AGN to H I reionization of protoclusters}",
      journal = {\aap},
         year = 2021,
        month = sep,
       volume = {653},
          eid = {A154},
        pages = {A154},
          doi = {10.1051/0004-6361/202037698},
archivePrefix = {arXiv},
       eprint = {2002.04045},
 primaryClass = {astro-ph.GA},
       adsurl = {https://ui.adsabs.harvard.edu/abs/2021A&A...653A.154T}
}

@ARTICLE{Lequeux:1979,
       author = {{Lequeux}, J. and {Peimbert}, M. and {Rayo}, J.~F. and {Serrano}, A. and {Torres-Peimbert}, S.},
        title = "{Chemical Composition and Evolution of Irregular and Blue Compact Galaxies}",
      journal = {\aap},
         year = 1979,
        month = dec,
       volume = {80},
        pages = {155},
       adsurl = {https://ui.adsabs.harvard.edu/abs/1979A&A....80..155L}
}

@ARTICLE{Tremonti:2004,
       author = {{Tremonti}, Christy A. and {Heckman}, Timothy M. and {Kauffmann}, Guinevere and {Brinchmann}, Jarle and {Charlot}, St{\'e}phane and {White}, Simon D.~M. and {Seibert}, Mark and {Peng}, Eric W. and {Schlegel}, David J. and {Uomoto}, Alan and {Fukugita}, Masataka and {Brinkmann}, Jon},
        title = "{The Origin of the Mass-Metallicity Relation: Insights from 53,000 Star-forming Galaxies in the Sloan Digital Sky Survey}",
      journal = {\apj},
         year = 2004,
        month = oct,
       volume = {613},
       number = {2},
        pages = {898-913},
          doi = {10.1086/423264},
archivePrefix = {arXiv},
       eprint = {astro-ph/0405537},
 primaryClass = {astro-ph},
       adsurl = {https://ui.adsabs.harvard.edu/abs/2004ApJ...613..898T}
}

@ARTICLE{Ellison:2008,
       author = {{Ellison}, Sara L. and {Patton}, David R. and {Simard}, Luc and {McConnachie}, Alan W.},
        title = "{Clues to the Origin of the Mass-Metallicity Relation: Dependence on Star Formation Rate and Galaxy Size}",
      journal = {\apjl},
         year = 2008,
        month = jan,
       volume = {672},
       number = {2},
        pages = {L107},
          doi = {10.1086/527296},
archivePrefix = {arXiv},
       eprint = {0711.4833},
 primaryClass = {astro-ph},
       adsurl = {https://ui.adsabs.harvard.edu/abs/2008ApJ...672L.107E}
}

@ARTICLE{Mannucci:2010,
       author = {{Mannucci}, F. and {Cresci}, G. and {Maiolino}, R. and {Marconi}, A. and {Gnerucci}, A.},
        title = "{A fundamental relation between mass, star formation rate and metallicity in local and high-redshift galaxies}",
      journal = {\mnras},
         year = 2010,
        month = nov,
       volume = {408},
       number = {4},
        pages = {2115-2127},
          doi = {10.1111/j.1365-2966.2010.17291.x},
archivePrefix = {arXiv},
       eprint = {1005.0006},
 primaryClass = {astro-ph.CO},
       adsurl = {https://ui.adsabs.harvard.edu/abs/2010MNRAS.408.2115M}
}

@ARTICLE{Yates:2012,
       author = {{Yates}, Robert M. and {Kauffmann}, Guinevere and {Guo}, Qi},
        title = "{The relation between metallicity, stellar mass and star formation in galaxies: an analysis of observational and model data}",
      journal = {\mnras},
         year = 2012,
        month = may,
       volume = {422},
       number = {1},
        pages = {215-231},
          doi = {10.1111/j.1365-2966.2012.20595.x},
archivePrefix = {arXiv},
       eprint = {1107.3145},
 primaryClass = {astro-ph.CO},
       adsurl = {https://ui.adsabs.harvard.edu/abs/2012MNRAS.422..215Y}
}

@ARTICLE{Lee:2006,
       author = {{Lee}, Henry and {Skillman}, Evan D. and {Cannon}, John M. and {Jackson}, Dale C. and {Gehrz}, Robert D. and {Polomski}, Elisha F. and {Woodward}, Charles E.},
        title = "{On Extending the Mass-Metallicity Relation of Galaxies by 2.5 Decades in Stellar Mass}",
      journal = {\apj},
         year = 2006,
        month = aug,
       volume = {647},
       number = {2},
        pages = {970-983},
          doi = {10.1086/505573},
archivePrefix = {arXiv},
       eprint = {astro-ph/0605036},
 primaryClass = {astro-ph},
       adsurl = {https://ui.adsabs.harvard.edu/abs/2006ApJ...647..970L}
}

@ARTICLE{Nakajima:2023,
       author = {{Nakajima}, Kimihiko and {Ouchi}, Masami and {Isobe}, Yuki and {Harikane}, Yuichi and {Zhang}, Yechi and {Ono}, Yoshiaki and {Umeda}, Hiroya and {Oguri}, Masamune},
        title = "{JWST Census for the Mass-Metallicity Star Formation Relations at z = 4-10 with Self-consistent Flux Calibration and Proper Metallicity Calibrators}",
      journal = {\apjs},
         year = 2023,
        month = dec,
       volume = {269},
       number = {2},
          eid = {33},
        pages = {33},
          doi = {10.3847/1538-4365/acd556},
archivePrefix = {arXiv},
       eprint = {2301.12825},
 primaryClass = {astro-ph.GA},
       adsurl = {https://ui.adsabs.harvard.edu/abs/2023ApJS..269...33N}
}

@ARTICLE{Chemerynska:2024,
       author = {{Chemerynska}, Iryna and {Atek}, Hakim and {Dayal}, Pratika and {Furtak}, Lukas J. and {Feldmann}, Robert and {Greene}, Jenny E. and {Maseda}, Michael V. and {Nanayakkara}, Themiya and {Oesch}, Pascal A. and {Fujimoto}, Seiji and {Labb{\'e}}, Ivo and {Bezanson}, Rachel and {Brammer}, Gabriel and {Cutler}, Sam E. and {Leja}, Joel and {Pan}, Richard and {Price}, Sedona H. and {Wang}, Bingjie and {Weaver}, John R. and {Whitaker}, Katherine E.},
        title = "{The Extreme Low-mass End of the Mass{\textendash}Metallicity Relation at z {\ensuremath{\sim}} 7}",
      journal = {\apjl},
         year = 2024,
        month = nov,
       volume = {976},
       number = {1},
          eid = {L15},
        pages = {L15},
          doi = {10.3847/2041-8213/ad8dc9},
archivePrefix = {arXiv},
       eprint = {2407.17110},
 primaryClass = {astro-ph.GA},
       adsurl = {https://ui.adsabs.harvard.edu/abs/2024ApJ...976L..15C}
}

@ARTICLE{Heintz:2023,
       author = {{Heintz}, Kasper E. and {Brammer}, Gabriel B. and {Gim{\'e}nez-Arteaga}, Clara and {Strait}, Victoria B. and {Lagos}, Claudia del P. and {Vijayan}, Aswin P. and {Matthee}, Jorryt and {Watson}, Darach and {Mason}, Charlotte A. and {Hutter}, Anne and {Toft}, Sune and {Fynbo}, Johan P.~U. and {Oesch}, Pascal A.},
        title = "{Dilution of chemical enrichment in galaxies 600 Myr after the Big Bang}",
      journal = {Nature Astronomy},
         year = 2023,
        month = dec,
       volume = {7},
        pages = {1517-1524},
          doi = {10.1038/s41550-023-02078-7},
archivePrefix = {arXiv},
       eprint = {2212.02890},
 primaryClass = {astro-ph.GA},
       adsurl = {https://ui.adsabs.harvard.edu/abs/2023NatAs...7.1517H}
}

@ARTICLE{Morishita:2024,
       author = {{Morishita}, Takahiro and {Stiavelli}, Massimo and {Grillo}, Claudio and {Rosati}, Piero and {Schuldt}, Stefan and {Trenti}, Michele and {Bergamini}, Pietro and {Boyett}, Kit and {Chary}, Ranga-Ram and {Leethochawalit}, Nicha and {Roberts-Borsani}, Guido and {Treu}, Tommaso and {Vanzella}, Eros},
        title = "{Diverse Oxygen Abundance in Early Galaxies Unveiled by Auroral Line Analysis with JWST}",
      journal = {\apj},
         year = 2024,
        month = aug,
       volume = {971},
       number = {1},
          eid = {43},
        pages = {43},
          doi = {10.3847/1538-4357/ad5290},
archivePrefix = {arXiv},
       eprint = {2402.14084},
 primaryClass = {astro-ph.GA},
       adsurl = {https://ui.adsabs.harvard.edu/abs/2024ApJ...971...43M}
}

@ARTICLE{Sarkar:2025,
       author = {{Sarkar}, Arnab and {Chakraborty}, Priyanka and {Vogelsberger}, Mark and {McDonald}, Michael and {Torrey}, Paul and {Garcia}, Alex M. and {Khullar}, Gourav and {Ferland}, Gary J. and {Forman}, William and {Wolk}, Scott and {Schneider}, Benjamin and {Bautz}, Mark and {Miller}, Eric and {Grant}, Catherine and {ZuHone}, John},
        title = "{Unveiling the Cosmic Chemistry: Revisiting the Mass{\textendash}Metallicity Relation with JWST/NIRSpec at 4 < z < 10}",
      journal = {\apj},
         year = 2025,
        month = jan,
       volume = {978},
       number = {2},
          eid = {136},
        pages = {136},
          doi = {10.3847/1538-4357/ad8f32},
archivePrefix = {arXiv},
       eprint = {2408.07974},
 primaryClass = {astro-ph.GA},
       adsurl = {https://ui.adsabs.harvard.edu/abs/2025ApJ...978..136S}
}

@ARTICLE{Marszewski:2024,
       author = {{Marszewski}, Andrew and {Sun}, Guochao and {Faucher-Gigu{\`e}re}, Claude-Andr{\'e} and {Hayward}, Christopher C. and {Feldmann}, Robert},
        title = "{The High-Redshift Gas-Phase Mass{\textendash}Metallicity Relation in FIRE-2}",
      journal = {\apjl},
         year = 2024,
        month = jun,
       volume = {967},
       number = {2},
          eid = {L41},
        pages = {L41},
          doi = {10.3847/2041-8213/ad4cee},
archivePrefix = {arXiv},
       eprint = {2403.08853},
 primaryClass = {astro-ph.GA},
       adsurl = {https://ui.adsabs.harvard.edu/abs/2024ApJ...967L..41M}
}

@ARTICLE{Langan:2020,
       author = {{Langan}, Ivanna and {Ceverino}, Daniel and {Finlator}, Kristian},
        title = "{Weak evolution of the mass-metallicity relation at cosmic dawn in the FirstLight simulations}",
      journal = {\mnras},
         year = 2020,
        month = may,
       volume = {494},
       number = {2},
        pages = {1988-1993},
          doi = {10.1093/mnras/staa880},
archivePrefix = {arXiv},
       eprint = {1910.11729},
 primaryClass = {astro-ph.GA},
       adsurl = {https://ui.adsabs.harvard.edu/abs/2020MNRAS.494.1988L}
}

@ARTICLE{Torrey:2019,
       author = {{Torrey}, Paul and {Vogelsberger}, Mark and {Marinacci}, Federico and {Pakmor}, R{\"u}diger and {Springel}, Volker and {Nelson}, Dylan and {Naiman}, Jill and {Pillepich}, Annalisa and {Genel}, Shy and {Weinberger}, Rainer and {Hernquist}, Lars},
        title = "{The evolution of the mass-metallicity relation and its scatter in IllustrisTNG}",
      journal = {\mnras},
         year = 2019,
        month = apr,
       volume = {484},
       number = {4},
        pages = {5587-5607},
          doi = {10.1093/mnras/stz243},
archivePrefix = {arXiv},
       eprint = {1711.05261},
 primaryClass = {astro-ph.GA},
       adsurl = {https://ui.adsabs.harvard.edu/abs/2019MNRAS.484.5587T}
}

@ARTICLE{Ucci:2023,
       author = {{Ucci}, Graziano and {Dayal}, Pratika and {Hutter}, Anne and {Kobayashi}, Chiaki and {Gottl{\"o}ber}, Stefan and {Yepes}, Gustavo and {Hunt}, Leslie and {Legrand}, Laurent and {Tortora}, Crescenzo},
        title = "{Astraeus V: the emergence and evolution of metallicity scaling relations during the epoch of reionization}",
      journal = {\mnras},
         year = 2023,
        month = jan,
       volume = {518},
       number = {3},
        pages = {3557-3575},
          doi = {10.1093/mnras/stac2654},
archivePrefix = {arXiv},
       eprint = {2112.02115},
 primaryClass = {astro-ph.GA},
       adsurl = {https://ui.adsabs.harvard.edu/abs/2023MNRAS.518.3557U}
}

@ARTICLE{Dubois:2021,
       author = {{Dubois}, Yohan and {Beckmann}, Ricarda and {Bournaud}, Fr{\'e}d{\'e}ric and {Choi}, Hoseung and {Devriendt}, Julien and {Jackson}, Ryan and {Kaviraj}, Sugata and {Kimm}, Taysun and {Kraljic}, Katarina and {Laigle}, Clotilde and {Martin}, Garreth and {Park}, Min-Jung and {Peirani}, S{\'e}bastien and {Pichon}, Christophe and {Volonteri}, Marta and {Yi}, Sukyoung K.},
        title = "{Introducing the NEWHORIZON simulation: Galaxy properties with resolved internal dynamics across cosmic time}",
      journal = {\aap},
         year = 2021,
        month = jul,
       volume = {651},
          eid = {A109},
        pages = {A109},
          doi = {10.1051/0004-6361/202039429},
archivePrefix = {arXiv},
       eprint = {2009.10578},
 primaryClass = {astro-ph.GA},
       adsurl = {https://ui.adsabs.harvard.edu/abs/2021A&A...651A.109D}
}

@ARTICLE{Pallottini:2022,
       author = {{Pallottini}, A. and {Ferrara}, A. and {Gallerani}, S. and {Behrens}, C. and {Kohandel}, M. and {Carniani}, S. and {Vallini}, L. and {Salvadori}, S. and {Gelli}, V. and {Sommovigo}, L. and {D'Odorico}, V. and {Di Mascia}, F. and {Pizzati}, E.},
        title = "{A survey of high-z galaxies: SERRA simulations}",
      journal = {\mnras},
         year = 2022,
        month = jul,
       volume = {513},
       number = {4},
        pages = {5621-5641},
          doi = {10.1093/mnras/stac1281},
archivePrefix = {arXiv},
       eprint = {2201.02636},
 primaryClass = {astro-ph.GA},
       adsurl = {https://ui.adsabs.harvard.edu/abs/2022MNRAS.513.5621P}
}

@ARTICLE{Kobayashi:2023,
       author = {{Kobayashi}, Chiaki and {Taylor}, Philip},
        title = "{Chemo-Dynamical Evolution of Galaxies}",
      journal = {arXiv e-prints},
         year = 2023,
        month = feb,
          eid = {arXiv:2302.07255},
        pages = {arXiv:2302.07255},
          doi = {10.48550/arXiv.2302.07255},
archivePrefix = {arXiv},
       eprint = {2302.07255},
 primaryClass = {astro-ph.GA},
       adsurl = {https://ui.adsabs.harvard.edu/abs/2023arXiv230207255K}
}

@ARTICLE{Arellano-Cordova:2022,
       author = {{Arellano-C{\'o}rdova}, Karla Z. and {Berg}, Danielle A. and {Chisholm}, John and {Arrabal Haro}, Pablo and {Dickinson}, Mark and {Finkelstein}, Steven L. and {Leclercq}, Floriane and {Rogers}, Noah S.~J. and {Simons}, Raymond C. and {Skillman}, Evan D. and {Trump}, Jonathan R. and {Kartaltepe}, Jeyhan S.},
        title = "{A First Look at the Abundance Pattern-O/H, C/O, and Ne/O-in z > 7 Galaxies with JWST/NIRSpec}",
      journal = {\apjl},
         year = 2022,
        month = nov,
       volume = {940},
       number = {1},
          eid = {L23},
        pages = {L23},
          doi = {10.3847/2041-8213/ac9ab2},
archivePrefix = {arXiv},
       eprint = {2208.02562},
 primaryClass = {astro-ph.GA},
       adsurl = {https://ui.adsabs.harvard.edu/abs/2022ApJ...940L..23A}
}

@ARTICLE{Arellano-Cordova:2025,
       author = {{Arellano-C{\'o}rdova}, K.~Z. and {Cullen}, F. and {Carnall}, A.~C. and {Scholte}, D. and {Stanton}, T.~M. and {Kobayashi}, C. and {Martinez}, Z. and {Berg}, D.~A. and {Barrufet}, L. and {Begley}, R. and {Donnan}, C.~T. and {Dunlop}, J.~S. and {Hamadouche}, M.~L. and {McLeod}, D.~J. and {McLure}, R.~J. and {Rowlands}, K. and {Shapley}, A.~E.},
        title = "{The JWST EXCELS survey: direct estimates of C, N, and O abundances in two relatively metal-rich galaxies at z ≃ 5}",
      journal = {\mnras},
         year = 2025,
        month = jul,
       volume = {540},
       number = {4},
        pages = {2991-3007},
          doi = {10.1093/mnras/staf855},
archivePrefix = {arXiv},
       eprint = {2412.10557},
 primaryClass = {astro-ph.GA},
       adsurl = {https://ui.adsabs.harvard.edu/abs/2025MNRAS.540.2991A}
}

@ARTICLE{Cameron:2023c,
       author = {{Cameron}, Alex J. and {Katz}, Harley and {Rey}, Martin P. and {Saxena}, Aayush},
        title = "{Nitrogen enhancements 440 Myr after the big bang: supersolar N/O, a tidal disruption event, or a dense stellar cluster in GN-z11?}",
      journal = {\mnras},
         year = 2023,
        month = aug,
       volume = {523},
       number = {3},
        pages = {3516-3525},
          doi = {10.1093/mnras/stad1579},
archivePrefix = {arXiv},
       eprint = {2302.10142},
 primaryClass = {astro-ph.GA},
       adsurl = {https://ui.adsabs.harvard.edu/abs/2023MNRAS.523.3516C}
}

@ARTICLE{Nomoto:2006,
       author = {{Nomoto}, Ken'ichi and {Tominaga}, Nozomu and {Umeda}, Hideyuki and {Kobayashi}, Chiaki and {Maeda}, Keiichi},
        title = "{Nucleosynthesis yields of core-collapse supernovae and hypernovae, and galactic chemical evolution}",
      journal = {\nphysa},
         year = 2006,
        month = oct,
       volume = {777},
        pages = {424-458},
          doi = {10.1016/j.nuclphysa.2006.05.008},
archivePrefix = {arXiv},
       eprint = {astro-ph/0605725},
 primaryClass = {astro-ph},
       adsurl = {https://ui.adsabs.harvard.edu/abs/2006NuPhA.777..424N}
}

@ARTICLE{Pilyugin:2009,
       author = {{Pilyugin}, L.~S. and {Mattsson}, L. and {V{\'\i}lchez}, J.~M. and {Cedr{\'e}s}, B.},
        title = "{On the electron temperatures in high-metallicity HII regions}",
      journal = {\mnras},
         year = 2009,
        month = sep,
       volume = {398},
       number = {1},
        pages = {485-496},
          doi = {10.1111/j.1365-2966.2009.15182.x},
archivePrefix = {arXiv},
       eprint = {0907.0084},
 primaryClass = {astro-ph.CO},
       adsurl = {https://ui.adsabs.harvard.edu/abs/2009MNRAS.398..485P}
}

@ARTICLE{Rickards-Vaught:2024,
       author = {{Rickards Vaught}, Ryan J. and {Sandstrom}, Karin M. and {Belfiore}, Francesco and {Kreckel}, Kathryn and {M{\'e}ndez-Delgado}, J. Eduardo and {Emsellem}, Eric and {Groves}, Brent and {Blanc}, Guillermo A. and {Dale}, Daniel A. and {Egorov}, Oleg V. and {Glover}, Simon C.~O. and {Grasha}, Kathryn and {Klessen}, Ralf S. and {Neumann}, Justus and {Williams}, Thomas G.},
        title = "{Investigating the Drivers of Electron Temperature Variations in H II Regions with Keck-KCWI and VLT-MUSE}",
      journal = {\apj},
         year = 2024,
        month = may,
       volume = {966},
       number = {1},
          eid = {130},
        pages = {130},
          doi = {10.3847/1538-4357/ad303c},
archivePrefix = {arXiv},
       eprint = {2309.17440},
 primaryClass = {astro-ph.GA},
       adsurl = {https://ui.adsabs.harvard.edu/abs/2024ApJ...966..130R}
}

@ARTICLE{Mendez-Delgado:2023,
       author = {{M{\'e}ndez-Delgado}, J.~E. and {Esteban}, C. and {Garc{\'\i}a-Rojas}, J. and {Arellano-C{\'o}rdova}, K.~Z. and {Kreckel}, K. and {G{\'o}mez-Llanos}, V. and {Egorov}, O.~V. and {Peimbert}, M. and {Orte-Garc{\'\i}a}, M.},
        title = "{Density biases and temperature relations for DESIRED H II regions}",
      journal = {\mnras},
         year = 2023,
        month = aug,
       volume = {523},
       number = {2},
        pages = {2952-2973},
          doi = {10.1093/mnras/stad1569},
archivePrefix = {arXiv},
       eprint = {2305.13136},
 primaryClass = {astro-ph.GA},
       adsurl = {https://ui.adsabs.harvard.edu/abs/2023MNRAS.523.2952M}
}

@ARTICLE{Endsley:2024,
       author = {{Endsley}, Ryan and {Stark}, Daniel P. and {Whitler}, Lily and {Topping}, Michael W. and {Johnson}, Benjamin D. and {Robertson}, Brant and {Tacchella}, Sandro and {Alberts}, Stacey and {Baker}, William M. and {Bhatawdekar}, Rachana and {Boyett}, Kristan and {Bunker}, Andrew J. and {Cameron}, Alex J. and {Carniani}, Stefano and {Charlot}, Stephane and {Chen}, Zuyi and {Chevallard}, Jacopo and {Curtis-Lake}, Emma and {Danhaive}, A. Lola and {Egami}, Eiichi and {Eisenstein}, Daniel J. and {Hainline}, Kevin and {Helton}, Jakob M. and {Ji}, Zhiyuan and {Looser}, Tobias J. and {Maiolino}, Roberto and {Nelson}, Erica and {Pusk{\'a}s}, D{\'a}vid and {Rieke}, George and {Rieke}, Marcia and {Rix}, Hans-Walter and {Sandles}, Lester and {Saxena}, Aayush and {Simmonds}, Charlotte and {Smit}, Renske and {Sun}, Fengwu and {Williams}, Christina C. and {Willmer}, Christopher N.~A. and {Willott}, Chris and {Witstok}, Joris},
        title = "{The star-forming and ionizing properties of dwarf z 6-9 galaxies in JADES: insights on bursty star formation and ionized bubble growth}",
      journal = {\mnras},
         year = 2024,
        month = sep,
       volume = {533},
       number = {1},
        pages = {1111-1142},
          doi = {10.1093/mnras/stae1857},
archivePrefix = {arXiv},
       eprint = {2306.05295},
 primaryClass = {astro-ph.GA},
       adsurl = {https://ui.adsabs.harvard.edu/abs/2024MNRAS.533.1111E}
}

@ARTICLE{Katz:2024b,
       author = {{Katz}, Harley and {Cameron}, Alex J. and {Saxena}, Aayush and {Barrufet}, Laia and {Choustikov}, Nichloas and {Cleri}, Nikko J. and {de Graff}, Anna and {Ellis}, Richard S. and {Fosbury}, Robert A.~E. and {Heintz}, Kasper E. and {Maseda}, Michael and {Matthee}, Jorryt and {McConachie}, Ian and {Oesch}, Pascal A.},
        title = "{21 Balmer Jump Street: The Nebular Continuum at High Redshift and Implications for the Bright Galaxy Problem, UV Continuum Slopes, and Early Stellar Populations}",
      journal = {The Open Journal of Astrophysics},
         year = 2025,
        month = jul,
       volume = {8},
          eid = {104},
        pages = {104},
          doi = {10.33232/001c.142570},
archivePrefix = {arXiv},
       eprint = {2408.03189},
 primaryClass = {astro-ph.GA},
       adsurl = {https://ui.adsabs.harvard.edu/abs/2025OJAp....8E.104K}
}

@ARTICLE{Kimm:2022,
       author = {{Kimm}, Taysun and {Bieri}, Rebekka and {Geen}, Sam and {Rosdahl}, Joakim and {Blaizot}, J{\'e}r{\'e}my and {Michel-Dansac}, L{\'e}o and {Garel}, Thibault},
        title = "{A Systematic Study of the Escape of LyC and Ly{\ensuremath{\alpha}} Photons from Star-forming, Magnetized Turbulent Clouds}",
      journal = {\apjs},
         year = 2022,
        month = mar,
       volume = {259},
       number = {1},
          eid = {21},
        pages = {21},
          doi = {10.3847/1538-4365/ac426d},
archivePrefix = {arXiv},
       eprint = {2110.02975},
 primaryClass = {astro-ph.GA},
       adsurl = {https://ui.adsabs.harvard.edu/abs/2022ApJS..259...21K}
}

@ARTICLE{Glazer:2025,
       author = {{Glazer}, Kelsey S. and {Jones}, Tucker and {Chen}, Yuguang and {Sanders}, Ryan L. and {Bradac}, Marusa and {Pahl}, Anthony J. and {Shapley}, Alice E. and {Ellis}, Richard S. and {Topping}, Michael W. and {Reddy}, Naveen A.},
        title = "{Stacking PANCAKEZ: sPectroscopic Analysis with NirspeC stAcKs in the Epoch of reioniZation. Weak ISM Absorption and Implications for Ionizing Photon Escape at $z\sim7$}",
      journal = {arXiv e-prints},
         year = 2025,
        month = apr,
          eid = {arXiv:2504.21080},
        pages = {arXiv:2504.21080},
          doi = {10.48550/arXiv.2504.21080},
archivePrefix = {arXiv},
       eprint = {2504.21080},
 primaryClass = {astro-ph.GA},
       adsurl = {https://ui.adsabs.harvard.edu/abs/2025arXiv250421080G}
}

@ARTICLE{Topping:2025b,
       author = {{Topping}, Michael W. and {Stark}, Daniel P. and {Senchyna}, Peter and {Chen}, Zuyi and {Zitrin}, Adi and {Endsley}, Ryan and {Charlot}, St{\'e}phane and {Furtak}, Lukas J. and {Maseda}, Michael V. and {Plat}, Adele and {Smit}, Renske and {Mainali}, Ramesh and {Chevallard}, Jacopo and {Molyneux}, Stephen and {Rigby}, Jane R.},
        title = "{Deep Rest-UV JWST/NIRSpec Spectroscopy of Early Galaxies: The Demographics of C IV and N-emitters in the Reionization Era}",
      journal = {\apj},
         year = 2025,
        month = feb,
       volume = {980},
       number = {2},
          eid = {225},
        pages = {225},
          doi = {10.3847/1538-4357/ada95c},
archivePrefix = {arXiv},
       eprint = {2407.19009},
 primaryClass = {astro-ph.GA},
       adsurl = {https://ui.adsabs.harvard.edu/abs/2025ApJ...980..225T}
}

@ARTICLE{Roberts-Borsani:2024,
       author = {{Roberts-Borsani}, Guido and {Treu}, Tommaso and {Shapley}, Alice and {Fontana}, Adriano and {Pentericci}, Laura and {Castellano}, Marco and {Morishita}, Takahiro and {Bergamini}, Pietro and {Rosati}, Piero},
        title = "{Between the Extremes: A JWST Spectroscopic Benchmark for High-redshift Galaxies Using {\ensuremath{\sim}}500 Confirmed Sources at z {\ensuremath{\geq}} 5}",
      journal = {\apj},
         year = 2024,
        month = dec,
       volume = {976},
       number = {2},
          eid = {193},
        pages = {193},
          doi = {10.3847/1538-4357/ad85d3},
archivePrefix = {arXiv},
       eprint = {2403.07103},
 primaryClass = {astro-ph.GA},
       adsurl = {https://ui.adsabs.harvard.edu/abs/2024ApJ...976..193R}
}

@ARTICLE{Calabro:2024,
       author = {{Calabr{\`o}}, Antonello and {Castellano}, Marco and {Zavala}, Jorge A. and {Pentericci}, Laura and {Arrabal Haro}, Pablo and {Bakx}, Tom J.~L.~C. and {Burgarella}, Denis and {Casey}, Caitlin M. and {Dickinson}, Mark and {Finkelstein}, Steven L. and {Fontana}, Adriano and {Llerena}, Mario and {Mascia}, Sara and {Merlin}, Emiliano and {Mitsuhashi}, Ikki and {Napolitano}, Lorenzo and {Paris}, Diego and {P{\'e}rez-Gonz{\'a}lez}, Pablo G. and {Roberts-Borsani}, Guido and {Santini}, Paola and {Treu}, Tommaso and {Vanzella}, Eros},
        title = "{Evidence of Extreme Ionization Conditions and Low Metallicity in GHZ2/GLASS-Z12 from a Combined Analysis of NIRSpec and MIRI Observations}",
      journal = {\apj},
         year = 2024,
        month = nov,
       volume = {975},
       number = {2},
          eid = {245},
        pages = {245},
          doi = {10.3847/1538-4357/ad7602},
archivePrefix = {arXiv},
       eprint = {2403.12683},
 primaryClass = {astro-ph.GA},
       adsurl = {https://ui.adsabs.harvard.edu/abs/2024ApJ...975..245C}
}

@ARTICLE{Mascia:2023,
       author = {{Mascia}, S. and {Pentericci}, L. and {Calabr{\`o}}, A. and {Treu}, T. and {Santini}, P. and {Yang}, L. and {Napolitano}, L. and {Roberts-Borsani}, G. and {Bergamini}, P. and {Grillo}, C. and {Rosati}, P. and {Vulcani}, B. and {Castellano}, M. and {Boyett}, K. and {Fontana}, A. and {Glazebrook}, K. and {Henry}, A. and {Mason}, C. and {Merlin}, E. and {Morishita}, T. and {Nanayakkara}, T. and {Paris}, D. and {Roy}, N. and {Williams}, H. and {Wang}, X. and {Brammer}, G. and {Brada{\v{c}}}, M. and {Chen}, W. and {Kelly}, P.~L. and {Koekemoer}, A.~M. and {Trenti}, M. and {Windhorst}, R.~A.},
        title = "{Closing in on the sources of cosmic reionization: First results from the GLASS-JWST program}",
      journal = {\aap},
         year = 2023,
        month = apr,
       volume = {672},
          eid = {A155},
        pages = {A155},
          doi = {10.1051/0004-6361/202345866},
archivePrefix = {arXiv},
       eprint = {2301.02816},
 primaryClass = {astro-ph.GA},
       adsurl = {https://ui.adsabs.harvard.edu/abs/2023A&A...672A.155M}
}

@ARTICLE{Sanders:2023,
       author = {{Sanders}, Ryan L. and {Shapley}, Alice E. and {Topping}, Michael W. and {Reddy}, Naveen A. and {Brammer}, Gabriel B.},
        title = "{Excitation and Ionization Properties of Star-forming Galaxies at z = 2.0-9.3 with JWST/NIRSpec}",
      journal = {\apj},
         year = 2023,
        month = sep,
       volume = {955},
       number = {1},
          eid = {54},
        pages = {54},
          doi = {10.3847/1538-4357/acedad},
archivePrefix = {arXiv},
       eprint = {2301.06696},
 primaryClass = {astro-ph.GA},
       adsurl = {https://ui.adsabs.harvard.edu/abs/2023ApJ...955...54S}
}

@ARTICLE{Simmonds:2024,
       author = {{Simmonds}, C. and {Tacchella}, S. and {Hainline}, K. and {Johnson}, B.~D. and {Pusk{\'a}s}, D. and {Robertson}, B. and {Baker}, W.~M. and {Bhatawdekar}, R. and {Boyett}, K. and {Bunker}, A.~J. and {Cargile}, P.~A. and {Carniani}, S. and {Chevallard}, J. and {Curti}, M. and {Curtis-Lake}, E. and {Ji}, Z. and {Jones}, G.~C. and {Kumari}, N. and {Laseter}, I. and {Maiolino}, R. and {Maseda}, M.~V. and {Rinaldi}, P. and {Stoffers}, A. and {{\"U}bler}, H. and {Villanueva}, N.~C. and {Williams}, C.~C. and {Willott}, C. and {Witstok}, J. and {Zhu}, Y.},
        title = "{Ionizing properties of galaxies in JADES for a stellar mass complete sample: resolving the cosmic ionizing photon budget crisis at the Epoch of Reionization}",
      journal = {\mnras},
         year = 2024,
        month = dec,
       volume = {535},
       number = {4},
        pages = {2998-3019},
          doi = {10.1093/mnras/stae2537},
archivePrefix = {arXiv},
       eprint = {2409.01286},
 primaryClass = {astro-ph.GA},
       adsurl = {https://ui.adsabs.harvard.edu/abs/2024MNRAS.535.2998S}
}

@ARTICLE{Atek:2024,
       author = {{Atek}, Hakim and {Labb{\'e}}, Ivo and {Furtak}, Lukas J. and {Chemerynska}, Iryna and {Fujimoto}, Seiji and {Setton}, David J. and {Miller}, Tim B. and {Oesch}, Pascal and {Bezanson}, Rachel and {Price}, Sedona H. and {Dayal}, Pratika and {Zitrin}, Adi and {Kokorev}, Vasily and {Weaver}, John R. and {Brammer}, Gabriel and {Dokkum}, Pieter van and {Williams}, Christina C. and {Cutler}, Sam E. and {Feldmann}, Robert and {Fudamoto}, Yoshinobu and {Greene}, Jenny E. and {Leja}, Joel and {Maseda}, Michael V. and {Muzzin}, Adam and {Pan}, Richard and {Papovich}, Casey and {Nelson}, Erica J. and {Nanayakkara}, Themiya and {Stark}, Daniel P. and {Stefanon}, Mauro and {Suess}, Katherine A. and {Wang}, Bingjie and {Whitaker}, Katherine E.},
        title = "{Most of the photons that reionized the Universe came from dwarf galaxies}",
      journal = {\nat},
         year = 2024,
        month = feb,
       volume = {626},
       number = {8001},
        pages = {975-978},
          doi = {10.1038/s41586-024-07043-6},
archivePrefix = {arXiv},
       eprint = {2308.08540},
 primaryClass = {astro-ph.GA},
       adsurl = {https://ui.adsabs.harvard.edu/abs/2024Natur.626..975A}
}

@ARTICLE{Simmonds:2024b,
       author = {{Simmonds}, C. and {Tacchella}, S. and {Hainline}, K. and {Johnson}, B.~D. and {McClymont}, W. and {Robertson}, B. and {Saxena}, A. and {Sun}, F. and {Witten}, C. and {Baker}, W.~M. and {Bhatawdekar}, R. and {Boyett}, K. and {Bunker}, A.~J. and {Charlot}, S. and {Curtis-Lake}, E. and {Egami}, E. and {Eisenstein}, D.~J. and {Hausen}, R. and {Maiolino}, R. and {Maseda}, M.~V. and {Scholtz}, J. and {Williams}, C.~C. and {Willott}, C. and {Witstok}, J.},
        title = "{Low-mass bursty galaxies in JADES efficiently produce ionizing photons and could represent the main drivers of reionization}",
      journal = {\mnras},
         year = 2024,
        month = jan,
       volume = {527},
       number = {3},
        pages = {6139-6157},
          doi = {10.1093/mnras/stad3605},
archivePrefix = {arXiv},
       eprint = {2310.01112},
 primaryClass = {astro-ph.GA},
       adsurl = {https://ui.adsabs.harvard.edu/abs/2024MNRAS.527.6139S}
}

@ARTICLE{Laseter:2025,
       author = {{Laseter}, Isaac H. and {Maseda}, Michael V. and {Simmonds}, Charlotte and {Endsley}, Ryan and {Stark}, Daniel and {Bunker}, Andrew J. and {Bhatawdekar}, Rachana and {Boyett}, Kristan and {Cameron}, Alex J. and {Carniani}, Stefano and {Curti}, Mirko and {Ji}, Zhiyuan and {Rinaldi}, Pierluigi and {Saxena}, Aayush and {Tacchella}, Sandro and {Willott}, Chris and {Witstok}, Joris and {Zhu}, Yongda},
        title = "{Efficient Ionizers with Low H{\ensuremath{\beta}} + [O III] Equivalent Widths: JADES Spectroscopy of a Peculiar High-redshift Population}",
      journal = {\apj},
         year = 2025,
        month = jul,
       volume = {988},
       number = {1},
          eid = {73},
        pages = {73},
          doi = {10.3847/1538-4357/adddb5},
archivePrefix = {arXiv},
       eprint = {2412.04542},
 primaryClass = {astro-ph.GA},
       adsurl = {https://ui.adsabs.harvard.edu/abs/2025ApJ...988...73L}
}

@ARTICLE{Heintz:2024,
       author = {{Heintz}, Kasper E. and {Watson}, Darach and {Brammer}, Gabriel and {Vejlgaard}, Simone and {Hutter}, Anne and {Strait}, Victoria B. and {Matthee}, Jorryt and {Oesch}, Pascal A. and {Jakobsson}, P{\'a}ll and {Tanvir}, Nial R. and {Laursen}, Peter and {Naidu}, Rohan P. and {Mason}, Charlotte A. and {Killi}, Meghana and {Jung}, Intae and {Hsiao}, Tiger Yu-Yang and {Abdurro'uf} and {Coe}, Dan and {Arrabal Haro}, Pablo and {Finkelstein}, Steven L. and {Toft}, Sune},
        title = "{Strong damped Lyman-{\ensuremath{\alpha}} absorption in young star-forming galaxies at redshifts 9 to 11}",
      journal = {Science},
         year = 2024,
        month = may,
       volume = {384},
       number = {6698},
        pages = {890-894},
          doi = {10.1126/science.adj0343},
archivePrefix = {arXiv},
       eprint = {2306.00647},
 primaryClass = {astro-ph.GA},
       adsurl = {https://ui.adsabs.harvard.edu/abs/2024Sci...384..890H}
}

@ARTICLE{Marques-Chaves:2024,
       author = {{Marques-Chaves}, R. and {Schaerer}, D. and {Kuruvanthodi}, A. and {Korber}, D. and {Prantzos}, N. and {Charbonnel}, C. and {Weibel}, A. and {Izotov}, Y.~I. and {Messa}, M. and {Brammer}, G. and {Dessauges-Zavadsky}, M. and {Oesch}, P.},
        title = "{Extreme N-emitters at high redshift: Possible signatures of supermassive stars and globular cluster or black hole formation in action}",
      journal = {\aap},
         year = 2024,
        month = jan,
       volume = {681},
          eid = {A30},
        pages = {A30},
          doi = {10.1051/0004-6361/202347411},
archivePrefix = {arXiv},
       eprint = {2307.04234},
 primaryClass = {astro-ph.GA},
       adsurl = {https://ui.adsabs.harvard.edu/abs/2024A&A...681A..30M}
}

@ARTICLE{Nakajima:2025,
       author = {{Nakajima}, Kimihiko and {Ouchi}, Masami and {Harikane}, Yuichi and {Vanzella}, Eros and {Ono}, Yoshiaki and {Isobe}, Yuki and {Nishigaki}, Moka and {Tsujimoto}, Takuji and {Nakamura}, Fumitaka and {Xu}, Yi and {Umeda}, Hiroya and {Zhang}, Yechi},
        title = "{An Ultra-Faint, Chemically Primitive Galaxy Forming at the Epoch of Reionization}",
      journal = {arXiv e-prints},
         year = 2025,
        month = jun,
          eid = {arXiv:2506.11846},
        pages = {arXiv:2506.11846},
          doi = {10.48550/arXiv.2506.11846},
archivePrefix = {arXiv},
       eprint = {2506.11846},
 primaryClass = {astro-ph.GA},
       adsurl = {https://ui.adsabs.harvard.edu/abs/2025arXiv250611846N}
}

@ARTICLE{Tang:2025b,
       author = {{Tang}, Mengtao and {Stark}, Daniel P. and {Plat}, Ad{\`e}le and {Feltre}, Anna and {Katz}, Harley and {Senchyna}, Peter and {Mason}, Charlotte A. and {Whitler}, Lily and {Chen}, Zuyi and {Topping}, Michael W.},
        title = "{JWST/NIRSpec Observations of High Ionization Emission Lines in Galaxies at High Redshift}",
      journal = {arXiv e-prints},
         year = 2025,
        month = may,
          eid = {arXiv:2505.06359},
        pages = {arXiv:2505.06359},
          doi = {10.48550/arXiv.2505.06359},
archivePrefix = {arXiv},
       eprint = {2505.06359},
 primaryClass = {astro-ph.GA},
       adsurl = {https://ui.adsabs.harvard.edu/abs/2025arXiv250506359T}
}

@ARTICLE{Steidel:2014,
       author = {{Steidel}, Charles C. and {Rudie}, Gwen C. and {Strom}, Allison L. and {Pettini}, Max and {Reddy}, Naveen A. and {Shapley}, Alice E. and {Trainor}, Ryan F. and {Erb}, Dawn K. and {Turner}, Monica L. and {Konidaris}, Nicholas P. and {Kulas}, Kristin R. and {Mace}, Gregory and {Matthews}, Keith and {McLean}, Ian S.},
        title = "{Strong Nebular Line Ratios in the Spectra of z \raisebox{-0.5ex}\textasciitilde 2-3 Star Forming Galaxies: First Results from KBSS-MOSFIRE}",
      journal = {\apj},
         year = 2014,
        month = nov,
       volume = {795},
       number = {2},
          eid = {165},
        pages = {165},
          doi = {10.1088/0004-637X/795/2/165},
archivePrefix = {arXiv},
       eprint = {1405.5473},
 primaryClass = {astro-ph.GA},
       adsurl = {https://ui.adsabs.harvard.edu/abs/2014ApJ...795..165S}
}

@ARTICLE{Sanders:2016,
       author = {{Sanders}, Ryan L. and {Shapley}, Alice E. and {Kriek}, Mariska and {Reddy}, Naveen A. and {Freeman}, William R. and {Coil}, Alison L. and {Siana}, Brian and {Mobasher}, Bahram and {Shivaei}, Irene and {Price}, Sedona H. and {de Groot}, Laura},
        title = "{The MOSDEF Survey: Electron Density and Ionization Parameter at z \raisebox{-0.5ex}\textasciitilde 2.3}",
      journal = {\apj},
         year = 2016,
        month = jan,
       volume = {816},
       number = {1},
          eid = {23},
        pages = {23},
          doi = {10.3847/0004-637X/816/1/23},
archivePrefix = {arXiv},
       eprint = {1509.03636},
 primaryClass = {astro-ph.GA},
       adsurl = {https://ui.adsabs.harvard.edu/abs/2016ApJ...816...23S}
}

@ARTICLE{Erb:2006,
       author = {{Erb}, Dawn K. and {Shapley}, Alice E. and {Pettini}, Max and {Steidel}, Charles C. and {Reddy}, Naveen A. and {Adelberger}, Kurt L.},
        title = "{The Mass-Metallicity Relation at z>\raisebox{-0.5ex}\textasciitilde2}",
      journal = {\apj},
         year = 2006,
        month = jun,
       volume = {644},
       number = {2},
        pages = {813-828},
          doi = {10.1086/503623},
archivePrefix = {arXiv},
       eprint = {astro-ph/0602473},
 primaryClass = {astro-ph},
       adsurl = {https://ui.adsabs.harvard.edu/abs/2006ApJ...644..813E}
}

@ARTICLE{Hainline:2009,
       author = {{Hainline}, Kevin N. and {Shapley}, Alice E. and {Kornei}, Katherine A. and {Pettini}, Max and {Buckley-Geer}, Elizabeth and {Allam}, Sahar S. and {Tucker}, Douglas L.},
        title = "{Rest-Frame Optical Spectra of Three Strongly Lensed Galaxies at z \raisebox{-0.5ex}\textasciitilde 2}",
      journal = {\apj},
         year = 2009,
        month = aug,
       volume = {701},
       number = {1},
        pages = {52-65},
          doi = {10.1088/0004-637X/701/1/52},
archivePrefix = {arXiv},
       eprint = {0906.2197},
 primaryClass = {astro-ph.CO},
       adsurl = {https://ui.adsabs.harvard.edu/abs/2009ApJ...701...52H}
}

@ARTICLE{Berg:2022,
       author = {{Berg}, Danielle A. and {James}, Bethan L. and {King}, Teagan and {McDonald}, Meaghan and {Chen}, Zuyi and {Chisholm}, John and {Heckman}, Timothy and {Martin}, Crystal L. and {Stark}, Dan P. and {Aloisi}, Alessandra and {Amor{\'\i}n}, Ricardo O. and {Arellano-C{\'o}rdova}, Karla Z. and {Bayliss}, Matthew and {Bordoloi}, Rongmon and {Brinchmann}, Jarle and {Charlot}, St{\'e}phane and {Chevallard}, Jacopo and {Clark}, Ilyse and {Erb}, Dawn K. and {Feltre}, Anna and {Gronke}, Max and {Hayes}, Matthew and {Henry}, Alaina and {Hernandez}, Svea and {Jaskot}, Anne and {Jones}, Tucker and {Kewley}, Lisa J. and {Kumari}, Nimisha and {Leitherer}, Claus and {Llerena}, Mario and {Maseda}, Michael and {Mingozzi}, Matilde and {Nanayakkara}, Themiya and {Ouchi}, Masami and {Plat}, Adele and {Pogge}, Richard W. and {Ravindranath}, Swara and {Rigby}, Jane R. and {Sanders}, Ryan and {Scarlata}, Claudia and {Senchyna}, Peter and {Skillman}, Evan D. and {Steidel}, Charles C. and {Strom}, Allison L. and {Sugahara}, Yuma and {Wilkins}, Stephen M. and {Wofford}, Aida and {Xu}, Xinfeng and {Classy Team}},
        title = "{The COS Legacy Archive Spectroscopy Survey (CLASSY) Treasury Atlas}",
      journal = {\apjs},
         year = 2022,
        month = aug,
       volume = {261},
       number = {2},
          eid = {31},
        pages = {31},
          doi = {10.3847/1538-4365/ac6c03},
archivePrefix = {arXiv},
       eprint = {2203.07357},
 primaryClass = {astro-ph.GA},
       adsurl = {https://ui.adsabs.harvard.edu/abs/2022ApJS..261...31B}
}

@ARTICLE{Isobe:2023,
       author = {{Isobe}, Yuki and {Ouchi}, Masami and {Nakajima}, Kimihiko and {Harikane}, Yuichi and {Ono}, Yoshiaki and {Xu}, Yi and {Zhang}, Yechi and {Umeda}, Hiroya},
        title = "{Redshift Evolution of Electron Density in the Interstellar Medium at z   0-9 Uncovered with JWST/NIRSpec Spectra and Line-spread Function Determinations}",
      journal = {\apj},
         year = 2023,
        month = oct,
       volume = {956},
       number = {2},
          eid = {139},
        pages = {139},
          doi = {10.3847/1538-4357/acf376},
archivePrefix = {arXiv},
       eprint = {2301.06811},
 primaryClass = {astro-ph.GA},
       adsurl = {https://ui.adsabs.harvard.edu/abs/2023ApJ...956..139I}
}

@ARTICLE{James:2014,
       author = {{James}, Bethan L. and {Pettini}, Max and {Christensen}, Lise and {Auger}, Matthew W. and {Becker}, George D. and {King}, Lindsay J. and {Quider}, Anna M. and {Shapley}, Alice E. and {Steidel}, Charles C.},
        title = "{Testing metallicity indicators at z {\ensuremath{\sim}} 1.4 with the gravitationally lensed galaxy CASSOWARY 20}",
      journal = {\mnras},
         year = 2014,
        month = may,
       volume = {440},
       number = {2},
        pages = {1794-1809},
          doi = {10.1093/mnras/stu287},
archivePrefix = {arXiv},
       eprint = {1311.5092},
 primaryClass = {astro-ph.GA},
       adsurl = {https://ui.adsabs.harvard.edu/abs/2014MNRAS.440.1794J}
}

@ARTICLE{Reddy:2023,
       author = {{Reddy}, Naveen A. and {Topping}, Michael W. and {Sanders}, Ryan L. and {Shapley}, Alice E. and {Brammer}, Gabriel},
        title = "{A JWST/NIRSpec Exploration of the Connection between Ionization Parameter, Electron Density, and Star-formation-rate Surface Density in z = 2.7-6.3 Galaxies}",
      journal = {\apj},
         year = 2023,
        month = aug,
       volume = {952},
       number = {2},
          eid = {167},
        pages = {167},
          doi = {10.3847/1538-4357/acd754},
archivePrefix = {arXiv},
       eprint = {2303.11397},
 primaryClass = {astro-ph.GA},
       adsurl = {https://ui.adsabs.harvard.edu/abs/2023ApJ...952..167R}
}

@ARTICLE{Li:2025,
       author = {{Li}, Sijia and {Wang}, Xin and {Chen}, Yuguang and {Jones}, Tucker and {Treu}, Tommaso and {Glazebrook}, Karl and {He}, Xianlong and {Henry}, Alaina and {Meng}, Xiao-Lei and {Morishita}, Takahiro and {Roberts-Borsani}, Guido and {Yang}, Lilan and {Yu}, Hao-Ran and {Calabr{\`o}}, Antonello and {Castellano}, Marco and {Leethochawalit}, Nicha and {Metha}, Benjamin and {Nanayakkara}, Themiya and {Roy}, Namrata and {Vulcani}, Benedetta},
        title = "{Early Results from GLASS-JWST. XXV. Electron Density in the Interstellar Medium at 0.7 {\ensuremath{\lesssim}} z {\ensuremath{\lesssim}} 9.3 with NIRSpec High-resolution Spectroscopy}",
      journal = {\apjl},
         year = 2025,
        month = jan,
       volume = {979},
       number = {1},
          eid = {L13},
        pages = {L13},
          doi = {10.3847/2041-8213/ad9eac},
archivePrefix = {arXiv},
       eprint = {2412.08382},
 primaryClass = {astro-ph.GA},
       adsurl = {https://ui.adsabs.harvard.edu/abs/2025ApJ...979L..13L}
}

\appendix

\section{A. Aperture Effects in Predicted Emission Line Luminosities}
\label{app:rvir}

\begin{figure}
    \includegraphics[width=\textwidth,trim={0.0cm 0cm 0.0cm 0cm},clip]{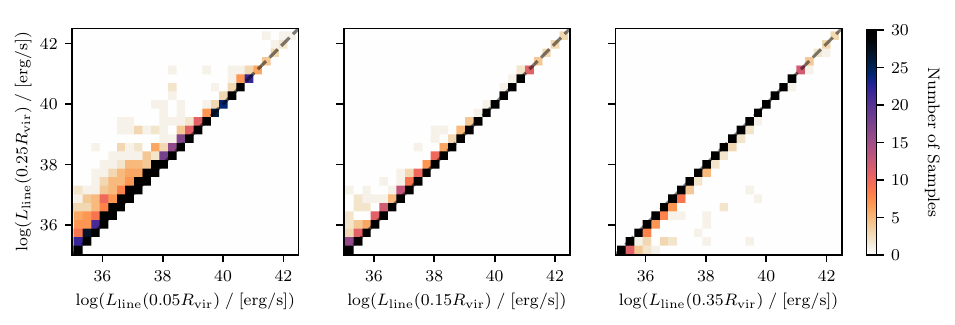}
    \caption{Intrinsic line luminosities computed for the 500 UV-brightest galaxies in the efficient star formation run at $z\sim 8.5$ computed within various regions. We compare a sphere of radius $0.25R_{\rm vir}$ (our fiducial), to those with radii of $0.05R_{\rm vir}$, $0.15R_{\rm vir}$, and $0.35R_{\rm vir}$. We find that line luminosities are well-converged at our chosen radius, and that a larger radius begins to include the presence of satellite galaxies.}
    \label{fig:rvir_cut}
\end{figure}

In order to predict the intrinsic spectrum of each galaxy, we have considered all gas cells within the central 25\% of the virial radius for each halo. This is designed to capture the bulk of the ISM in such a way as to exclude contributions from satellite galaxies. Furthermore, as we are not including the effects of scattering on resonant lines (e.g. $\rm Ly\alpha$, $\rm C~{\small IV}~\lambda\lambda1548,1550$), we can further ignore the impact of the CGM \citep[e.g.][]{Blaizot:2025}. However, doing so can underestimate contributions to emission lines from fluorescence in the CGM \citep[e.g.][]{O'Sullivan:2020}, or miss the presence of extended halos (e.g. for $\rm [C~{\small II}]$) which have been observed in the high-redshift Universe \citep{Fujimoto:2019,Meyer:2022}. In order to test the validity of this approach, Figure \ref{fig:rvir_cut} shows 2D distributions of various line luminosities (including $\rm N~{\small IV}]~\lambda\lambda1485$, $\rm C~{\small IV}~\lambda\lambda1549$, $\rm He~{\small II}~\lambda1640$, $\rm O~{\small III}~\lambda\lambda1663$, $\rm N~{\small II}~\lambda\lambda1750$, $\rm C~{\small III}]~\lambda\lambda1907$, $\rm [O~{\small II}]~\lambda\lambda3727$, $\rm [Ne~{\small III}]~\lambda\lambda3968$, $\rm [O~{\small III}]~\lambda\lambda4363$, $\rm [O~{\small III}]~\lambda\lambda5007$, $\rm H\alpha$, and $\rm [O~{\small II}]~\lambda\lambda7325$) for the 500 UV-brightest galaxies in the efficient star formation simulation at $z\sim8.5$ computed within a sphere of radius 0.25$R_{\rm vir}$, compared to those measured within radii of 0.05$R_{\rm vir}$, 0.15$R_{\rm vir}$, and 0.35$R_{\rm vir}$. We find that in each case, increasing the size of the considered region can increase line luminosities (shown by a tail on the side of the larger radius). However, line fluxes are well converged for luminous lines by 0.25$R_{\rm vir}$). Furthermore, if a larger radius is used (e.g. 0.35$R_{\rm vir}$), then emission from satellite galaxies begins to be included. Therefore, we have chosen to use 0.25$R_{\rm vir}$.

\section{B. Gas-phase Metallicity Recovery with the Direct Method} 
\label{app:direct_method}

\begin{figure*}
    \includegraphics[width=\textwidth,trim={0.0cm 0cm 0.0cm 0cm},clip]{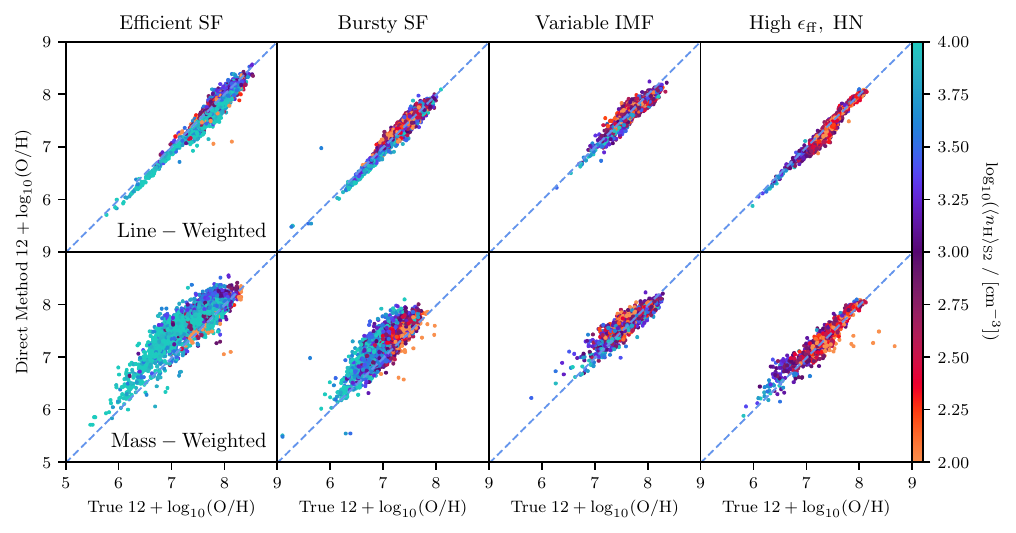}
    \caption{Recovery test for gas-phase oxygen abundance inferred through the direct method \citep[e.g.][]{Peimbert:1967,Cameron:2023} compared to the line-weighted (top) and mass-weighted (bottom) measured for each galaxy in the simulation. We find that generally, the direct method performs well at recovering the line-weighted oxygen abundance. In contrast, in both the efficient and bursty star formation simulations, the direct method tends to over-predict the mass-weighted oxygen abundance of each galaxy. This is primarily due to fluctuations in the temperature, density and abundance structure of the ISM.}
    \label{fig:direct_met}
\end{figure*}

In Figure~\ref{fig:direct_met}, we show the gas-phase metallicities inferred for galaxies in each \megatron\ simulation as a function of the line-weighted metallicity (top) and mass-weighted metallicity (bottom) taken from the simulation. We find that the direct method traces the line-weighted values very well, with residuals correlating with the S2-weighted gas density of each galaxy. This is due to the fact that under-estimating the density of the system tends to under-predict the overall abundance of the gas. In contrast, we find that the direct method tends to significantly over-predict oxygen abundances in galaxies with a denser ISM. This is due to the fact that in these systems, emission lines are particularly biased to dense, star-forming regions which may be unrepresentative of the galaxy as a whole.

\section{C. Stellar Mass Recovery with \texttt{BAGPIPES}}
\label{app:bagpipes}

\begin{figure*}
    \includegraphics[width=\textwidth,trim={0.0cm 0cm 0.0cm 0cm},clip]{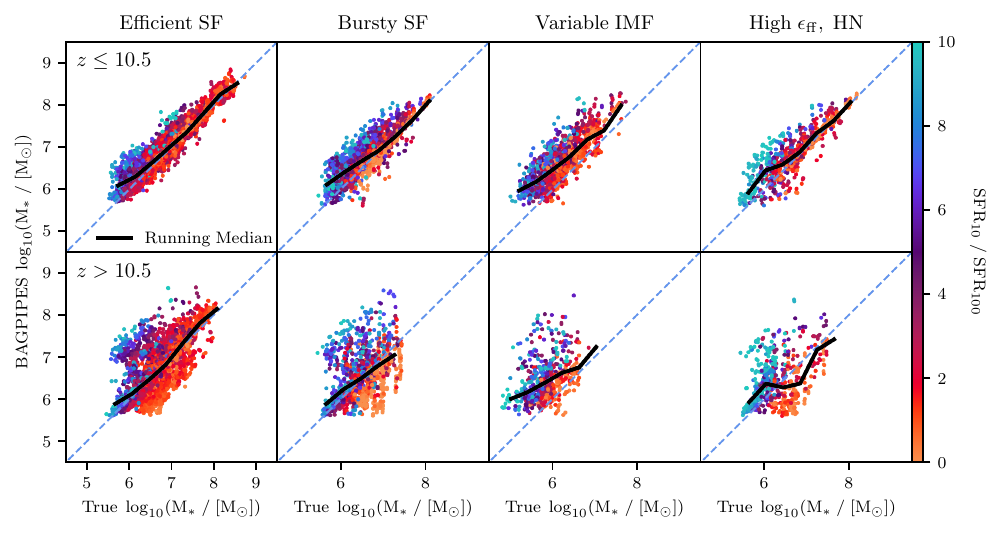}
    \caption{Recovery test for stellar masses inferred through SED-fitting by \texttt{BAGPIPES} \citep{bagpipes} compared to their values taken from each simulation in two redshift bins. Each point is coloured by the burstiness in the recent star formation history given by ${\rm SFH}_{10}/{\rm SFH}_{100}$. We also provide running medians and one-to-one relations in black and blue, respectively. We find that at lower redshifts, masses are recovered fairly well while at higher redshifts this approach struggles, due to the fact that the Balmer jump has shifted out of the last filter used. In all cases, we find that galaxy with more bursty recent star formation tend to have over-predicted stellar masses.}
    \label{fig:bagpipes_mass}
\end{figure*}

In Figure~\ref{fig:bagpipes_mass}, we show recovery tests for the stellar masses of \megatron\ galaxies, as measured using \texttt{BAGPIPES} \citep{bagpipes}. In particular, we show predicted stellar masses as a function of true (after mass loss) stellar masses for each simulation, coloured by $\rm SFR_{10} / SFR_{100}$ (an indicator of burstiness). We find that below $z \lesssim10.5$, \texttt{BAGPIPES} performs well, although it tends to over-predict the masses of the least massive galaxies. Above $z\gtrsim10.5$, it tends to struggle considerably more, with the residuals correlating with burstiness. This is primarily due to the fact that at these redshifts, the Balmer break/jump redshifts out of the red-most filter used -- suggesting that the inclusion of MIRI bands may help to alleviate this issue. 

\end{document}